# Space gravitational wave detection: Progress and outlook

Wei-Tou Ni[1,2*]

[0] *International Centre for Theoretical Physics Asia-Pacific, University of Chinese Academy of Sciences, Beijing 100190, China;*
[2] *Innovation Academy of Precision Measurement Science and Technology (APM), Wuhan Institute of Physics and Mathematics, Chinese Academy of Sciences, Wuhan 430071, China*

*Corresponding author (email: wei-tou.ni@wipm.ac.cn)

(*Original extended English abstract*) Space gravitational wave (GW) detection is to detect and measure the distance change between spacecraft/celestial bodies or status change intra spacecraft/celestial body according to the astrodynamical equations of general relativity or a specific gravitational theory. The basic method is using electromagnetic waves (including radio, microwave, light, X-ray, γ-ray, etc.) to Doppler track the spacecraft/celestial body and compare them with the two stable frequency standards (sources) at the emission end and the receiving end, e.g. microwave Doppler tracking, optical clock Doppler tracking, atom-interferometry GW detection, laser-interferomatic GW detection. If the emission phases of the electromagnetic waves are unknown, the statistic Doppler tracking method can be used as in the Pulsar Timing Arrays. If the frequency standards at the emission and the receiving ends are not stable enough in the desired detection frequency band, then it is necessary to exploit the generalized Michelson interferometry based on two paths each consisting of multi-segments of Doppler tracking. In this case, the phase (length metrology) noise at the combination end is proportional to the product of laser source frequency noise times the pathlength difference of the two paths, and the two paths need to be carefully designed and evaluated. Each set of two paths is called a TDI (Time-Delay Interferomerty) configuration. The study of TDI configurations together with the orbit design and the noise requirement at each optical link and the final spacecraft is called TDI interferometry. The final mission products for scientists to use are TDI phase (range) sequences/spectra. These products are also useful for other gravity measurements or testing specific gravitational theories, e.g. measuring gravitomagnetic effects. The current projects under construction and/or study are mainly using this method of generalized Michelson laser-interferometry which includes AMIGO (Astrodynamical Middle-frequency Gravitational Observatory), BBO (Big Bang Observer), B-DECIGO, DECIGO (Deci-Hertz Gravitational Observatory) and DO (Deci-Hertz Observatory) in middle frequency band (0.1−10 Hz), LISA (Laser-Interferometric Space Antenna) and TAIJI/TianQ in the mHz low frequency band (0.1−100 mHz), and ASTROD-GW (Astrodynamical Space Test of Relativity using Optical Devices dedicated for Gravitational wave detection), Folkner's mission, LISAmax, μAries and Super-ASTROD (0.1− 100 μHz) in the μHz low-frequency band. In this article, we review the current status quo of these space detection methods and present an outlook.

**doi:** 10.1360/SSPMA-2024-0186 (Open Access)

**CONTENTS of Space gravitational detection: Progress and outlook**

**Translation of the Chinese Abstract**

Space-based gravitational wave detection is based on the astrodynamical equations derived from gravitational theory to detect changes in distance between spacecraft/celestial bodies and/or their state changes caused by gravitational waves. The fundamental method involves using electromagnetic waves (including radio waves, microwaves, light waves, X-rays, gamma rays, etc.) for Doppler tracking and comparing to the stable frequency standards (sources) at both the transmitting and receiving ends. Examples include microwave Doppler tracking, optical clock gravitational wave detection, atom interferometry gravitational wave detection, and laser interferometry gravitational wave detection. If the frequency sources at both ends are not sufficiently stable, a generalized dual-path Michelson interferometer based on Doppler tracking combinations is needed. Currently, the main space-based gravitational wave detectors under construction or planning are laser interferometers, which cover medium frequency (0.1-10 Hz) and low-frequency (millihertz 0.1-100 mHz and microhertz 0.1-100 μHz) gravitational wave detection bands. This article reviews the current status and prospects of these gravitational wave detection methods.

## 1. Introduction

Space laser interferometric gravitational wave detectors are fundamentally based on the generalized Michelson interferometer principle. Michelson interferometry splits the wave front of a light beam into two parts, creating two beams that travel along different paths and interfere when they recombine at a same location. Since its invention in the nineteenth century, Michelson interferometry has been a highly precise measurement method, requiring that the path length difference between the two paths be less than the coherence length. Gravitational waves affect the optical paths differently, and space laser interferometric gravitational wave detection utilizes Michelson interferometry to detect this path difference to measure the incoming gravitational waves.

This article reviews and discusses the sensitivity required for space gravitational wave detection using generalized Michelson interferometry, as well as the demands on path (orbit) design and core noise (laser metrology and inertial sensing). It also explores the potential for detecting various gravitational wave sources. Space-based gravitational wave detection essentially measures the changes in distance between spacecraft/celestial bodies caused by gravitational waves arriving in the solar system. To distinguish the effects of various factors on distance changes, astronomical dynamical equations are used, with the measurements implemented by space missions focused on astronomical dynamics.

Currently, the main space-based gravitational wave detectors under construction or planning are laser interferometers using generalized Michelson interferometry, covering medium frequency (0.1-10 Hz) and low-frequency (millihertz 0.1-100 mHz and microhertz 0.1-100 μHz) gravitational wave detection bands. Generalized Michelson interferometry can be considered a combination of Doppler tracking methods. Microwave Doppler tracking has been realized and constrains the strength of arriving gravitational waves. The basic principle behind detecting very low-frequency (nanohertz 0.3-100 nHz) background gravitational waves with pulsar timing arrays is based on statistical Doppler tracking methods.

Optical clock gravitational wave detection and atom interferometry gravitational wave detection methods have development and competitive potential, as their principles are also based on Doppler tracking. Space starlight interferometers have the potential to detect gravitational waves too. This article briefly reviews and discusses the current research status and prospects of these gravitational wave detection methods.

### 1.1 Historical Background

The detection of gravitational waves is inseparable from the development of astronomical observation and astrometry. Before Galileo Galilei (1564-1642) [1] and others used telescopes to observe celestial bodies and stars in 1609, humans observed the positions and movements of celestial bodies with the naked eye and mechanical instruments (such as armillas (armillary spheres) and simplified armillas). Guo Shoujing (1231-1316) invented twelve new instruments, including the simplified armilla and the giant gnomon, to observe celestial phenomena. He also participated in the compilation of the "Shoushi Calendar" (1276-1280), edited by Xu Heng (1209-1281) et al., completing an accurate and applicable calendar of the time ([Ming] Song Lian. Yuan History · Astronomical Records [2]). Tycho Brahe (1546-1601) established Uraniborg and Stjerneborg observatories on the island of Hven in the Øresund and continuously improved and invented many new instruments for long-term observation of the stars [3]. His assistant, Johannes Kepler, used this observational data to develop the three laws of planetary motion (Kepler's laws) [4]. Kepler's laws, Galileo's law of motion on inclined planes [5], and the development of astronomy in the first half of the 17th century led to the formulation of Newton's three laws and his system of the world. During this period, Cassini and Picard [6] in Paris in 1671, and Richer in South America measured the distance to Mars using triangulation, determining the scale of the solar system. In 1676, Rømer discovered the finite speed of light by observing the time differences of Io's eclipses at different times and calculating the changes in distance between Jupiter and Earth, thus estimating the speed of light [7]. This was a highlight in the history of astrometry.

The completion of Newton's system of the world [8] and the extensive astronomical observations of the 17th and 18th centuries further propelled the development of astronomy. The discovery and study of solar spectrum lines in the early 19th century marked the beginning of astrophysics [9]. In the field of astrometry during this period: Halley predicted in 1705 that the great comet observed in 1680 would return in 1758. Indeed, on December 25, 1758, German amateur astronomer Johann Georg Palitzsch observed its return [10]. This comet was later named Halley's Comet, which will next return in 2061. In 1781, William Herschel discovered Uranus. Continued observations revealed discrepancies between its orbit and Newtonian mechanics calculations. In 1846, Le Verrier predicted the presence of another planet causing orbital perturbations at a specific location. That same year, German astronomers Galle and d'Arrest discovered a new planet within one degree of the predicted position, which was named Neptune [10]. This was the peak of Newtonian system development. In 1859, Le Verrier discovered that Mercury's perihelion advanced at a rate of 38" per century faster than calculated by Newton's law of universal gravitation [11,12]. This discrepancy, known as the anomaly in Mercury's perihelion precession, prompted many new gravitational theories in the latter half of the 19th century [12].

**1.2 The Difference Between the Gravitational Wave Amplitude Predicted by General Relativity and the Detectable Precision of the Technology at That Time Historical Background**

In 1915, Einstein's General Theory of Relativity ultimately explained the anomaly in Mercury's perihelion precession [12-14], and theoretically predicted the existence and radiation intensity of gravitational waves [14,15]. However, at that time, Einstein estimated the intensity of gravitational wave radiation to be extremely small compared to the detectable levels, asserting that while gravitational wave radiation exists, it would be technically impossible to measure it.

In 1910, the astronomical community determined the size and mass of exceptionally dim white dwarfs [16,17]. Binary white dwarfs in the Milky Way galaxy can produce millihertz gravitational waves, and the foreground radiation formed by the superposition of these waves is a significant target for current space-based gravitational wave detection. This also represented an existing source of gravitational waves at the time. The characteristic strain caused by this radiation in the solar system would be approximately $10^{-19}$–$10^{-20}$, resulting in a millihertz variation of about $10^{-19}$–$10^{-20}$ radians for celestial bodies from our line of sight to the solar system. At that time, astronomical observations did not employ adaptive optics, and angle observations were limited by atmospheric seeing, approximately 1 arcsecond (about $5\times10^{-6}$ radians), a difference of 14-15 orders of magnitude from the detectable precision. Thus, it was indeed technically impossible to measure gravitational waves in the 1910s [18].

After a century of technological development and persistent efforts by the gravitational wave research community, including Weber's experiments using resonant bars to detect gravitational waves, the LIGO and Virgo teams finally detected gravitational waves from a black hole merger in 2015 using the LIGO detectors with 4 km arms [19,20]. Subsequently, in 2017, gravitational waves from a neutron star merger were detected using the LIGO and Virgo detectors, with the Virgo detector having 3 km arms [21]. The strain amplitude was around $10^{-20}$.

In 1915, the precision of measuring strain with a Michelson interferometer was about 10 ppm, which was 15 orders of magnitude less precise than the necessary detection level. Due to the influence of ground and atmospheric mass variations and vibrations, detecting millihertz gravitational waves on the ground or underground is almost impossible. To detect millihertz gravitational waves produced by binary white dwarfs or gravitational waves from supermassive black hole mergers, it is necessary to conduct observations in space.

**1.3 The Origin of Space Gravitational Wave Detection**

In 1957, the Soviet Union launched humanity's first satellite, marking the beginning of the space age. The invention of lasers in the 1960s, along with the implementation of satellite laser ranging and lunar laser ranging during the same decade, allowed for astronomical ranging precision to reach meter-level and later centimeter-level, millimeter-level, and currently progressing towards sub-millimeter accuracy [22,23]. These advancements, combined with developments in radio and microwave ranging, significantly improved the precision of lunar and planetary ephemerides [24-26]. Regarding astronomical angular observations, adaptive optics technology has been realized on the ground [27], and stellar interferometers have been constructed [28,29]. In space, the proposed Space Interferometry Mission (SIM) and the successful Gaia mission [30,31] have achieved precisions at sub-milliarcsecond and sub-microarcsecond (sub-μas) levels, respectively. Very Long Baseline Interferometry (VLBI) has achieved angular precision of 1 μas, moving towards sub-microarcsecond accuracy [32,33].

The development of lunar laser ranging and the 1970s research on drag-free space flight, combined with the proposal of ground-based kilometer-scale laser gravitational wave detection concepts, prompted the development of space-based laser interferometry concepts for gravitational wave detection between 1977 and 1989 [34-40]. The first public proposals for space-based laser interferometers for gravitational wave detection were presented at the following two conferences: (i) at the Active Optical Devices and Applications conference held in Washington in 1980 by Decker, Randall, Bender, and Faller [35]; (ii) at the Second International Conference on Precision Measurement and Fundamental Constants (PMFC-II) held in Gaithersburg from June 8-12, 1981, by Faller and Bender [36,37]. This pioneering proposal [36,37] introduced the concept of a gravitational wave mission using laser interferometry in space. It mentioned two essential elements: drag-free spacecraft to compensate for disturbances and laser interferometry to enhance measurement sensitivity.

In 1989, Faller, Bender, and their colleagues [40] selected the spacecraft formation orbit shown in Figure 1 as the mission orbit for their space-based laser gravitational wave observatory (LAGOS: LAser Gravitational-wave Observatory in Space). The formation consists of three spacecraft trailing the Earth's orbit by about 30 degrees, forming a triangular configuration with 1 million-kilometer arms. The central spacecraft orbits the sun with a period of one year, and the plane of the three spacecraft forms a 60-degree angle with the ecliptic, creating an isosceles right triangle. The end spacecraft receive 530 nm wavelength, 1 W laser beams emitted by the central spacecraft. After receiving the weak light, they use phase-locked local lasers (equivalent to amplification) to retransmit it back to the central spacecraft, generating interference. If a gravitational wave passes through, the interference fringes will change and can be detected (Figures 1 and 2). The orbits of the end spacecraft have an inclination of 1/300 and an eccentricity of $\sqrt{3}/300$ relative to the central spacecraft's circular orbit with a radius of 1 million kilometers and a one-year period (refer to Section 10.2). The expected strain sensitivity in the $10^{-3}$-$10^{-1}$ Hz range is $1\times10^{-21}/\sqrt{Hz}$. The primary scientific objectives are to detect continuous gravitational waves from numerous binary star systems and burst gravitational waves from the galaxy formation era.

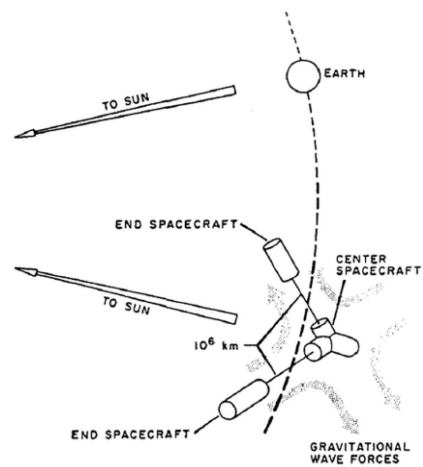

Figure 1 (LAGO's) Laser heterodyne gravitational wave antenna. The figure is from ref. [40].

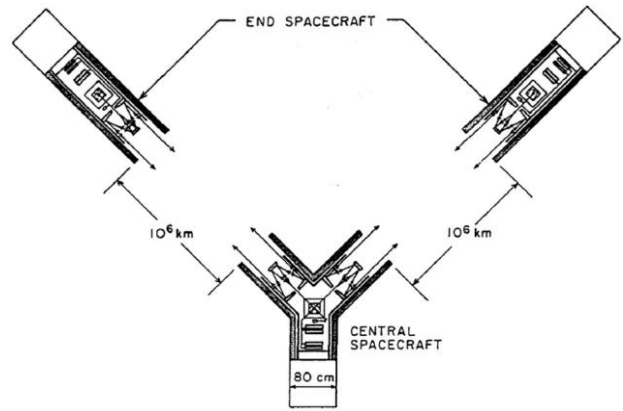

Figure 2 (LAGO's) Laser heterodyne gravitational wave antenna showing spacecraft detail. The figure is from ref. [40].

Bender and Faller, who had dedicated their time to lunar laser ranging and measuring free-fall acceleration using interferometry, naturally proposed this space-based gravitational wave detection experiment. In fact, in the interferometric measurement of Earth's gravitational acceleration, the motion of a test mass freely falling within a vacuum drop vessel can be considered as drag-free motion in space [41]. The discrepancies observed in the 1990 comparison of absolute gravimeters at the International Bureau of Weights and Measures (BIPM) could be partially explained by corrections related to the finite speed of light in interferometric measurements [42]. In spacecraft tracking, the finite speed of light is always accounted for. The test masses in gravitational wave missions and gravimeters can both be regarded as free-falling within the solar system, tracked using celestial dynamics equations. Thus, we see a connection between space geodesy, Galileo's equivalence principle (universality of free-fall) experiments, and gravitational wave detection missions. An example of this connection is the contribution of the LISA (Laser Interferometer Space Antenna) space laser interferometric gravitational wave detection team to the continuous laser ranging gravity gradient measurement of the GRACE follow-on space mission [43], and how developing this sub-project contributes to LISA technology. Another example is the adjustment of the Taiji-1 technology demonstrator satellite for observing the Earth's geoid after completing its primary mission [44].

**1.4 LISA (Laser Interferometric Space Antenna for Gravitational Wave Detection)**

A significant advancement in space-based gravitational wave detection was the selection of LISA (Laser Interferometer Space Antenna, with an arm length of 5 million kilometers) by the European Space Agency (ESA) in 1993 for an M3 assessment study, and its subsequent inclusion as the third cornerstone mission of "Horizon 2000 Plus" [45]. After 2000, LISA became a joint ESA-NASA mission until NASA withdrew in 2011. In 1998, LISA Pathfinder was chosen as the second mission in ESA's Small Missions for Advanced Research in Technology (SMART) program, aimed at developing and testing extremely precise drag-free spacecraft technology. In commemoration of the centenary of General Relativity, ESA launched LISA Pathfinder on December 3, 2015, from the European Spaceport in Kourou, French Guiana, using a Vega rocket, successfully demonstrating the drag-free technology necessary for gravitational wave observation [46,47].

Simultaneously, ESA sponsored a common bus technology reference study for fundamental physics, completed in 2008 [48]. In 2011, the LISA team proposed the New Gravitation-wave Observatory (NGO)/eLISA (evolved LISA) [49] to accommodate budget changes, reducing the target arm length from 5 million kilometers to 1 million kilometers, which received favorable evaluations. In November 2013, ESA announced the selection of scientific themes for the L2 and L3 mission opportunities—L2's "The hot and energetic universe" and L3's "The gravitational universe" (ESA Media Relations Office/Communication Department. The hot and energetic Universe and the search for elusive gravitational waves, http://www.esa.int/Our_Activities/Space_Science/ESA_s_new_vision_to_study_the_invisible_Universe). At that time, a potential launch opportunity for ESA's L3 mission was in 2034. Following the successful launch of the ESA-NASA collaborative project LISA Pathfinder in December 2015 and the achievement of its technical objectives in 2016, ESA officially selected LISA as its third large-class space mission (L3) in June 2017 [50]. NASA agreed to participate, contributing 20% of the mission's total cost. The LISA science team then selected an arm length of 2.5 million kilometers (2.5 Gm), with the detection formation consisting of three spacecraft forming a precise equilateral triangle within 1% accuracy (Figure 3), capable of performing multiple Time Delay Interferometry (TDI) [50]. The definition study report was completed in September 2023 [51].

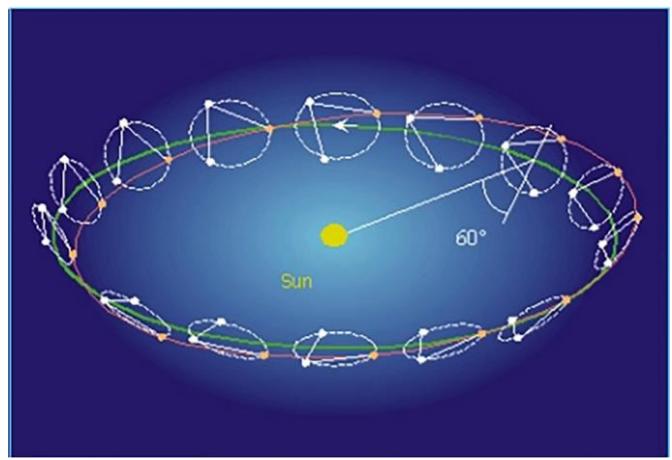

Figure 3 (Color online) Schematic of LISA/LISA-type orbit configuration in Earth-like solar orbit. The figure is from ref. [45].

On January 25 of this year, ESA's Science Programme Committee approved the LISA space mission [52]. This marks the first time a scientific effort to detect and study gravitational waves from space has received formal adoption, recognizing that the mission concept and technology are sufficiently advanced. The committee has authorized the construction of the instruments and spacecraft, with work set to begin in January 2025, once a European industrial contractor is selected. LISA is scheduled to launch in 2035 on an Ariane 6 rocket and will take a year to transition to its scientific orbit, with the science phase commencing in 2036. The LISA space mission aims to detect gravitational waves produced by the inspiral and merger of numerous binary black hole systems with masses ranging from $10^4$ to $10^7$ $M_\odot$, enabling the study of the co-evolution of black holes and galaxies. It will also utilize the characteristic distance scale of binary black hole gravitational wave sources to accurately measure the luminosity distances of cosmic objects, aiding in the verification of cosmic expansion rates determined by other methods. By observing the distribution of numerous binary white dwarfs and neutron stars within the Milky Way, LISA will enhance the galactic structure model obtained from ESA's Gaia astrometric space mission and test the MOdified Newtonian Dynamics (MOND) theory. Additionally, LISA will explore gravitational waves from the early universe and test General Relativity and cosmological theories.

## 1.5 ASTROD: Astrodynamical Space Tests of Relativity Using Optical Devices and the ASTROD-GW Laser Interferometric Space Gravitational Wave Detection Mission

The overall concept of the Astrodynamical Space Test of Relativity using Optical Devices (ASTROD) involves using a set of spacecraft in drag-free flight within the solar system. These spacecraft would use optical devices to measure distances between each other, aiming to detect the gravitational field of the solar system, measure related solar system parameters, test relativistic gravity, observe solar oscillations (low-$l$ g-modes, f-modes, and p-modes), and detect gravitational waves. The foundational ASTROD scheme was proposed in 1993 and has since been in the stages of concept, simulation, and laboratory research [53-70].

In 1996, the concept of ASTROD I (also called Mini-ASTROD) was proposed. This involved a mission using two-way ranging between ground stations and the ASTROD I spacecraft to test general relativity and map the gravitational field of the solar system [53]. Simulation studies indicated that for the Eddington parameter $\gamma$, the precision of solar system relativistic gravity tests could reach $10^{-9}$, improving current test precision by 3 to 4 orders of magnitude [71-75].

In early 2009, responding to the Chinese Academy of Sciences' call for gravitational wave mission research, the ASTROD-GW (ASTROD optimized for gravitational waves) mission concept was proposed. This concept involved three spacecraft operating near the Sun-Earth Lagrange points L3, L4, and L5, using 260 Gm arm-length laser interferometry to detect gravitational waves [76-81], as shown in Figure 4 [78]. To achieve sensitivity in the direction of the ecliptic pole, the spacecraft formation must have some inclination. The schematic diagram of the ASTROD-GW orbital configuration with inclination is shown in Figure 5 [82].

The arm length of the ASTROD-GW space mission is 100 times that of LISA, providing two orders of magnitude better sensitivity at low frequencies. This will enable the detection of gravitational waves produced by the spin and merger of binary black holes with masses ranging from $10^5$ to $10^{10}$ $M_\odot$, allowing for further study of the co-evolution of black holes and galaxies. The observation of characteristic scales of binary black hole gravitational wave sources will allow for more accurate measurements of the luminosity distances of cosmic objects, helping to verify and improve the measurements of cosmic expansion determined by LISA and other methods. Additionally, ASTROD-GW will explore microhertz gravitational waves from the early universe and provide a broader range of tests for general relativity and cosmology.

Before the ASTROD-GW scheme was proposed, the concept of Super-ASTROD was introduced in 1996 [53]. This concept involved three spacecraft in triangular formation in Jupiter-like orbits, one spacecraft near the Sun-Earth Lagrange point L1, and another in a Jupiter-like orbit with a large inclination. In 2008, dual scientific goals for gravitational wave detection and cosmological model/relativistic gravity testing were studied [83]. With the proposal of ASTROD-GW, the basic gravitational wave detection configuration for Super-ASTROD involved placing three of its 4-5 spacecraft near the Sun-Jupiter Lagrange points L3, L4, and L5. The arm length of the Super-ASTROD space mission is five times that of ASTROD, offering five times better sensitivity at low frequencies. This would enable the detection of the spin and merger of larger black holes, further exploring the cosmic gravitational wave background radiation and testing cosmological theories.

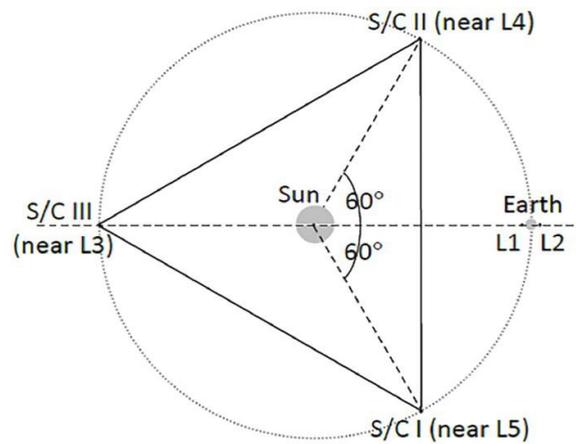

Figure 4 Schematic of ASTROD-GW orbit configuration [78].

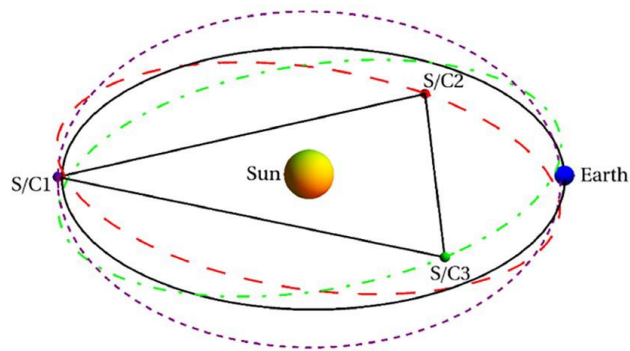

Figure 5 (Color online) Schematic of ASTROD-GW orbit configuration with inclination [82]. 3D view with the scale of the vertical axis multiplied tenfold. Its projection on the ecliptic plane looks basically like Figure 4.

## 1.6 DECIGO and BBO Laser Interferometric Space Gravitational Wave Detection Missions

DECIGO (DECI-Hz Gravitational Observatory), a sub-hertz interferometer gravitational wave observatory, was proposed in 2001 [84]. Its objective is to detect background gravitational waves from the early universe in the mid-frequency observational band between the high-frequency band of ground-based detectors and the low-frequency band of other space-based gravitational wave detectors at that time. DECIGO will employ the Fabry-Perot method similar to ground-based interferometers but with an arm length of 1000 km and a finesse of 10. As a successor to LISA, the United States proposed BBO (Big Bang Observer) with an arm length of 50,000 km [85], also aiming for similar goals. As depicted in Figure 6, DECIGO/BBO has two potential configurations: one with 3 spacecraft in a group of 4 groups and another with 4 spacecraft in a group of 3 groups, totaling 12 spacecraft each. They will directly measure the stochastic background of gravitational waves through correlation analysis [86]. Their primary goal is to detect primordial gravitational waves. Following the detection of gravitational waves by ground-based detectors, the DECIGO team proposed B-DECIGO as a precursor to DECIGO. It is designed as a cluster of three spacecraft separated by 100 km, derived from a simplified design of DECIGO. The mirrors have a diameter of 0.3 m, mass of 30 kg, finesse of 100, and sufficient noise sensitivity to detect many mid-frequency gravitational wave sources. B-DECIGO represents a feasible planning phase between the first and second stages [87].

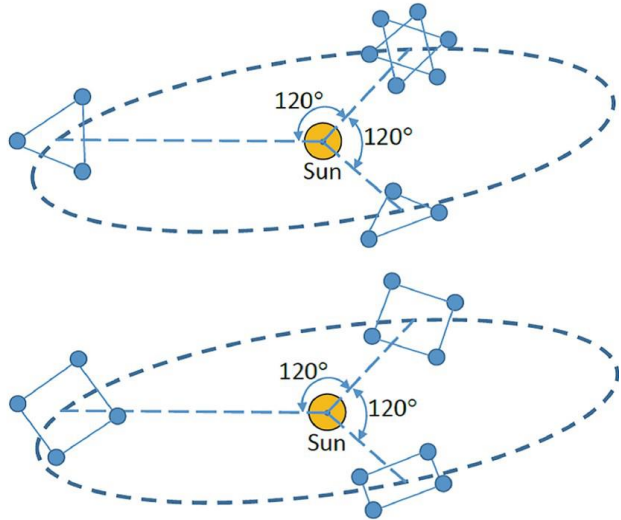

Figure 6 (Color online) Two schematic configurations of BBO and DECIGO in Earth-like solar orbits.

## 1.7 Taiji and TianQin Laser Interferometric Space Gravitational Wave Detection Missions

In 2015, following the successful direct detection of gravitational waves from binary black hole mergers three times using the aLIGO twin 4 km ground-based detectors, gravitational wave detection became a hot topic. The successful launch of LISA Pathfinder and achievement of its technical goals [46] further fueled interest in space-based gravitational wave detection. In 2016, the Taiji team advanced the ALIA-descope space gravitational wave detection plan [88] proposed in 2014, featuring an arm length of 3 million kilometers and using a LISA-like orbit. This initiative was formally named the Taiji mission [89], capable of joint observations with LISA to enhance scientific objectives [90-93], particularly improving the localization of gravitational wave sources. The orbital configuration is illustrated in Figure 7 [93].

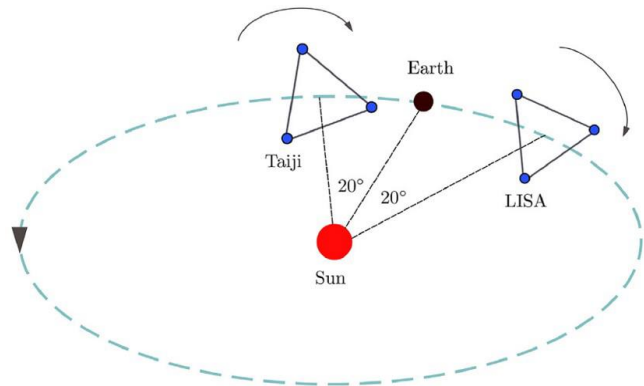

Figure 7 (Color online) Configuration of the LISA-Taiji network. The LISA constellation follows the Earth by 20°, while the Taiji constellation leads the Earth by 20°. The figure is from ref. [93].

The TianQin team proposed the TianQin mission with an arm length of 170,000 kilometers, using a geocentric orbit [94]. The orbital configuration is illustrated in Figure 8 [94,95]. TianQin primarily operates in the millihertz frequency band and can participate in networked joint observations to enhance scientific objectives [96,97]. In 2018, the Taiji team launched the Taiji-1 technology satellite [98], and the TianQin team launched the TianQin-1 technology satellite [99], completing the first phase of technical development and testing. Both teams are actively engaged in conceptual and technical research to complete their definition study reports.

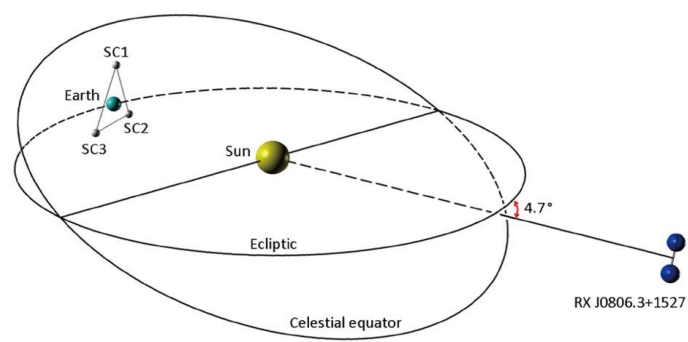

Figure 8 (Color online) Schematic of TianQin (S/C1, S/C2 & S/C3) orbit configuration. The direction to J0806 is shown. The figure is from refs. [94,95].

**1.8 Mid-Frequency Gravitational Wave Detection Planning**

Due to theoretical gaps in black hole mass distribution, in the first decade of this century, many believed that sources of astrophysical gravitational waves from black hole mergers in the mid-frequency band were extremely rare. Consequently, there was a push to detect low-amplitude primordial gravitational waves in this frequency band, leading to the proposal and study of two mission concepts—DECIGO [84] and BBO (Big Bang Observer) [85]. Because of the high sensitivity required for these mission concepts, they fall under the category of second or third-generation space-based gravitational wave missions. Their most sensitive detection band lies between the 35-500 Hz gravitational waves detected on the ground in 2015 and the millihertz gravitational waves that LISA aims to detect, specifically in the sub-hertz mid-frequency band (0.1 Hz - 10 Hz).

The LIGO-Virgo detections of binary black hole mergers often observed sources with masses around 30 solar masses, whose gravitational waves generated from pre-merger inspiral traverse the entire mid-frequency band. In 2019, LIGO-Virgo detected an event with a total mass of about 150 solar masses [100], further enhancing expectations for detectable sources in the mid-frequency gravitational wave range.

Prior to 2015, methods for mid-frequency gravitational wave detection included the aforementioned DECIGO and BBO space-based laser interferometer gravitational wave detection concepts, along with various ground-based gravitational wave detection concepts [101]. Post-2016, further research has been conducted on these detectors, as well as proposals for various timely detectors [102,103].

For the latest plans and developments in lunar-based gravitational wave detectors, please refer to the website of the 1st Sino-European Workshop on Lunar-Based Gravitational Waves Detection, April 27-30, 2024, Beijing: https://swlgwd2024.scimeeting.cn/en/web/index/22418_1878717__ .

Regarding space-based laser interferometer detection, projects like AMIGO (Astrodynamical Middle-frequency Interferometer GW Observatory) [104-106] and B-DECIGO [87] are crucial. AMIGO and B-DECIGO aim to network observations and detect intermediate-mass black hole binaries. AMIGO is technically classified as a first-generation middle-frequency laser interferometer gravitational wave detector.

**1.9 Joint Observations with Gravitational Wave Detector Networks**

The network collaboration of LIGO-Virgo has significantly enhanced the precision of gravitational wave source localization, promoting the realization of multi-messenger observations.

Space-based gravitational wave detectors participating in network observations offer several advantages, including rapid localization of gravitational wave sources, in-depth joint tests of relativistic gravity, polarization testing of gravitational waves, potential measurement of non-zero cosmic polarization rotation, better determination of galactic gravitational wave backgrounds, and increased detection rates of stellar-mass binary black hole systems and massive binary black hole system gravitational wave events, among others.

The Taiji space gravitational wave detection plan, with an arm length of 3 million kilometers, adopts a detection scheme similar to LISA. If joint observations can be conducted during the LISA detection period (after 2036), it will rapidly improve the localization precision of low-frequency gravitational wave sources in the 0.1 mHz to 0.1 Hz frequency range [90-93], facilitating multi-messenger observations.

Using LISA-Taiji as an example, possible network configuration choices are illustrated (Figure 9) [92,93]: (i) LISA-Taiji-c, where Taiji-c is co-located with LISA and coplanar. Due to its good optical overlap with LISA, this configuration is most sensitive for random gravitational wave observations. (ii) LISA-Taiji-p (baseline configuration), where Taiji-p is offset from Earth by about 20 degrees and the formation plane is inclined +60 degrees relative to the ecliptic plane, similar to LISA. (iii) LISA-Taiji-m, where Taiji-m is offset from Earth by about 20 degrees and the formation plane is inclined –60 degrees relative to the ecliptic plane.

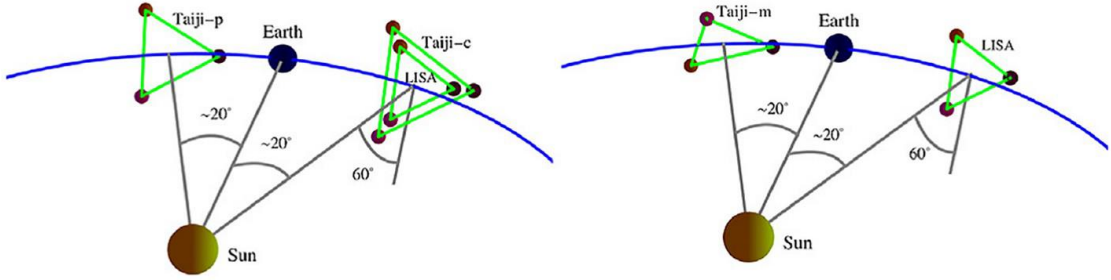

Figure 9 (Color online) Diagrams of the LISA and alternative Taiji mission orbit deployments. The left plot shows the LISA mission, which trailing the Earth by approximately 20° and is inclined by +60° with respect to the ecliptic plane. The Taiji-p leads the Earth by ∼20° and is inclined by +60°. The Taiji-c is co-located and coplanar with LISA. The right panel shows the LISA and Taiji-m, which lead the Earth by ∼20° and –60° inclined. The angle between the LISA and Taiji-m constellation planes is ∼71°, and the angle between LISA and Taiji-p formation planes is ∼34.5°. The figure is from refs. [92,93].

Currently, the LISA-Taiji-p configuration serves as the baseline configuration. The Taiji team will timely decide on the final configuration to adopt.

In linearized phenomenological models or template-based least squares (Gaussian, Kalman, Fisher matrix, etc.) parameter estimation using independent detector networks, the order or sequence of observational data does not affect the final results. For example, in a network composed of detectors A, B, and C: (i) A observes and obtains data block A1 during time interval $[t_{A1}, T_{A1}]$, and data block A2 during time interval $[t_{A2}, T_{A2}]$; (ii) B observes and obtains data block B1 during time interval $[t_{B1}, T_{B1}]$, and data block B2 during time interval $[t_{B2}, T_{B2}]$; (iii) C observes and obtains data block C1 during time interval $[t_{C1}, T_{C1}]$, and data block C2 during time interval $[t_{C2}, T_{C2}]$. The least squares parameter estimation method does not differentiate whether the total data blocks are used or estimated in different orders, or even sequentially in any order; the final results will be the same.

Due to similar sensitivity targets between LISA and Taiji, treating LISA and Taiji independently would increase sensitivity by approximately 1.4 times. However, with a LISA-Taiji network, there are additional observable quantities related to relative positions and orientations, crucial for determining the angular position and direction of gravitational sources. With these, angular positions and directions can be rapidly determined, similar to the LIGO-Virgo-KAGRA team's capabilities in ground-based gravitational wave observations. In a space-based LISA-Taiji network, precise determination of the relative distances between LISA and Taiji, and their directions, are crucial observable quantities for angular position and direction determination. These improve the estimation of other parameters quickly. LISA and Taiji-p/Taiji-m will be separated in orbit by approximately 100 Gm or about 40 degrees, which is 30-40 times the size of LISA and Taiji individually.

Most observable gravitational wave sources are (quasi-)periodic. For these sources, the sky localization area over two months can be improved by two orders of magnitude compared to a single detector. After six months or a year, a single LISA or Taiji detector rotates around its orbit, while these sources are (quasi-)periodic, so a single LISA or Taiji is equivalent to multiple detectors in different positions and times, making it possible to detect (quasi-)periodic sources with good sky positioning. The network will only have a greater than 2-fold improvement in solid angle positioning resolution ($2^{1/2} \times 2^{1/2} \times$ the enhancement factor due to the relative distance information between LISA and Taiji). Nevertheless, early rapid improvement in network positioning resolution is crucial for multi-messenger observations.

The LIGO-Virgo network and the LISA-Taiji network are examples of simultaneous and co-frequency band observations. In addition to rapid localization, simultaneous or non-simultaneous multi-band observations increase the number of observable events. If phenomenological/theoretical templates are robust, observing at different frequencies, either simultaneously or non-simultaneously, will effectively improve parameter estimation. For example, using a single binary star merger template across millihertz (mHz), decihertz (deci-Hz), and decahertz (deca-Hz) frequencies, the space-based gravitational wave detector network (LT-AMIGO-ET-CE) comprising LISA, TAIJI, AMIGO, ET, and CE can improve parameter estimation accuracy by two orders of magnitude (Table 1) [107]. Joint observations by LT-AMIGO-ET-CE enhance the localization of gravitational wave sources by two to three orders of magnitude, with the precision of determining the position of most detectable binary black hole gravitational wave sources at Δ90% ranging from $7 \times 10^{-7}$ to $2 \times 10^{-3}$ deg$^2$ (median Δ90% is $4.6 \times 10^{-5}$), facilitating the identification of electromagnetic/neutrino sources associated with gravitational wave sources and promoting active multi-messenger astronomy observations. The results of multi-messenger enhanced observations will strengthen the distinguishability of various gravitational wave source models, advancing understanding of stellar formation, black holes, and galaxy co-evolution. The multi-band capability of LISA-TAIJI-AMIGO-ET-CE to locate a large proportion of binary black hole systems within a sky area of $10^{-3}$ deg$^2$, with a relative error in luminosity distance measurement of $\sigma_{d_L}/d_L \leq 0.001$, where the host galaxy can likely be directly determined through gravitational wave observations. Together with redshift measurements, precision in determining cosmic model parameters and the Hubble constant can reach one thousandth or even better, significantly advancing our understanding of the universe and dark energy.

In this simulation study [107], to investigate the prospects of multi-band gravitational wave observations of binary black hole systems evolving from the inspiral phase to the final merger phase, four different models were considered, taking into account different formation channels and constraints from LVK O3 observations [108]: The local merger rate density $R(0)$ in the first three models is normalized to the latest constraints provided by LVK after its first three observation runs [108]. Based on the redshift evolution of merger rate density in each model, the primary mass distribution of binary black hole systems, and the distribution of mass ratios, simulated binary black hole systems were generated using Monte Carlo methods. Considering the uncertainty of LVK's constraints on the local merger rate density, each model generated 100 simulated samples, including three different $R(0)$ values: 19.1, 10.6, and 27.5 $\text{Gpc}^{-3}\,\text{yr}^{-1}$, representing the median and 90% confidence interval of the constrained local merger rate density. The four models are: (i) EMBS model: Assumes all binary black hole systems originate from the evolution of massive binary stars in galactic fields (referred to as the EMBS [Evolution of Massive Binary Stars in galactic field] channel) [109]. (ii) Dynamical model: Assumes all binary black hole systems form through dynamical interactions in environments like dense star clusters (referred to as the dynamical channel, using the A023 GC channel model from reference [110]). (iii) Mixed model: Considers a combination of the EMBS formation channel and the dynamical formation channel, with 75% of binary black hole systems originating from the EMBS channel and the remaining 25% from the dynamical channel. (iv) GWTC-3 model: Based on the latest observational constraints from LVK collaboration [108]. The redshift-dependent merger rate density scales proportionally with $(1 + z)^{\kappa}$, where at low redshifts (i.e., $z \leq 1$), $\kappa = 2.7^{+1.8}_{-1.9}$.

Table 1 Median values of the distributions of parameter estimation uncertainties for multiband BBHs[a] [107]

| GW detector | $\Delta\Omega_{90\%}$ median | $\sigma_{d_L}/d_L$ median | $\sigma_{M_c}/M_c$ median | $\sigma_\eta$ median |
|---|---|---|---|---|
| LT (LISA-Taiji) | $8.2\times10^{-1}$ | $1.1\times10^{-1}$ | $3.4\times10^{-6}$ | $5.5\times10^{-3}$ |
| AMIGO | $1.1\times10^{-1}$ | $7.5\times10^{-2}$ | $4.5\times10^{-7}$ | $4.7\times10^{-4}$ |
| ET-CE | $5.7\times10^{-3}$ | $1.8\times10^{-3}$ | $1.6\times10^{-3}$ | $2.0\times10^{-3}$ |
| LT-AMIGO | $5.4\times10^{-1}$ | $5.8\times10^{-2}$ | $1.5\times10^{-7}$ | $3.0\times10^{-4}$ |
| LT-ET-CE | $1.5\times10^{-4}$ | $1.3\times10^{-3}$ | $4.7\times10^{-8}$ | $1.5\times10^{-4}$ |
| AMIGO-ET-CE | $4.9\times10^{-5}$ | $1.1\times10^{-3}$ | $2.4\times10^{-7}$ | $7.9\times10^{-5}$ |
| LT-AMIGO-ET-CE | $4.6\times10^{-5}$ | $1.1\times10^{-3}$ | $2.9\times10^{-8}$ | $6.1\times10^{-5}$ |

a) The first column indicates the name of the gravitational wave detectors or their different combinations. The second column shows the median value of the localization accuracy $\Delta\Omega_{90\%}$ distribution for 100 multi-band binary black hole systems. The third, fourth, and last columns respectively display the median values of the distribution for the relative error in luminosity distance measurement $\sigma_{d_L}/d_L$, the relative error in the chirp mass ratio $\Delta_{M_c}/M_c$, and the relative error in the symmetric mass ratio $\sigma_\eta$. For the lower and upper bounds of the 68% confidence interval for each of these quantities, please refer to reference [107].

Figure 10 depicts the evolutionary trajectories of characteristic amplitudes of simulated binary black hole systems (mock BBHs) across sensitivity spectra of various missions.

In this simulation, it is assumed that LISA, Taiji, and AMIGO begin observing simultaneously and each has a continuous observation period of 4 years, although this may differ from actual future scenarios. If AMIGO begins observing after the observation periods of LISA/Taiji have ended, some binary black hole systems that merge during the LISA/Taiji observation period would not be observable by AMIGO before it begins. However, some binary black hole systems detected in the low-frequency band with merger timescales greater than 4 years could evolve into the mid-frequency band and eventually be observable by AMIGO.

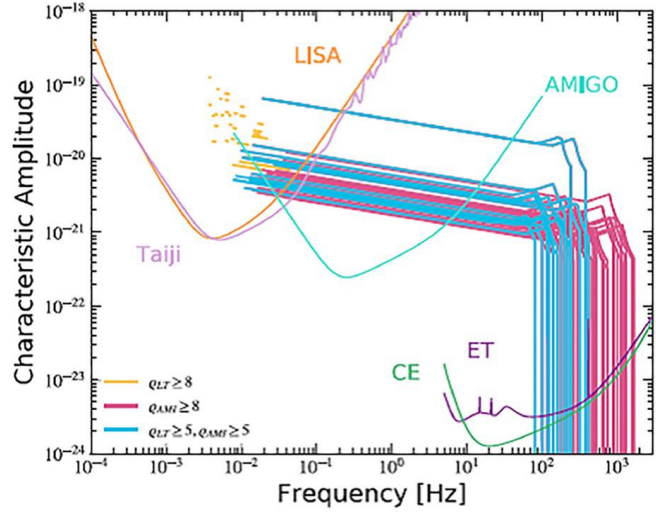

Figure 10 (Color online) Evolution tracks of characteristic amplitudes for the mock BBHs which are detectable by LISA-Taiji-AMIGO (i.e. S/N ratios $\rho_{\text{LISA-Taiji}} \geq 5$ and $\rho_{\text{AMIGO}} \geq 5$, blue lines), by LISA-Taiji (i.e. $\rho_{\text{LISA-Taiji}} \geq 8$, yellow lines), and by AMIGO (i.e. $\rho_{\text{AMIGO}} \geq 8$, magenta lines), respectively. All the BBHs plotted here are assumed to be observed for a continuous period of 4 yr. Orange, pink, sky blue, green, and purple curves represent the sensitivity curves for LISA, Taiji, AMIGO, CE, and ET, respectively [107].

Simulation studies of detectable binary black hole systems observed by different gravitational wave detectors at different frequency bands serve as demonstrations and planning for multi-band gravitational wave observations.

## 1.10 Micro-Hertz Gravitational Wave Detection Planning, U.S. Astro2020 APC White Paper, and ESA Post-LISA Space Gravitational Wave Detection Preliminary Solicitation White Paper

Last year the pulsar timing array detected background gravitational waves in the nanohertz frequency band, while LISA and Taiji aim to detect gravitational waves in the millihertz low-frequency band (0.1 mHz - 100 mHz). Between these two bands lies the low-frequency microhertz band (0.1 μHz - 100 μHz) of gravitational waves. For the detection of low-frequency microhertz gravitational waves, there are astrodynamical space laser interferometric gravitational wave detection concepts as described in Section 1.5 [76-82]. Their primary objective is to detect gravitational waves produced by the merger and coalescence of binary black holes with masses ranging from $10^5$ to $10^{10}$ $M_\odot$, studying the coevolution of black holes and galaxies, and accurately measuring distances to cosmic objects using binary black hole gravitational wave sources.

The experimental technology development required for space low-frequency microhertz band (0.1 μHz - 100 μHz) laser interferometric gravitational wave detectors has matured to a level comparable to LISA and Taiji: their strain sensitivity is limited by inertial sensor/accelerometer noise. Below 1 millihertz, astronomical unit (AU)-scale arm length gravitational wave detectors are two orders of magnitude more sensitive compared to Gm-scale arm length detectors like LISA/Taiji. Constraints from confusion backgrounds still exist above 0.1 millihertz, but below 0.01 millihertz, AU-scale gravitational wave detectors are not constrained by these. The sky resolution of AU-scale arm length gravitational wave detectors is on average a factor of $(S/N)^2$ better than Gm-scale arm length detectors for source localization, i.e., four orders of magnitude. In such finely resolved sky regions, averaging less than one galaxy, enabling clear identification of host galaxies of gravitational wave sources through multi-messenger observations, with redshift measurements becoming a routine observable. Cosmodynamics can be established through observations, providing detection evidence for "dark energy" and "dark matter" and their evolution.

The European Space Agency (ESA) and NASA collaborated on LISA Pathfinder (LISA Precursor), successfully launched in December 2015, achieving its technical objectives by 2016. In June 2017, ESA formally designated LISA as its L3 large space mission [50], and after NASA's participation, the United States solicited the Astro2020 APC White Paper and ESA solicited the LISA Space Gravitational Wave Detection White Paper.

In 2019, the Baker et al. [112] white paper proposed two possible space gravitational wave schemes, one of which has sensitivity to the gravitational wave's most sensitive frequency band (1 mHz - 1 Hz) higher by an order of magnitude than LISA's sensitivity to the same frequency band, and the other has sensitivity to the gravitational wave's most sensitive frequency band in microhertz (0.1 - 100 μHz); the former is illustrated by the ALIA [113], and the latter is illustrated by the Folkner low-frequency space mission concept [114], and suggested as a third possibility such as gLISA [115-118] scheme. The two space gravitational wave schemes correspond to the AMIGO and ASTROD-GW space gravitational wave space detection plans proposed in China.

The European Space Agency has called for white papers on space missions from 2035 to 2050. Four white papers [119, 120, 121, 122] on gravitational wave astronomy and gravitational wave detection primarily discuss and propose scientific objectives, and briefly propose corresponding possible space mission concepts. The first white paper concerns mid-frequency gravitational wave observations [119]; the second white paper concerns low-frequency gravitational wave observations in millihertz [120]; the third white paper concerns low-frequency microhertz gravitational wave observations [121]; the fourth white paper concerns high angular resolution gravitational wave astronomy [122], with high-resolution angular observations and high signal-to-noise ratio observations both capable of providing high angular resolution. These four white papers propose three gravitational wave detection schemes: (i) Decihertz Observatories (DOs) [119], corresponding to our AMIGO; (ii) ESA AMIGO [120], which correspondig to a more sensitive mission after the solar-orbit Taiji mission; (iii) μAres [121], with three spacecraft on the Martian orbit. Recently, ESA has made a mission concept study on LISAmax [123] relying on three spacecraft in Earth-like orbit, corresponding to our ASTROD-GW.

There is also a related white paper [124] on detecting dark matter near the solar system, using space laser gravitational wave detection technology, which is also within the scope considered by ESA.

In the previous section, we saw an example of multi-band gravitational wave observation simulation using LISA-TAIJI-AMIGO-ET-CE, simulating the observation of a group of stellar-mass binary black hole inspirals and mergers statistically inferred by the discoveries by ground-based gravitational wave (GW) observatories, demonstrating the importance and effectiveness of such observations (simulations). Such observations (simulations) can be applied to any other combination of mid-frequency gravitational waves, such as PTA-LISAmax-LT or PTA-ASTROD-GW-LT, etc.

Table 2 lists the orbital configurations, arm lengths, orbital periods, planned number of spacecraft, acceleration noise requirements, laser ranging noise requirements, and gravitational wave-sensitive frequency bands for space-based laser interferometric gravitational wave detection missions that have been approved, launched technology demonstration satellites, or are currently under study.

**Table 2** A compilation of laser-interferometric GW Mission Proposals

| Mission concept | S/C configuration | Arm length (Gm) | Orbit period | S/C # | Acceleration noise (fm/(s$^2$ Hz$^{1/2}$)) | Laser metrology noise (pm/Hz$^{1/2}$) | Sensitive frequency band |
|---|---|---|---|---|---|---|---|
| *Solar-Orbit GW Mission Proposals* | | | | | | | |
| LISA[45,50,51] | Earth-like solar orbits with 20° lag | 2.5 | 1 year | 3 | 3 | 10 (15)[51] | 100 μHz–1 Hz |
| Taiji (ALIA-descope)[98] | Earth-like solar orbits with 20° advance | 3 | 1 year | 3 | 3 | 8 | 100 μHz–1 Hz |
| ALIA[113] | Earth-like solar orbits | 0.5 | 1 year | 3 | 0.3 | 0.06 | 1 mHz–10 Hz |
| ASTROD-GW[81] | Near Sun-Earth L3, L4, L5 points | 260 | 1 year | 3 | 3 | 1000 | 100 nHz–10 mHz |
| aASTROD-GW[81,111] | Near Sun-Earth L3, L4, L5 points | 260 | 1 year | 3 | 0.3 | 100 | 100 nHz–10 mHz |
| Folkner's mission[114] | Earth-like solar orbits | 260 | 1 year | 3 | 3 | 1000 | 100 nHz–10 mHz |
| LISAmax[123] | Earth-like solar orbits | 260 | 1 year | 3 | 3 | 100 | 100 nHz–10 mHz |
| μAries[121] | Mars-like solar orbits | 395 | 1.88 year | 3+3 | 1 | 50 | 100 nHz–10 mHz |
| Big Bang Observer[87] | Earth-like solar orbits | 0.05 | 1 year | 12 | 0.03 | $1.4 \times 10^{-5}$ | 10 mHz–10 Hz |
| DECIGO[84] | Earth-like solar orbits | 0.001 | 1 year | 12 | 0.0004 | $2 \times 10^{-6}$ | 10 mHz–10 Hz |
| B-DECIGO[84] | Earth-like solar orbits | 0.0001 | 1 year | 3 | 0.01 | $2 \times 10^{-4}$ | 10 mHz–10 Hz |
| AMIGO[104,106] | Earth-like solar orbits | 0.01 | 1 year | 3 | 3 | 0.0038 | 10 mHz–10 Hz |
| b-AMIGO[104,106] | Earth-like solar orbits | 0.01 | 1 year | 3 | 3 | 0.012 | 10 mHz–10 Hz |
| e-AMIGO[104,106] | Earth-like solar orbits | 0.01 | 1 year | 3 | 3 | 0.0005 | 10 mHz–10 Hz |
| DO[119] | Earth-like solar orbits | 0.01 | 1 year | 3 | 3 | 0.0005 | 10 mHz–10 Hz |
| Super-ASTROD[83] | Near Sun-Jupiter L3, L4, L5 points (3 S/C), near Sun-Earth L1 point (1 S/C), and (optional) Jupiter-like solar orbit with inclination (1 S/C), | 1300 | 11 year | 4 or 5 | 0.3 | 5000 | 100 nHz–1 mHz |
| *Earth-Orbit GW Mission Proposals* | | | | | | | |
| TianQin[99] | 0.057 Gm height orbit | 0.17 | 44 h | 3 (+3) | 1 | 1 | 1 mHz–10 Hz |
| B-DECIGO[87] | Earth orbit | 0.0001 | tbd | 3 | 0.01 | $2 \times 10^{-4}$ | 10 mHz–10 Hz |
| gLISA[115–118] (GADFLI[118]/ GEOGRAWI[116])a) | Geostationary orbit | 0.073 | 24 h | 3 | 3 | 0.3 | 1 mHz–10 Hz |

a) gLISA [115] was formed by merging two independent but similar proposals—GADFLI and GEOGRAWI.

## 1.11 Doppler Tracking, Optical Clock Gravitational Wave Detection, Atom Interferometry Gravitational Wave Detection, and Pulsar Timing Array Detection

The above discussion pertains to laser interferometric gravitational wave detection using dual-path generalized Michelson interferometry based on Doppler tracking combinations. The principle of Doppler tracking for measuring gravitational waves is illustrated in Figure 11 [125]. Single-frequency electromagnetic waves emitted from ground-based tracking stations are received by Doppler tracking on spacecraft, which then relay the signals back to ground stations for tracking and recording time. The integrated frequency difference corresponds to a phase difference, which in turn corresponds to a time delay. If a gravitational wave passes through, the time at which electromagnetic waves are received changes, allowing the detection of gravitational waves. The key lies in whether the clocks at the recording time can detect the time difference caused by gravitational waves.

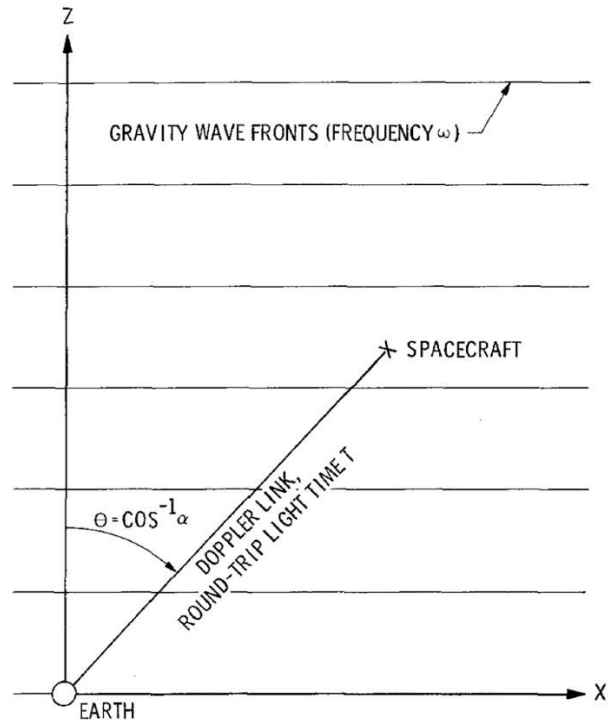

Figure 11 Early diagram for tracking geometry. The figure is from ref. [125].

If the stability of the spacecraft clock/frequency source is insufficient, round-trip timing must be used, known as two-way ranging. If ground-based hydrogen maser is used, the precision can reach $10^{-15}$ to $10^{-16}$, corresponding to the accuracy achieved in current microwave Doppler tracking gravitational wave detections. See Section 3 for details.

To date, the primary alternative approach has been using dual-path generalized Michelson interferometry based on Doppler tracking combinations, ensuring that the optical path length difference is below the requirement of frequency noise. See Section 7 for details.

The stability and accuracy of optical clocks have reached levels of $10$ , making them an option for astronomical unit-scale microhertz gravitational wave detectors. The development of atomic interferometers has also made them feasible for decihertz gravitational wave detectors. See Section 5 for details.

Millisecond pulsars emit stable pulses that can serve as clocks; the arrival time of these pulses at receiving stations is affected by gravitational waves passing through the line of sight. Although the initial phase of pulsar pulses is unknown, a group of millisecond pulsars can statistically detect gravitational wave events and backgrounds. See Section 4 for details.

From the above discussion, we understand that the fundamental equation for space-based gravitational wave detection is the Doppler tracking equation (see Section 3).

## 1.12 Strain Power Spectral Density Amplitude, Characteristic Strain, and Normalized Gravitational Wave Energy Spectral Density

When discussing gravitational wave sensitivity and signal strength, three commonly used plotting methods include: characteristic strain $h_c(f)$ vs. frequency $f$, square root of power spectral density $[S_h(f)]^{1/2}$ vs. frequency $f$, and normalized gravitational wave energy density $\Omega_{gw}$ vs. frequency $f$. For detailed definitions of these quantities, please refer to references [126,127].

Figures 12-14 respectively depict the relationship between gravitational wave detector strain sensitivity and strain power spectral density amplitude $[S_h(f)]^{1/2}$ vs. frequency for the low-frequency and mid-frequency ranges, as well as plots of characteristic strain $h_c(f)$ vs. frequency and normalized gravitational wave energy density $\Omega_{gw}$ vs. frequency. The relationships of these 3 ordinates are as follows:

$$h_c(f) = f^{1/2} [S_h(f)]^{1/2}; \ \Omega_{gw}(f) = (2\pi^2/3H_0^2) f^3 S_h(f) = (2\pi^2/3H_0^2) f^2 h_c^2(f). \tag{1}$$

The detailed explanations and interpretations of Figures 12-14 can be found in the following sections and in references [126]. A significant portion of these figures is derived from the corresponding low-frequency and mid-frequency sections of Figure 2-4 in reference [126], which itself draws from the cited literature, particularly [127].

This review updates and expands upon a review of space-based gravitational wave detection written in commemoration of the centenary of the theory of general relativity in 2016 [128]. The progress made over these eight years has been extensive and robust.

## 2 Basic Principles of Astrodynamical Missions and Space Gravitational Wave Detection

### 2.1 Gravitational and Orbital Observations/Experiments in the Solar System

From a historical perspective, with the improvement in accuracy and precision, the orbits and gravitational observations/experiments within the solar system have always been crucial resources for the development of fundamental physical laws. This holds true for both the Newtonian world system and the development of Einstein's general relativity [129-131]. With significant improvements in orbit and gravitational measurements, we are currently in an era of tremendous progress in testing and developing these fundamental laws. The gravitational field within the solar system is determined by three factors: the dynamic distribution of matter within the solar system, the dynamic distribution of matter outside the solar system (Galactic and cosmic), and gravitational waves propagating through the solar system. Different relativistic gravitational theories and cosmologies make different predictions for the solar system's gravitational field. Therefore, precise measurements of the solar system's gravitational field not only enable gravitational wave observations, determine the distribution of matter within the solar system, and assess observable (testable) effectson our galaxy and the universe, but also serve to test these relativistic theories and cosmologies. To measure the solar system's gravitational field, we measure/monitor the distances between different natural and/or artificial celestial bodies. In the solar system, the equations of motion for celestial bodies or spacecraft are given by the celestial mechanics equations:

$$\mathbf{a} = \mathbf{a}_N + \mathbf{a}_{1PN} + \mathbf{a}_{2PN} + \mathbf{a}_{Gal\text{-}Cosm} + \mathbf{a}_{GW} + \mathbf{a}_{nongrav}, \tag{2}$$

where $\mathbf{a}$ is the acceleration of celestial bodies or spacecraft, $\mathbf{a}_N$ is the acceleration caused by Newtonian gravitational forces within the solar system, $\mathbf{a}_{1PN}$ is the acceleration due to first-order post-Newtonian effects within the solar system, $\mathbf{a}_{2PN}$ is the acceleration due to second-order post-Newtonian effects within the solar system, $\mathbf{a}_{\text{Gal–Cosm}}$ is the acceleration due to Galactic and cosmic gravitational forces, $\mathbf{a}_{\text{GW}}$ is the acceleration due to gravitational waves, and $\mathbf{a}_{\text{nongrav}}$ is the acceleration due to all other non-gravitational sources [78]. The distances between spacecraft primarily depend on solar system gravity (including gravitational oscillations caused by the Sun) and incoming gravitational waves, and are also affected by gravitational waves generated by solar oscillations and the motion of the Milky Way. Predicting these distances as a function of time relies on correct gravitational theories and cosmologies. Accurate measurements of these distances over time will help determine the reasons for distance variations and validate the correct gravitational theory/cosmology. Ideally, a fleet of drag-free spacecraft navigating within the solar system using optical (or other sensitive) devices to measure mutual distances would map the solar system's gravitational field, measure relevant solar system parameters, test relativistic gravity, observe solar low-$l$ g, f, p mode oscillations, and detect gravitational waves [78, 132]. In practice, certain orbital configurations are conducive to testing relativistic gravity; others are suitable for measuring solar system parameters; and still others are advantageous for detecting gravitational waves. These orbital configurations are integral parts of mission designs for various purposes [78, 132].

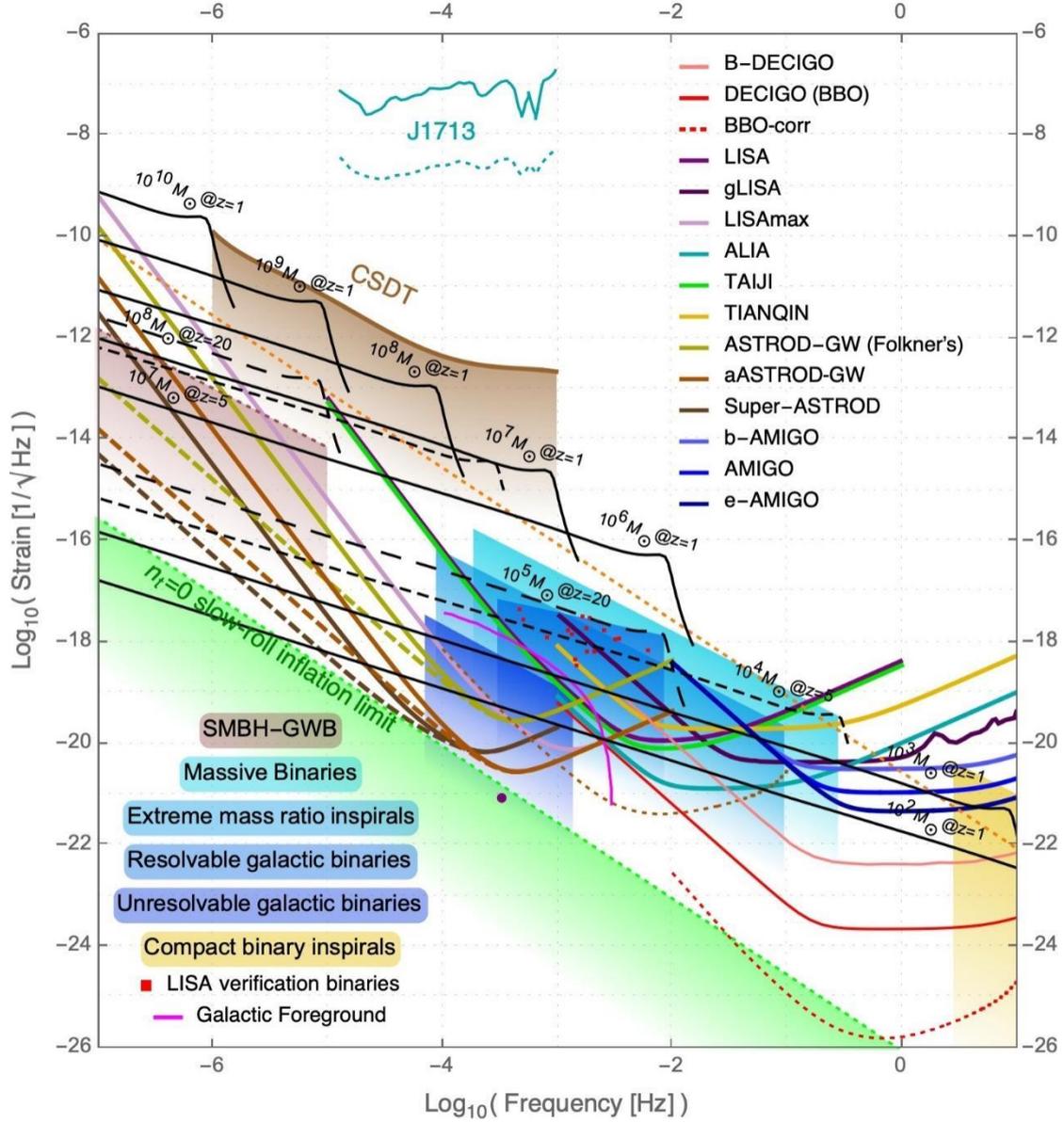

Figure 12 (Color online) Strain power spectral density amplitude (asd) vs. frequency for various GW detectors and GW sources. The black lines show the inspiral, coalescence and oscillation phases of GW emission from various equal-mass black-hole binary mergers in circular orbits at various redshift: Solid line, $z = 1$; dashed line, $z = 5$; long-dashed line $z = 20$. See text for more explanation. CSDT: Cassini Spacecraft Doppler Tracking; SMBH-GWB: Supermassive Black Hole-GW Background.

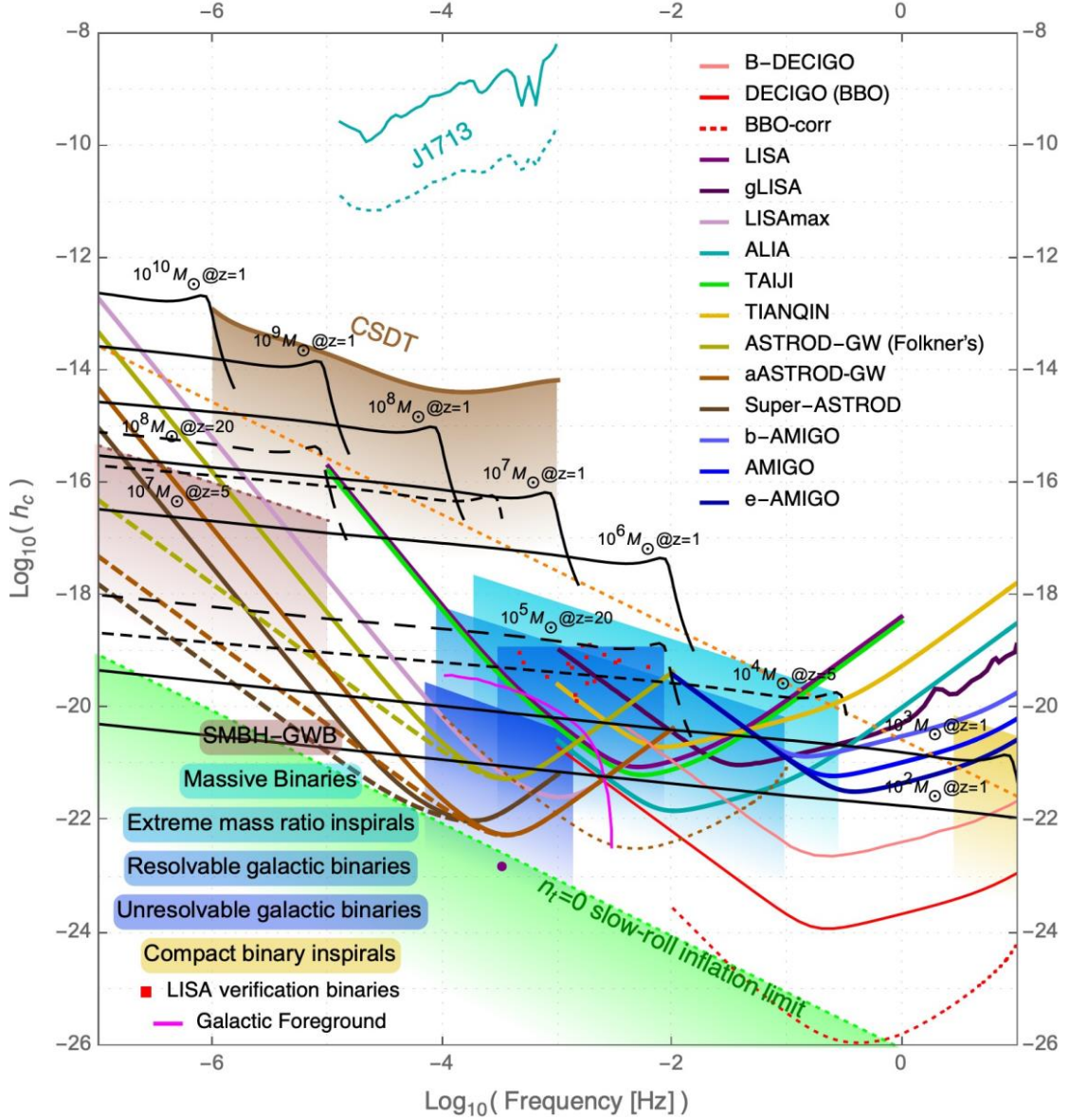

Figure 13 (Color online) Characteristic strain $h_c$ vs. frequency for various GW detectors and sources. The black lines show the inspiral, coalescence and oscillation phases of GW emission from various equal-mass black-hole binary mergers in circular orbits at various redshift: Solid line, $z = 1$; dashed line, $z = 5$; long-dashed line $z = 20$. See text for more explanation. CSDT: Cassini Spacecraft Doppler Tracking; SMBH-GWB: Supermassive Black Hole-GW Background.

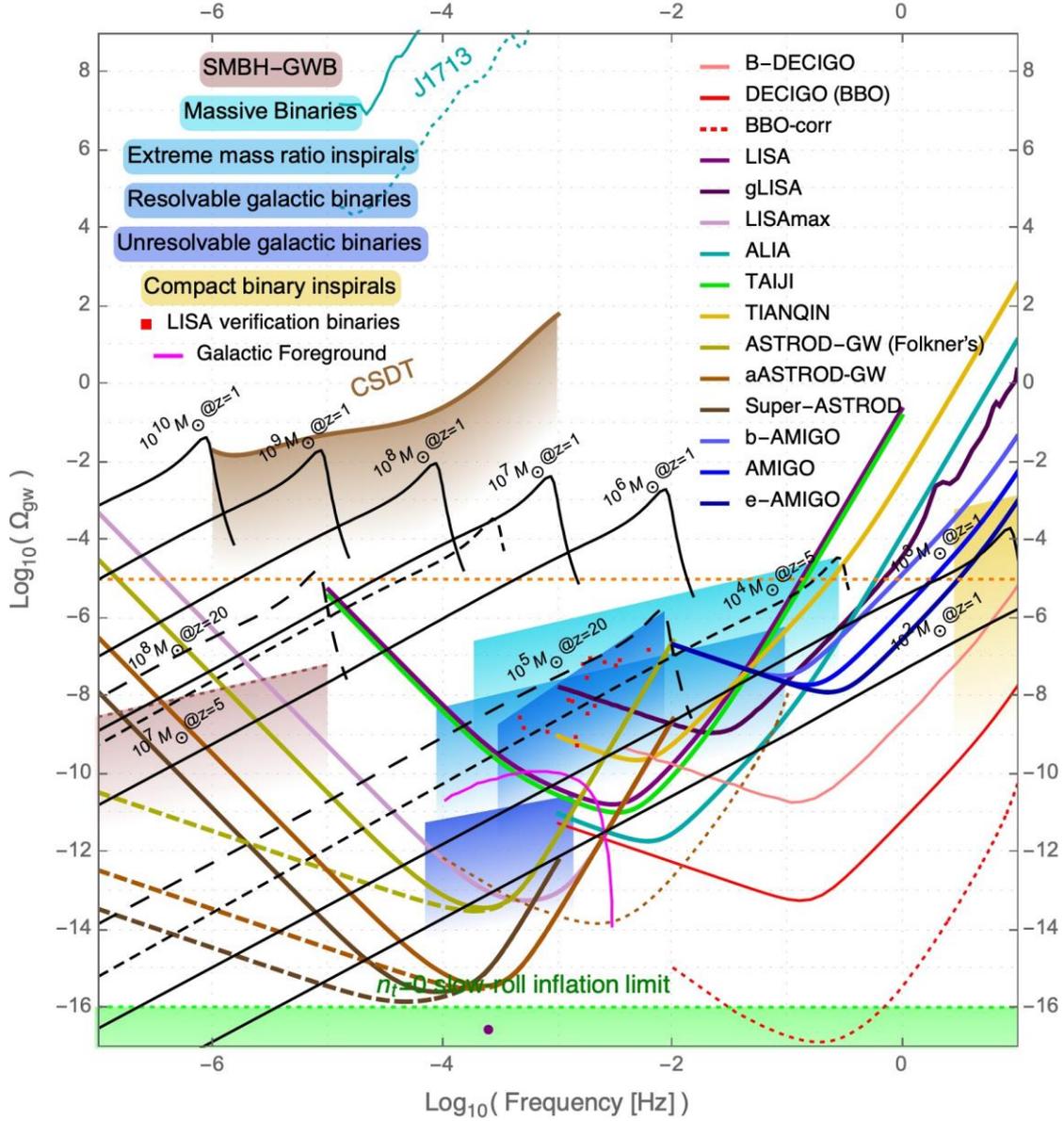

Figure 14 (Color online) Normalized GW energy spectral density $\Omega_{GW}$ vs. frequency for various GW detectors and GW sources. The black lines show the inspiral, coalescence and oscillation phases of GW emission from various equal-mass black-hole binary mergers in circular orbits at various redshift: Solid line, $z = 1$; dashed line, $z = 5$; long-dashed line $z = 20$. See text for more explanation. CSDT: Cassini Spacecraft Doppler Tracking; SMBH-GWB: Supermassive Black Hole-GW Background.

To test relativistic gravity, spacecraft need to enter inner solar orbits where the gravitational influence of the Sun is stronger, or send signals passing near the edge of the Sun to experience a stronger gravitational effect. An example of such a mission concept is ASTROD I, which used continuous laser ranging with 1 mm precision around superior solar conjunction to measure Shapiro delays of light and improve experimental errors in relativistic parameters [63, 71-73].

The BepiColombo spacecraft, a joint mission of the European Space Agency (ESA) and the Japan Aerospace Exploration Agency (JAXA), launched on October 20, 2018 (https://en.wikipedia.org/wiki/BepiColombo; http://sci.esa.int/bepicolombo/) [133, 134]. On October 1, 2021, it made its first flyby of Mercury and is set to arrive in orbit around Mercury on December 5, 2025 (https://www.esa.int/Science_Exploration/Space_Science/BepiColombo). One of its radio science objectives is to test relativistic gravity. While in orbit around Mercury, it will measure the motion of Mercury's center of mass with accuracy several orders of magnitude higher than radar ranging to the surface of Mercury. This presents a good opportunity to measure the advance of Mercury's perihelion and Shapiro time delay, and improve post-Newtonian parameters by orders of magnitude [135].

To measure or improve solar and planetary parameters, spacecraft need to approach the object under study or have extremely high sensitivity. Typical examples include the NEAR Near-Earth Asteroid Rendezvous mission [136] and the MESSENGER Mercury probe [137].

Current activities in space gravitational wave detection primarily focus on the development and implementation of space-based laser interferometric gravitational wave detectors. Given that current laser stability reaches only $10^{-16}$, which does not meet the requirements of space gravitational wave detection at $10^{-21}$, it is necessary to use generalized Michelson interferometry methods to ensure that the optical path difference between two beams of light remains below a certain threshold.

Space-based laser interferometric gravitational wave detection is extremely sensitive within its detection frequency range and holds great potential for measuring gravitational magnetic effects. In the next section, we will discuss possible methods for measuring gravitational magnetic effects.

## 2.2 Gravitomagnetic Effect, and the Measurement of Solar Angular Momentum and Galactic Angular Momentum

When the charge is stationary, it produces an electric field; when it moves, it produces a magnetic field. When matter is stationary, it produces a gravitational (gravitoelectric) field; when it moves, it produces a gravitomagnetic field. Gravitomagnetic field can drag local reference frames with/around it it, which is an important aspect of relativistic gravity and also a manifestation of gravitomagnetism. After Lense-Thirring published their paper [138], more than 100 years of theoretical research and experimental efforts have resulted in the astrodynamical measurements of the Gravity Probe B and the LAGEOS-LARES mission achieving a precision of Earth's gravitomagnetic field at the level of 2% [139, 140, 141, 142]. Planned/proposed astrodynamical missions within the solar system would potentially measure the solar gravitomagnetic effect separately from other target quantities and further improve the experimental measurement accuracy of the gravitomagnetic effect. A specific series proposal for astrodynamical missions is the ASTROD series --- ASTROD I, ASTROD (ASTROD II), ASTROD-GW, and Super-ASTROD [72, 73, 74, 75, 78, 81, 82, 83]. LISA [45] and TAIJI [98] are two ongoing astrodynamical missions in the solar system, with their main goals being the detection of gravitational waves. However, due to their precision requirements, they have the potential to test solar system relativistic gravity and measure the Lense-Thirring effect. This potential is discussed here. In space astrodynamical missions, the Sagnac effect in the gravitational field is much larger than Lense-Thirring's, which must be clarified first before separate measurements can be made. In rotating platforms/frames, light circulating in one direction and the opposite direction takes different time intervals, which is commonly referred to as the Sagnac effect. For split light beams, the phase shift $\Delta\varphi$ derived from Sagnac interference fringes is:

$$\Delta\varphi \approx (8\pi/\lambda c)\boldsymbol{\omega} \cdot \mathbf{A}, \tag{3}$$

where $\boldsymbol{\omega}$ is the angular rotation speed of the platform/frame, $\mathbf{A}$ is the oriented area of the loop, and $\lambda$ is the wavelength of light. The phase shift is proportional to the dot product of the angular rotation speed and the oriented area of the frame. Converted into a time difference $\Delta t$ for two light paths, equation (3) transforms into

$$\Delta t \approx 4\boldsymbol{\omega} \cdot \boldsymbol{A} / c^2. \tag{4}$$

For mobile formations, in a gravitational field (or where the gravitational field can be neglected in calculation), the time difference in light propagation between closed and reverse paths can also be referred to as the Sagnac effect.

Table 3 lists the numerical optical path differences for the Sagnac-α, β, γ and Sagnac-α2, β2, γ2 Time Delay Interferometry (TDI) configurations of spacecraft formations during scientific simulation missions of LISA and Taiji (refer to Section 7.2 (48) formula and explanation), as well as the average optical path differences for α, β, γ at different starting points [145]. The 10$^{th}$ row of Table 3 shows the Sagnac effect calculated analytically under a triangular formation with an angle of 60 degrees to the ecliptic plane. The calculation is as follows (Ni W-T, Ciufolini I, Wang G. LISA, Taiji, Sagnac effect, Lense-Thirring effect, the solar angular momentum and the Galactic angular momentum, in preparation).

In the reference frame where the center of mass of the LISA/Taiji formation rotates around the sun, if the LISA/Taiji formation is stationary, the Sagnac time difference between the two optical paths of Sagnac-α obtained from formula (4) is

$$\Delta t_1 \approx 4\boldsymbol{\Omega} \cdot \mathbf{A} / c^2 = 3^{1/2}\Omega(L^2/c^2)\cos 60° + O(e) = (3^{1/2}/2)\Omega(L^2/c^2) + O(e), \tag{5}$$

where e is the eccentricity of the spacecraft orbit, see Section 10.1.

The additional delay caused by the LISA/Taiji formation rotating clockwise about its center is

$$\Delta t_2 \approx 4\boldsymbol{\Omega} \cdot \boldsymbol{A} / c^2 = -3^{1/2}\Omega(L^2/c^2)\cos 0° = -3^{1/2}\Omega(L^2/c^2) + O(e). \tag{6}$$

Hence, the total Sagnac time difference is

$$\Delta t_{\text{Sagnac}} = \Delta t_1 + \Delta t_2 = -(3^{1/2}/2c^2)\Omega(L^2/c^2) + O(e). \tag{7}$$

**Table 3** Compilation of the RMS (root mean square) path length differences together with their Sagnac part and Lense-Thirring part of first generation Sagnac-α, Sagnac-β and Sagnac-γ, and second generation Sagnac-α2, Sagnac-β2 and Sagnac-γ2 TDI configurations for LISA and Taiji[a] [145,153]

| Sagnac TDI configuration | LISA TDI path difference RMS average [min, max], RMS deviation | Taiji TDI path difference RMS average [min, max], RMS deviation |
|---|---|---|
| Sagnac-α | 11911 [−12309, −11551], 153 ns | 17151 [−17759, −16623], 234 ns |
| Sagnac-β | 11915 [−12262, −11624], 144 ns | 17156 [−17666, −16749], 216 ns |
| Sagnac-γ | 11906 [−12199, −11593], 125 ns | 17145 [−17611, −16661], 192 ns |
| Average | 11912 ns | 17151 ns |
| Sagnac-α2 | 1.3 [−2.7, −3.0] ps | 2.2 [−4.8, −5.2] ps |
| Sagnac-β2 | 1.2 [−3.0, −2.6] ps | 2.1 [−5.2, −4.4] ps |
| Sagnac-γ2 | 1.2 [−2.9, −2.7] ps | 2.1 [−5.0, −4.9] ps |
| Average | 1.23 ps | 2.13 ps |
| Sagnac part (equalateral model with 60° inclination) | 11989 ns | 17265 ns |
| Lense-Thirring effect due to the Sun | 3.52$\chi\cos\lambda'$ as (1.05$\chi\cos\lambda'$ nm) | 5.04$\chi\cos\lambda'$ as (1.52$\chi\cos\lambda'$ nm) |
| Lense-Thirring effect due to the Galaxy | 0.075$\chi'\cos\lambda''$ as (0.022$\chi'\cos\lambda''$ nm) | 0.107$\chi'\cos\lambda''$ as (0.032$\chi'\cos\lambda''$ nm) |
| Laser metrology noise of the detector @ 1 mHz | ~41 pm Hz$^{-1/2}$ | ~33 pm Hz$^{-1/2}$ |
| Laser metrology noise of the detector @100 μHz | ~4000 pm Hz$^{-1/2}$ | ~3200 pm Hz$^{-1/2}$ |
| Inertial sensor/accelerometer noise @ 1 mHz | 3 fm s$^{-2}$ Hz$^{-1/2}$ | 3 fm s$^{-2}$ Hz$^{-1/2}$ |
| Inertial sensor/accelerometer noise @ 100 μHz | 13 fm s$^{-2}$ Hz$^{-1/2}$ | 13 fm s$^{-2}$ Hz$^{-1/2}$ |

a) $\chi$ is a parameter with a value of approximately 1, determined by Lense-Thirring measurements. $\lambda'$ is the angle between the orbital plane normal and the direction of the Sun's angular momentum. $\chi'$ is a parameter determined by Lense-Thirring measurements, with a value of around or greater than 1. $\lambda''$ is the angle between the orbital plane normal and the direction of the Galaxy's angular momentum.

For LISA, $L_{\text{LISA}}$ = 2.5 Gm, $\Omega$ = 1.9909865788600 × 10$^{-7}$ Hz, we have

$$\Delta t_{\text{Sagnac-LISA}} = - (3^{1/2} / 2c^2)\Omega(L_{\text{LISA}}^2 / c^2) + O(e) = - 1.199054 \times 10^{-5} \text{ s} \times [1 + O(e)]. \tag{8}$$

For Taiji, $L_{\text{Taiji}}$ = 3 Gm, $\Omega$ = 1.9909865788600 × 10$^{-7}$ Hz, we have

$$\Delta t_{\text{Sagnac-Taiji}} = - (3^{1/2} / 2c^2)\Omega(L_{\text{TAIJI}}^2 / c^2) + O(e) = - 1.726638 \times 10^{-5} \text{ s} \times [1 + O(e)]. \tag{9}$$

The next approximation needs to consider the linear correction term for eccentricity *e*. Shown on the right of Figure 9, if the formation plane is selected at an angle of −60 ° to the ecliptic plane, the sign of $\Delta t_2$ changes, and the Sagnac time difference becomes −3 times that of formulas (8) and (9).

In rotating bodies' gravitational fields, local inertial frames are rotated by the rotating body. If the rotational speed of the rotating body is of the order of $\Omega$, the local inertial frame is rotated at a speed of the order $(U/c^2) \Omega$, where *U* is the Newtonian potential of the rotating body at that point. This effect is a type of gravitomagnetic effect, commonly known as frame dragging or the aforementioned Lense-Thirring effect.

The Lense-Thirring effect is very important in astrophysics and in the determination and implementation of ultra-precision astronomical reference frames. The Gravity Probe B and Lageos-Lares laser ranging experiments have measured Earth's Lense-Thirring effect to 2%. Several authors have also proposed methods to measure the Lense-Thirring effect caused by the Sun (e.g., references [146]) and proposed the possibility of measuring the Lense-Thirring effect of the Milky Way (e.g., references [147]).

In a quasi-Minkowski coordinate system, for weak field situations, as a first approximation, we express the metric $g_{\alpha\beta}$ as

$$g_{\alpha\beta} = \eta_{\alpha\beta} + h_{\alpha\beta}, |h_{\alpha\beta}| \ll 1, \eta_{\alpha\beta} = \mathrm{diag}(1, -1, -1, -1), \tag{10}$$

$$h_{00} = -2U/c^2 \approx -2(GM_\odot/c^2), \tag{11}$$

$$h_{0i} = -2(G/c^3)(\mathbf{J} \times \mathbf{x})_i/\mathbf{r}^3. \tag{12}$$

Since $h_{\alpha\beta}$ is a small quantity in the solar system's gravitational field ($< 4 \times 10^{-6}$), we expand and linearize $h_{\alpha\beta}$, obtaining the linear (weak field) approximation of light propagation equations under the physical metric (10-12). Let $\mathbf{r} = \mathbf{r}(t) = (x(t), y(t), z(t))$ be the trajectory of light. Light propagation follows the geodesic of metric $g_{\alpha\beta}$, with trajectory $\mathbf{r}(t)$ satisfying

$$0 = ds^2 = g_{\alpha\beta}dx^\alpha dx^\beta = (1 + h_{00})c^2 dt^2 + 2h_{0i}cdx^i dt - (\delta_{ij} - h_{ij})dx^i dx^j. \tag{13}$$

Under the post-Minkowski approximation, we represent $dx^i/dt$ as

$$dx^i/dt = (dx^i/dt)^{(0)i} + O(h) = cn^{(0)i} + cn^{(1)i} + O(h^2), \text{ with } \sum_i (n^{(0)i})^2 = 1, \tag{14}$$

where $cn^{(0)i}$ are constants, $cn^{(1)i}$ are functions of the trajectory, and are of order $O(h)$. Substituting (14) into (13), we solve for $|d\mathbf{r}/dt|$, obtaining the light propagation equation of order $O(h)$:

$$|d\mathbf{r}/dt| = c[1 + (1/2)h_{00} + h_{0i}n^{(0)i} + (1/2)h_{ij}n^{(0)i}n^{(0)j} + O(h^2)]. \tag{15}$$

Thus, the light travel time $\Delta t_{\mathrm{TT}}$ (time delay) between two observers (spacecraft) is

$$\Delta t_{\mathrm{TT}} = (1/c)\int |d\mathbf{r}|[1 - (1/2)h_{00} - h_{0i}n^{(0)i} - (1/2)h_{ij}n^{(0)i}n^{(0)j} + O(h^2)]. \tag{16}$$

For the solar system with quadrupole moment $J_2$ and angular momentum $\mathbf{J}$ [$= (J_x, J_y, J_z)$],

$$g_{00} = 1 - 2U/c^2 + 2\beta U^2/c^4 + O(v^5/c^5),$$
$$g_{0i} = -2(G/c^3)(\mathbf{J} \times \mathbf{x})_i/\mathbf{r}^3 + O(v^5/c^5),$$
$$g_{ij} = -(1 + 2\gamma U/c^2)\delta_{ij} + O(v^4/c^4).$$
$$\tag{17}$$

Choosing the z-axis along the initial light propagation direction from $z_1$ to $z_2$, i.e., $n^{(0)i} = (0, 0, 1)$, we obtain

$$\Delta t_{\mathrm{TT}} = \int dt = (1/c)\int dz[1 + \tfrac{1}{2}(1+\gamma)U + h_{0i}n^{(0)i} + O(h^2)] = \Delta t^N + [(1+\gamma)/2]\Delta t_S^{\mathrm{GR}} + \Delta t_{L\text{-}T}^{\mathrm{GR}} + O(h^2)]$$
$$= (1/c)(z_2 - z_1) + (1+\gamma)(GM_\odot/c^3)\ln\{[(z_2^2 + b^2)^{1/2} + z_2]/[(z_1^2 + b^2)^{1/2} + z_1]\}$$
$$+ (2/c^3)G_N J \cos\lambda' \cdot \{2/b - 1/(z_2^2 + b^2)^{1/2} - 1/(z_1^2 + b^2)^{1/2}\} + O(h^2), (z_1 < 0, z_2 > 0), \tag{18}$$

where the first term is the Newtonian light travel time $\Delta t^N$ (Römer[7] delay), the second term is the relativistic Shapiro time delay $\Delta t_S^{GR}$ [148,149], and the third term is the Lense-Thirring effect on light travel time $\Delta t_{TT-LT}^{GR}$. $b$ is the minimum distance of the light trajectory from the Sun (impact parameter), and $\lambda'$ is the angle between the orbital plane normal and the solar angular momentum direction.

The gravitatomagnetic field $h_{0i}$ effect in equation (12), namely the Lense-Thirring effect on light travel time, is

$$\Delta t_{TT-LT} = (1/c)\int |d\mathbf{r}|\,[h_{0i}n^{(0)i} + O(h^2)] = (1/c)\int d\mathbf{r}\cdot\mathbf{h} + O(h^2). \tag{19}$$

Here $\mathbf{h} = (h_{0i})$ is a 3-vector. According to Stokes' theorem, for a closed path

$$\Delta t_{TT-LT} = (1/c)\int_S \nabla\times\mathbf{h}\cdot d\mathbf{S} + O(h^2), \tag{20}$$

where, $\mathbf{S}$ denotes an oriented surface enclosed by oriented closed paths. For local space networks like LISA/TAIJI, the integral of $\Delta t_{TT-LT}$ can be approximated using the inner product of $\nabla\times\mathbf{h}$ at the center of the detection array and the oriented surface area $\mathbf{S}$:

$$\Delta t_{TT-LT} = (1/c)\,\nabla\times\boldsymbol{h}\cdot\boldsymbol{S} = (1/c)\,|\nabla\times\boldsymbol{h}|S\cos\lambda', \tag{21}$$

where $\lambda'$ is the angle between $\nabla\times\mathbf{h}$ at the center of the detection array and the oriented surface area $\mathbf{S}$, and $S$ represents the absolute value of $\mathbf{S}$.

Several determinations of the solar rotational inertia and angular momentum obtained through helioseismology observations are approximately around $2\times10^{41}$ kg m$^2$ s$^{-1}$. Here, we use the parameter $\chi$ to represent an approximately determined value close to 1, i.e., $J_{sun} = 2\chi\times10^{41}$ kg m$^2$ s$^{-1}$ (a modern value for solar spin angular momentum is $1.8838\times10^{41}$ kg m$^2$ s$^{-1}$ [150], and Allen's Astrophysical Quantities, 3rd edition, page 161 gives a value of $1.63\times10^{41}$ kg m$^2$ s$^{-1}$). Due to the deviation of solar angular momentum from the vertical direction relative to the ecliptic plane by 7°, for missions like LISA and Taiji, $\lambda'$ varies approximately between 60°±7° with an annual cycle.

For TAIJI, aside from factor of cos the cos $\lambda'$, $\Delta t_{TT-LT}$ = 0.758 nm = 2.52 as, for LISA, $\Delta t_{TT-LT}$=0.526 nm = 1.76 as. The Lense-Thirring effect in time delay interferometry (TDI) involves twice the time (optical path) difference in Sagnac-α2, Sagnac-β2, Sagnac-γ2 (positive loop minus negative loop, see Section 7.2). Thus, for LISA, it amounts to 3.52 as (1.05 nm), and for TAIJI, it amounts to 5.04 as (1.52 nm) (converted to strain amplitude of 0.4-0.5 × 10$^{-18}$ at a 0.03 μHz annual frequency). These numerical values are listed in Table 3, sixth row from the bottom.

For the inner Galaxy angular momentum $J^{inner}_{Galaxy}$, a benchmark model is

$$J^{inner}_{Galaxy} = M^{inner}_{Galaxy} \times R_{\text{sun-to-Galaxy}} \times V_{\text{solar-system-relative-to-Galaxy}}. \tag{22}$$

Choosing $M^{inner}_{Galaxy}$ = mass inside Sun's orbit around Galactic Center = $2 \times 10^{11}\,M_{sun}$ (Galactic total mass: ~$10^{12}\,M_{sun}$; Galactic virial mass: $1.3\pm0.3 \times 10^{12}\,M_{sun}$); $R_{\text{sun-to-Galaxy}}$ = 8.2 kpc (8.2±0.1 kpc, Bland-Hawthorn and Gerhard 2017 [151]）；$V_{\text{solar-system-relative-to-Galaxy}}$ = 200 km/s  (250 km/s, Bland-Hawthorn and Gerhard 2017 [151]), we make a conservative estimate of $J^{inner}{}_{\text{Galaxy}}$ to be

$$J^{inner}{}_{\text{Galaxy}} = \chi' \ 2 \times 10^{11} \ M_\odot \times 2 \times 10^{30} \text{ kg}/M_\odot \times 8.2 \text{ kpc} \times 3.086 \times 10^{19} \text{ m/kpc} \times 200 \text{ km/s}$$
$$= \chi' \ 2 \times 10^{67} \text{ kg m}^2/\text{s}. \tag{23}$$

Because the Solar System is very close to the Galactic equatorial plane, the Galactic gravitational field is approximately

$$\nabla \times \boldsymbol{h} \sim 2(G/c^3) \, (\boldsymbol{J}/r^3) = 0.0626 \times 10^{-28} \text{ m}^{-1} \cdot \chi', \tag{24}$$

therefore, on the Earth's orbit around the Sun, the ratio of the Galactic to Solar gravitational field is

$$0.0626 \times 10^{-28} \text{ m}^{-1} \cdot \chi' / (0.292 \times 10^{-27} \text{ m}^{-1} \cdot \chi) = 0.0214 \, \chi'/\chi = (1/47) \, (\chi'/\chi). \tag{25}$$

Because the normal of the LISA and Taiji spacecraft formation plane and the Solar angular momentum angle $\lambda'$ varies between 60°±7° annually, the gravitational magnetic effect also has an annual variation with a frequency of 0.03 μHz. LISA and Taiji have an inertial-sensing noise requirement of 3 fm s$^{-2}$ Hz$^{-1/2}$ at 1 mHz, to achieve an accuracy of 0.4−0.5 × 10$^{-18}$ @ 1 mHz, requiring approximately 1 year. At 100 μHz, it is 10$^{-15}$ fm s$^{-2}$ Hz$^{-1/2}$, requiring approximately 10 years. However, at 0.03 μHz LISA/Taiji has no requirment, extraneous optimistic estimate with LISA/Taiji method, it might reach ~1 nm s$^{-2}$ Hz$^{-1/2}$, requiring approximately 10$^{7\text{-}8}$ years. In the fourth international workshop on gravitational magnetic and large-scale rotation measurements (GRM 2023: IV International Workshop on Gravitomagnetism and Large-Scale Rotation Measurement, June 14-16, 2023, Pisa) the consensus is that with the current LISA and Taiji space mission plans, the detection of solar gravitomagnetic effects is not forseen [152, 153] (papers being prepared: Ni W-T, Ciufolini I, Wang G. LISA, TAIJI, Sagnac effect, Lense-Thirring Effect, the Solar angular momentum, and the Galactic angular momentum.).

Equation (18) affects the change $\Delta t_{\text{TT-LT}}$ in light propagation time due to gravitomagnetic influence:

$$\Delta t_{\text{TT-LT}} = (2/c^3) \, G_N J \cos \lambda' \cdot \{2/b - 1/(z_2^2+b^2)^{1/2} - 1/(z_1^2+b^2)^{1/2}\} + O(h^2), \, (z_1 < 0, z_2 > 0), \tag{26}$$

For the propagation of laser beams between the ASTROD-GW spacecraft, where b=0.5 AU, $z_1$=−0.866 AU, $z_2$=0.866 AU. From this equation, the time $\Delta t_{\text{TT-LT}}$ affected by the solar gravitomagnetic field is calculated as:

$$\Delta t_{\text{TT-LT}} = 0.0439 \text{ ps} = 13.17 \text{ μm}/c. \tag{27}$$

For Sagnac-α, Sagnac-β, Sagnac-γ: $\Delta t_{\text{TT-LT}}$ = 0.0439 ps × 6 = 0.263 ps = 70.9 μm/c.

Table 4 lists the path differences in time-delay interferometry (TDI) configurations for various orbital inclinations during the ASTROD-GW scientific mission [refer to Section 7.2], rows 3-6, and the average path differences for different starting points α, β, γ [82]. Rows 7, 8, 9 of Table 4 respectively list the Sagnac effect of the formation under various orbital inclinations, contributions of orbital inclination, and Lense-Thirring effect, and rows 10-13 list the relevant noise of ASTROD-GW. For detailed analysis and discussion of whether the solar angular momentum and galactic angular momentum can be measured, please refer to [153, 154].

For laser/light passing near the edge of the Sun with b ≈ R☉, if z1 = -0.866 AU, z2 = 0.866 AU. From this (26) equation, the duration affected by the solar gravitomagnetic field is calculated as:

$$\Delta t_{TT\text{-}LT} = 9.42 \text{ ps} = 2.825 \text{ mm}/c. \tag{28}$$

The accuracy of pulse laser ranging can reach 1 mm or better. In the ASTROD-GW mission concept, if ground stations add dual-wavelength dual-unidirectional laser ranging with spacecraft near L3, the gravitomagnetic effect of the Sun can be measured, and thus the solar angular momentum can be determined. In the concepts of ASTROD-I (Mini-ASTROD) and ASTROD (ASTROD-II) space missions, lasers will pass near the edge of the Sun, enabling measurement of the solar gravitomagnetic effect and hence the determination of solar angular momentum [72, 73 and references therein].

**Table 4** Compilation of the RMS path length differences together with their Sagnac part and Lense-Thirring part of first generation Sagnac-α, Sagnac-β and Sagnac-γ TDI configurations for ASTROD-GW of various degrees of formation inclination (0°, 0.5°, 1°, 1.5°, 2°, 2.5°, and 3°) with respect to the ecliptic plane. $\chi$ is a parameter of value ≈ 1 to be determined by ASTROD-GW Lense-Thirring measurement a) [82,153,154]

| Sagnac TDI configuration | ASTROD-GW TDI path difference | | | | | | |
|---|---|---|---|---|---|---|---|
| | 0° (μs) | 0.5° (μs) | 1° (μs) | 1.5° (μs) | 2° (μs) | 2.5° (μs) | 3° (μs) |
| Sagnac-α | 257610 | 257590 | 257531 | 257432 | 257293 | 257115 | 256898 |
| Sagnac-β | 257608 | 257588 | 257529 | 257431 | 257294 | 257118 | 256902 |
| Sagnac-γ | 257607 | 257588 | 257530 | 257432 | 257297 | 257122 | 256909 |
| Average | 257608.3 | 257588.7 | 257530 | 257431.7 | 257294.7 | 257118.3 | 256903 |
| Sagnac part | 257608 | 257588 | 257530 | 257431 | 257294 | 257.118 | 256902 |
| Sagnac part due to $\lambda \neq 0$ | 0 | −19.62 | −78.47 | −176.56 | −313.88 | −490.44 | −706.24 |
| Lense-Thirring part | $0.263\chi\cos\lambda'$ ps | $0.263\chi\cos\lambda'$ ps | $0.263\chi\cos\lambda'$ ps | $0.263\chi\cos\lambda'$ ps | $0.263\chi\cos\lambda'$ ps | $0.263\chi\cos\lambda'$ ps | $0.263\chi\cos\lambda'$ ps |
| Laser metrology noise of the detector @ 100 μHz | ASTROD-GW ~3000 pm Hz$^{-1/2}$ | | | | | | |
| Laser metrology noise of the detector @ 0.03 μHz | ASTROD-GW ~30 μm Hz$^{-1/2}$ | | | | | | |
| Inertial sensor/accelerometer noise @ 100 μHz | ASTROD-GW ~4.2 fm s$^{-2}$ Hz$^{-1/2}$ | | | | | | |
| Inertial sensor/accelerometer noise @ 0.03 μHz | ASTROD-GW ~13 pm s$^{-2}$ Hz$^{-1/2}$ | | | | | | |

a) $\chi$ is a parameter with a value of approximately 1, determined by Lense-Thirring measurements. $\lambda'$ is the angle between the orbital plane normal and the direction of the Sun's angular momentum. $\chi'$ is a parameter determined by Lense-Thirring measurements, with a value around or greater than 1. $\lambda''$ is the angle between the orbital plane normal and the direction of the Galaxy's angular momentum. Upcoming paper: Ni W-T, Ciufolini I, Wang G. LISA, Taiji, Sagnac effect, Lense-Thirring effect, the solar angular momentum, and the Galactic angular momentum.

## 3 Spacecraft Doppler Tracking and Pulsed Laser Ranging

### 3.1 Spacecraft Electromagnetic Wave Doppler Tracking and Its Gravitational Wave Response

Radio Doppler tracking of spacecraft in space missions can be used to constrain/detect the magnitude of low-frequency gravitational waves. Such gravitational wave detection experiments involve ground-based Doppler tracking radio antennas and distant spacecraft. Doppler tracking measures changes in their relative distances. Through these measurements, the size of gravitational waves between test masses can be detected/constrained. In 1967, Braginsky and Gertsenshtein [155] first proposed the use of Doppler data from spacecraft tracking for gravitational wave searches. In 1971, Anderson [156] utilized existing data for this purpose. Davis [157] derived Doppler tracking gravitational wave responses under certain special conditions in 1974; Estabrook and Wahlquist [125] derived a general formula in 1975 for the effects of gravitational waves crossing the line of sight of a spacecraft on Doppler tracking frequency measurements (also see [158]).

In spacecraft Doppler tracking, highly stable clocks on Earth are used as references to control single-frequency radio wave transmissions to the spacecraft (uplink). When the transponder on the spacecraft receives the single-frequency radio wave, it phase-locks or frequency-offset locks its local oscillator and coherently retransmits the local oscillator signal to ground stations (downlink).

The one-way Doppler response $y(t)$ is defined as:

$$y(t) \equiv \delta v/v_0 \equiv (v_1(t) - v_0)/v_0, \tag{29}$$

where $v_0$ is the frequency of the transmitted signal and $v_1$ is the frequency of the received signal. In current experimental/observational scenarios, gravitational wave sources are very distant from us, and upon arrival, can be treated as plane waves with very small amplitudes. For small-amplitude plane waves propagating along the z-direction in general relativity, the spacetime metric is:

$$ds^2 = dt^2 - (\delta_{ij} + h_{ij}(ct - z))dx^i dx^j, \qquad |h_{ij}| \ll 1, \tag{30}$$

where Latin indices run from 1 to 3, and repeated indices imply summation. Estabrook and Wahlquist [125, 158] derived the one-way and two-way Doppler responses for plane gravitational waves in the weak-field approximation (Equations (10)) in a transverse-traceless gauge. For spacecraft S/C2 receiving signals from S/C1, the one-way Doppler response formula, using the notation from Armstrong, Estabrook, and Tinto [159], is:

$$y(t) = (1 - \underline{k} \cdot \underline{n}) \, [\Psi(t - (1 + \underline{k} \cdot \underline{n})L) - \Psi(t)], \tag{31}$$

where $\mathbf{k} = (k^i) = (k^1, k^2, k^3)$ is the unit vector in the direction of gravitational wave propagation, $\mathbf{n} = (n^i) = (n^1, n^2, n^3)$ is the unit vector along the spacecraft 1 to spacecraft 2 link, and $L$ is the path length of the Doppler link. The function $\Psi(t)$ is defined as:

$$\Psi(t) \equiv -n^i h_{ij}(t) n^j / \{2[1 - (\underline{k} \cdot \underline{n})^2]\}. \tag{32}$$

As noted and derived by Tinto and da Silva Alves [160], the Doppler response formulas (31) and (32) are also valid for any gravitational wave solution in the form of Equation (10) of any gravitational metric theory.

From the one-way Doppler response, the two-way and multi-link responses can be derived. The two-way Doppler response is derived as follows. The local oscillator transmits a beam aimed at the uplink spacecraft, and produces beats with signals returned from spacecraft and measures the phase. The phase (and frequency) measured as a function of time is analyzed. The Doppler response for a single link is given by (31). The two-way Doppler tracking [125, 158] response is given by:

$$y(t) = -(1 - \underline{k} \cdot \underline{n})\,\Psi(t) - 2(\underline{k} \cdot \underline{n})\,\Psi(t - (1 + \underline{k} \cdot \underline{n})L) + (1 + \underline{k} \cdot \underline{n})\,\Psi(t - 2L). \tag{33}$$

The three terms in equation (33) correspond respectively to the amplitude of gravitational waves when receiving Doppler tracking signals on Earth, when spacecraft retransmit, and when tracking signals are transmitted from Earth. Microwave Doppler tracking involves an ultra-stable oscillator on the spacecraft. Its precision is not as good as ground-based hydrogen-maser clocks, so the accuracy of microwave Doppler tracking depends on ground-based hydrogen-maser clocks (or other clocks used on the ground).

Doppler tracking of Viking S/C (S-band, 2.3 GHz) [161], Voyager I S/C (S-band uplink + coherent retransmission S-band and X-band (8.4 GHz) downlink) [162], Pioneer 10 (S-band) [163], and Pioneer 11 (S-band) [164] has been used for gravitational wave detection and provided constraints on the low-frequency gravitational wave background.

The latest measurements come from the Cassini spacecraft Doppler tracking (CSDT). Armstrong, Iess, Tortora, and Bertotti [165] used the Cassini multi-link radio system during the 2001-2002 flybys to improve observational constraints on the isotropic background of low-frequency gravitational waves. The Cassini multi-link radio system simultaneously receives signals from the X and Ka bands on two uplinks and transmits three downlink signals, where the X band is coherent with the X-band uplink, the Ka band is coherent with the X-band uplink, and the Ka band is coherent with the Ka-band uplink. Armstrong et al. [165] used the Cassini multi-link radio system with higher frequencies and advanced tropospheric calibration systems to reduce the impact of major noise sources (plasma and tropospheric scintillation) to levels below other noises. The data obtained were used to construct upper limits of isotropic background intensity in the frequency range of 1 μHz to 1 mHz [165]. The characteristic strain upper limit curve labeled CSDT in Figure 13 is a smoothed version of curve 4 in reference [165]. The corresponding CSDT curves in Figure 12 for strain spectral density amplitude and in Figure 14 for normalized spectral energy density were converted using formula (1). The minima on these curves are: (i) at several frequencies in the range of 0.2-0.7 mHz $[S_h(f)]^{1/2} < 8 \times 10^{-13}$; (ii) at a frequency of approximately 0.3 mHz, $h_c(f) < 2 \times 10^{-15}$; (iii) at a frequency of 1.2 μHz, $\Omega_{gw}(f) < 0.03$.

If equipped with starborne optical clocks, the gravitational wave sensitivity of spacecraft Doppler tracking can be improved by at least 1-2 orders of magnitude [166].

In spacecraft radio tracking, the frequency of the received signal is tracked. Its integral is the phase. In spacecraft radio ranging, the phase of the received signal is measured. The derivative of the phase is the frequency. For coherent retransmission, apart from determining a constant addition, the accumulated measured phase is essentially relative distance.

**3.2 Pulsed Laser Ranging**

Another method for measuring range and distance is using pulse timing. This is what satellite laser ranging and lunar laser ranging do. For ranging through the Earth's atmosphere, the optimal method to find atmospheric delay is to use two colors (two wavelengths) to measure the atmospheric delay and subtract the effect. The ranging accuracy of dual-color (two-wavelength) satellite laser ranging has reached sub-millimeter levels. With the implementation of the new generation of lunar laser ranging [167, 168], the precision of lunar distance measurements has also reached millimeter levels and is approaching sub-millimeter levels. The laser link time transfer (T2L2) event timer on the Jason 2 satellite has achieved a starborne timing accuracy of 3 ps (0.9 mm) [169, 170]. Based on these advancements, the technological capability of single-directional ranging systems across the entire solar system can achieve millimeter or even sub-millimeter precision. With this level of accuracy and an expanded range of 20 astronomical units, the ability to probe the fundamental laws of spacetime and measure the solar system's gravitational field will be significantly enhanced [72-75]. For 1 millimeter in 20 astronomical units, the fractional uncertainty is $3\times10^{-16}$. Achieving this fractional uncertainty level requires matched laser stability and clock accuracy; this precision has been achieved in the laboratory and will be realized in space. One of the goals of ASTROD I [71-75, 128] using onboard precise clocks is to enhance the gravitational wave sensitivity of the Cassini spacecraft Doppler tracking by an order of magnitude. In fact, optical clock stability and quasi-precision have already reached the $10^{-19}$ level. When space optical clocks achieve this level, pulse laser ranging combined with drag-free technology will become an important alternative for microhertz low-frequency gravitational wave detection.

The fundamental principles of spacecraft Doppler tracking, spacecraft laser ranging, space laser interferometry, and pulsar timing arrays (PTAs) for gravitational wave detection are similar. In the development of gravitational wave detection methods, spacecraft Doppler tracking and pulse laser ranging have played important roles. The methods using space laser interferometry and pulsar timing arrays have become two important approaches for detecting gravitational waves. The following sections will discuss pulsar timing gravitational wave detection, clock and atomic interferometry gravitational wave detection, and core noise in space laser interferometry gravitational wave detection.

**4 Pulsar Timing Array Gravitational Wave Detection**

In 1967, pulsars were discovered. In 1978, Sazhin proposed using pulsar timing to detect gravitational waves from binary supermassive stars with masses between $10^8$ and $10^{10}$ $M_\odot$, which have orbital periods of years [171]. In 1979, Detweiler analyzed published data from pulsar pulse emission and gave an upper limit on the energy density of random gravitational waves with a period of 1 year as the critical density of the universe [172]. In 1982, Backer et al. discovered stable millisecond pulsars [173], increasing the precision of timing measurements.

When gravitational waves pass through the line of sight to pulsar observations, the arrival time of pulses is affected. This effect can be used to observe gravitational waves. The phase when pulsar pulses arrive at radio observatories can be observed, but their initial time and phase are unknown and must be statistically averaged over initial phases. In 1983, Hellings and Downs derived the correlation $c_{HD}$ of timing residuals from two pulsars with an observing angle $\Theta$ for an isotropic random gravitational wave background, which depends only on $\Theta$ and not on the individual angular positions of the two pulsars on the celestial sphere. The formula is [174]:

$$c_{HD} = 1/2 + (3/4)(1 - \cos\Theta)[\ln(1 - \cos\Theta) - 1/6]. \quad (34)$$

This formula is known as the Hellings-Downs formula, and the curve it describes is called the Hellings-Downs curve, as shown by the black dashed line in Figure 15. Because the background is assumed to be isotropic, the correlation should only depend on the angle and not on the individual angular positions of the pulsars. Because gravitational waves in general relativity are quadrupolar, the formula takes the above form. At that time, it was used to constrain the energy density of gravitational waves in the frequency range of 4-10 nHz to less than $1.4 \times 10^{-4}$ of the critical density of the universe. It also constrained the precision and stability of pulsars as timing signals.

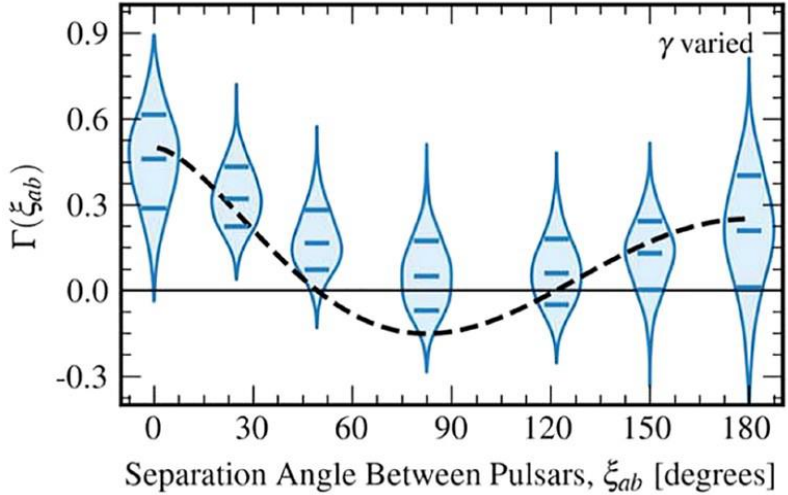

Figure 15 (Color online) Bayesian reconstruction of normalized interpulsar correlations, modeled as a cubic spline within a variable-exponent power-law model. The violins plot the marginal posterior densities (plus median and 68% credible values) of the correlations at the knots. The knot positions are fixed and are chosen on the basis of features of the Hellings-Downs curve (also shown as a dashed black line for reference): they include the maximum and minimum angular separations, the two zero-crossings of the Hellings-Downs curve, and the position of minimum correlation. See section 3 of ref. [185] for more details. The figure is from ref. [185].

In 1990, Foster and Backer [175] proposed using an array of stable millisecond pulsars for: (i) providing a long-term standard; (ii) detecting disturbances in the Earth's orbit; (iii) detecting cosmic background radiation. The millisecond pulsar array subsequently became the primary method for detecting nanohertz gravitational waves. In 1996 and 2002, McHugh et al. [176] and Lommen [177], respectively, derived upper limits on the gravitational wave background in the frequency range of 4-40 nHz as $\Omega_{gw} \leq 10^{-7}$, and at $6 \times 10^{-8}$ Hz as $\Omega_{gw} \leq 4 \times 10^{-9}$ from timing observations of millisecond pulsars. In 2002, the National Observatory of CAS also proposed constructing a 50 m radio telescope array to detect extremely low-frequency gravitational waves or provide better constraints using an array observing approximately 10 millisecond pulsars [178].

By 2015, there were three major pulsar timing arrays (PTA) engaged in long-term observations: the European Pulsar Timing Array (EPTA, http://www.epta.eu.org/), the North American Nanohertz Observatory for Gravitational Waves (NANOGrav, http://nanograv.org/), and the Parkes Pulsar Timing Array (PPTA, http://www.atnf.csiro.au/research/pulsar/ppta/) [179]. Reference [179] discusses pulsar timing arrays and their sensitivity.

Table 5 lists the upper limits on isotropic stochastic backgrounds given by EPTA, PPTA, and NANOGrav in 2015 [180-182]. These constraints assume a relationship between the characteristic strain $h_c(f)$ of the gravitational wave background and frequency:

$$h_c(f) = A_{yr} [f / (1 \text{ yr}^{-1})]^\alpha. \quad [\alpha = -(2/3)] \tag{35}$$

**Table 5** Upper limits on the isotropic stochastic background from 3 pulsar timing arrays [180–182]

| | No. of pulsars included | No. of years observed | Observation radio band (MHz) | Constraint on $A_y$ ($f=10^{-9}$–$10^{-7}$ Hz) |
|---|---|---|---|---|
| EPTA | 6 | 18 | 120–3000 | $A_{yr} < 3 \times 10^{-15}$ |
| PPTA | 4 | 11 | 3100 | $A_{yr} < 1 \times 10^{-15}$ |
| NANOGrav | 27 | 9 | 327–2100 | $A_{yr} < 1.5 \times 10^{-15}$ |

If α takes values other than –(2/3), the constraints on $A_{yr}$ are of the same order of magnitude. These constraints have ruled out predictions from many models at the time, excluding most models of supermassive black hole formation.

After improvements in timing accuracy, there are now six pulsar timing arrays (PTAs): the addition of the Chinese Pulsar Timing Array (CPTA), the Indian Pulsar Timing Array (InPTA) [183], and the South African Pulsar Timing Array (SAPTA) [184]. The International Pulsar Timing Array (IPTA, http://www.ipta4gw.org/) is an alliance of regional PTAs aimed at achieving better sensitivity to gravitational waves. Currently, PPTA, EPTA, NANOGrav, and InPTA are formal members of IPTA, while SAPTA and CPTA are observers within IPTA. In recent years, these six pulsar timing arrays have significantly improved the accuracy of gravitational wave detection.

By 2015, the resolution had nearly reached the level of detecting gravitational waves. Systematic errors need to be clarified first. Following improvements in solar system ephemerides, NANOGrav, EPTA/InPTA, PPTA, and CPTA submitted four papers (arXiv numbers: 2306.16213 to 2306.16216) on June 28, 2023, reporting the discovery of nanohertz background gravitational waves [185-188]. The compiled amplitudes of the nanohertz background gravitational waves from their observations and analyses are presented in Table 6. The theoretical models in Table 6 correspond to Equations (34) and (35). The angle-dependent observational analysis results from NANOGrav are shown in Figure 15 [185].

**Table 6** Compilation of the amplitude of background GW from NANOGrav, EPTA/InPTA, PPTA and CPTA observations [185–188]

| | No. of pulsars included | No. of years observed | Observation radio band (MHz) | Constraint on $A_{yr}[10^{-15}]$ ($f=10^{-9}$–$10^{-7}$ Hz) |
|---|---|---|---|---|
| NANOGrav | 67 | 15 | 327–2100 | $A_{yr} = 2.4^{+0.7}_{-0.6}$ |
| EPTA/ InPTA | 25+InPTA10 | 25.4, 24.7, 10.3, 11 | 120–3000 | $A_{yr} = 2.5 \pm 0.7$ |
| PPTA | 30 | 18 | 3100 | $A_{yr} = 2.04^{+0.25}_{-0.22}$ |
| CPTA | 57 | 3.5 | 1000–1500 | $\log A_{yr} = 0.3^{+0.9}_{-1.9}$ [$A_{yr} = 2^{\times 8}_{+80}$] |

If all contributions from supermassive binary black holes (SMBHBs) evolve purely through circular orbital energy loss to gravitational radiation, then the resulting gravitational wave background (GWB) spectrum can be well described by Equation (35), namely the simple $f^{-2/3}$ characteristic strain power-law [189]. However, gravitational wave background signals not arising from this mechanism may also exist in the nanohertz band. These signals include those from inflationary primordial gravitational waves [190,191], gravitational waves induced by scalar fields, vector fields [192], and various processes induced by cosmological phase transitions [193-197] (such as vacuum bubble collisions after phase transitions, sound waves, turbulence), as well as various defects such as cosmic strings [198-200] or domain walls [201,202] decay. Establishing these requires more precise observations and careful model analyses.

Looking ahead over the next hundred years, we [126] adopt and extrapolate estimates from Moore, Taylor, and Gair [203]. The sensitivity of Pulsar Timing Arrays (PTAs) to single-frequency gravitational waves primarily depends on timing precision, including the modeled timing residuals (root-mean-square deviation of time residuals). The bandwidth depends on the sampling frequency and the duration of data span. For observations every $\Delta t$ time interval over an observation span of $T$, the bandwidth $f$ is $[1/T, 1/\Delta t]$. The sensitivity $h_c(f)$ has a frequency dependence linearly related to $f$:

$$h_c(f) = B_{yr} (f / \text{yr}^{-1}), \quad (1/T) < f < (1/\Delta t). \tag{36}$$

We assume $B_{yr}$ is proportional to timing precision, inversely proportional to the duration of observations, and inversely proportional to the number of pulsars in the PTA. Moore, Taylor, and Gair [203] set a canonical PTA (MTG canonical PTA), assuming a random distribution of 36 pulsars in the sky, observed every two weeks with a timing precision of 100 ns over 5 years; the sensitivity of this canonical PTA is (36) with $\mathbf{B}_{yr} = 4 \times 10^{-16}$, roughly equivalent to the OPEN 1 simulated dataset in the IPTA Data Challenge (http://ipta4gw.org/?page_id=89). In Table 7, we estimate the sensitivity of IPTA,

FAST107, and SKA108 under observation spans of 20 years, 50 years, and 100 years, listing the basic simulation assumptions for IPTA, FAST, and SKA.

**Table 7** Sensitivities of IPTA, FAST and SKA to monochromatic GWs [126]

|  | No. of pulsars | No. of years of observation | Timing accuracy (ns) | Sensitivity in characteristic strain $h_c(f)$ [$=B_{yr}(f/yr^{-1})$] for monochromatic GWs |
|---|---|---|---|---|
| IPTA | 36 | 20 | 100 | $B_{yr}=1\times 10^{-16}$ |
| FAST | 50 | 50 | 50 | $B_{yr}=1.5\times 10^{-17}$ |
| SKA | 100 | 100 | 20 | $B_{yr}=1.5\times 10^{-18}$ |

Rough estimated sensitivity curves for IPTA, FAST, and SKA are [126]:

$$h_c(f) = 1\times 10^{-16}\,(f/\mathrm{yr}^{-1}),\ 1.58\times 10^{-9}\mathrm{Hz} < f < 8.27\times 10^{-7}\mathrm{Hz},\ \text{for IPTA-20}, \tag{37}$$

$$h_c(f) = 1.5\times 10^{-17}\,(f/\mathrm{yr}^{-1}),\ 6.34\times 10^{-10}\mathrm{Hz} < f < 8.27\times 10^{-7}\mathrm{Hz},\ \text{for FAST-50}, \tag{38}$$

$$h_c(f) = 1.5\times 10^{-18}\,(f/\mathrm{yr}^{-1}),\ 3.17\times 10^{-10}\mathrm{Hz} < f < 8.27\times 10^{-7}\mathrm{Hz},\ \text{for SKA-100}. \tag{39}$$

We note the corresponding sensitivity curves plotted in Figures 2, 3, and 4 of reference [126]. The $\Omega_{gw}(f)$ sensitivity of SKA for single-frequency gravitational waves reaches $10^{-22}$ around a frequency of approximately $3.17\times 10^{-10}$ Hz. These curves are denoted as IPTA-20, FAST-50, and SKA-100.

Global 24-hour observations of gravitational waves from J1713+0747 provide upper limits in the frequency range of $10^{-5}$-$10^{-3}$ Hz [204]; solid lines in Figures 12-14 represent limits for random directions, while dashed lines indicate limits for pulsar directions [126].

**5 Optical Clock and Atom Interferometry Gravitational Wave Detection**

In Section 3, we discussed spacecraft Doppler tracking and pulse laser ranging, using microwave atomic clocks as time and frequency references. In the previous section, pulsar array gravitational wave detection was discussed, utilizing pulsar regularity in pulse intervals as the basis for time. With the improvement in the accuracy and stability of optical clocks, optical clocks can also serve as time and frequency standards between two spacecraft (or between ground and a spacecraft) [126].

In addition to using optical clocks instead of hydrogen maser clocks for microwave Doppler tracking in gravitational wave detection, there are currently two specific proposals for optical clock-based Doppler tracking systems. One such system uses a strontium optical lattice clock, as shown in Figure 16 [205]. Another system known as the Interplanetary Network of Optical lattice clocks (INO) also employs optical lattice clocks, as depicted in Figure 17 [206].

Optical clocks have advanced to unprecedented levels of stability, precision, and sensitivity [207–209], with uncertainties now reaching below $10^{-18}$ and approaching the $10^{-19}$ level [207]. Further miniaturization is underway, and optical clocks are becoming competitive with laser interferometry for arm lengths equal or greater than astronomical unit scales.

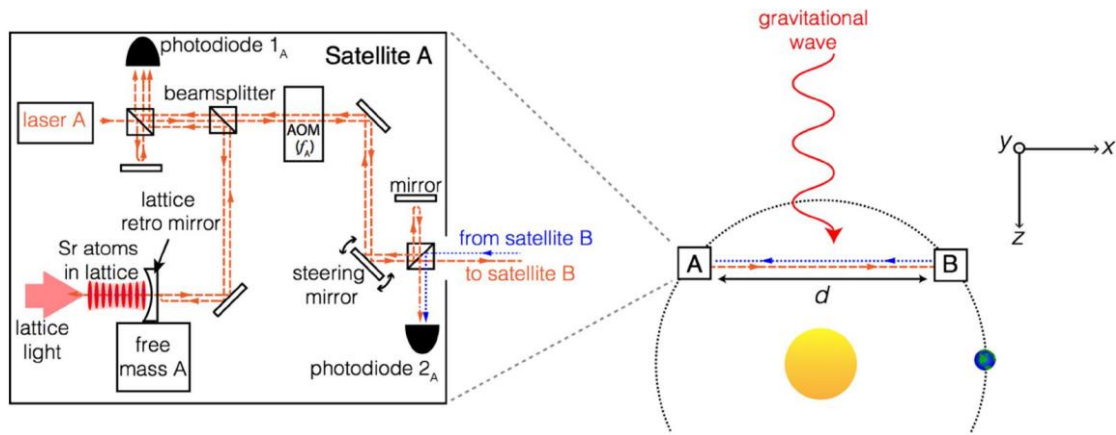

Figure 16 (Color online) Proposed optical-clock gravitational wave detector. The detector consists of two identical drag-free spacecraft, A and B, separated from each other by a distance $d$ along the $x$ axis. Each satellite contains a free-floating reference mass, an ultrastable laser, and a strontium optical lattice clock. A mirror is mounted on the free mass and is used to define the standing wave of light forming the optical lattice and confining the Sr atoms. Some of the laser light from spacecraft A (orange, dashed line) is sent to spacecraft B. The light first passes through an acousto-optic modulator driven at frequency $f_A$, which offsets the frequency of the light reaching photodiode 2B in spacecraft B and enables the phase locking of laser B to laser A through heterodyne detection. Vibrations and thermal drifts of the optics on each satellite can be corrected locally by feeding back on the beat notes at $2f_{A;B}$ on photodiodes $1_{A;B}$. Light from laser B (blue, dotted line) is sent back to satellite A to verify the phase lock, to maintain pointing stability, and to enable operation in the reverse mode, with laser A locked to laser B. A plus-polarized gravitational wave propagating along the $z$ axis induces relative motion between the two free masses, which can be detected using a clock comparison measurement protocol. The figure is from ref. [205].

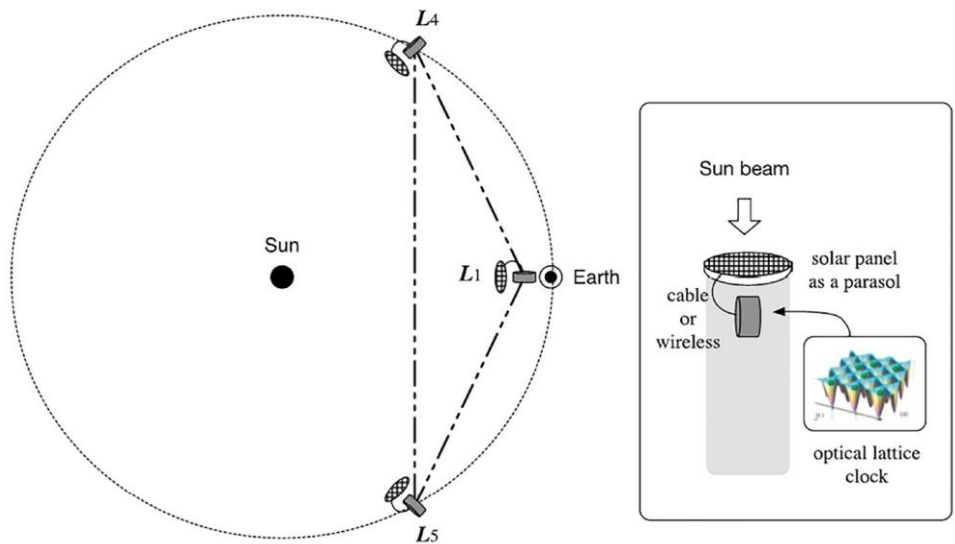

Figure 17 (Color online) A planned location of the spacecraft: Lagrangian points $L_1$, $L_4$, and $L_5$ of the Sun-Earth orbit. The $L_1$ is at 1/100 AU from the Earth, while $L_4$ and $L_5$ form equilateral/isosceles triangle with the Sun/the Earth, respectively; the distance between $L_1$–$L_4$ ($L_5$) is 1 AU, while that of $L_4$–$L_5$ is $3^{1/2}$ AU. Two-frequency radio or light will be used for communication between spacecraft. The inset explains that the solar panel of the spacecraft is separated as a parasol from the main body, in order to prevent acceleration noise due to solar wind. The figure is from ref. [206].

Since the successful demonstration of AIs from the teams of Pritchard [210], Bordé [211], and Zhu [212] in 1991, atomic interferometry (AI) technology has been in rapid and promising development phase. It has made significant contributions to precision measurements and fundamental physics. Atomic interferometric gravimeters now rival optical interferometric gravimeters. From August 21 to September 29, 2023, the preliminary draft results of the 11th International Comparison of Absolute Gravimeters (ICAG 2023) showed that the atomic interferometric gravimeter WAG-H5-2 from the Innovation Academy of Precision Measurement Science and Technology (APM) of CAS, exhibited consistency with the FG5 from JILA at the Table Mountain Geophysical Observatory (TMGO), Boulder, Colorado, with uncertainties comparable [213].

In 2011, Stanford University proposed using atomic interferometers as an alternative method on the LISA band for detecting gravitational waves [214-218]. However, achieving LISA-like sensitivity with this proposal requires further precision improvements and addressing several challenges [216, 217]. Since the proposal, significant efforts have been made in laboratory research and development, demonstrating high technological content and application potential.

In 2015, the SYRTE at the Paris Observatory launched the first phase of its MIGA (Matter-wave laser Interferometric Gravitation Antenna) project, constructing a 200-meter optical cavity in the LSBB (Laboratoire Souterrain à Bas Bruit) underground laboratory in Rustrel for studying atomic interferometers [219]. In the project's second phase, MIGA will focus on scientific experiments and data analysis to explore the spatiotemporal structure of the local gravity field in the LSBB area. Meanwhile, MIGA is also evaluating the potential future applications of atomic interferometers in the intermediate frequency band (0.1-10 Hz) for gravitational wave detection.

Currently, there are five large fully-funded prototype projects under construction: a 10-meter fountain project in Stanford, USA [220] and the MAGIS-100 project at FNAL [221] in the US; the MIGA project in France [222]; the VLBAI project in Hannover, Germany [223]; and the AION-10 project in Oxford, UK [224], potentially with a 100-meter site in Boulby, under investigation at CERN; a 10-meter fountain project [225] and the ZAIGA project in China [226]. These projects will demonstrate the feasibility of long-arm atomic interferometric detection, paving the way for future ground-based kilometer-scale experiments. Discussions are already underway for constructing multiple kilometer-scale detector projects, including the ELGAR project in Europe [227], the MAGIS-km project at the Sanford Underground Research Facility (SURF) in the US [221], the AION-km project at the STFC Boulby Facility in the UK [224], and the advanced ZAIGA project in China [226]. The goal is to have at least one kilometer-scale detector operational around 2035. These kilometer-scale experiments will systematically explore the intermediate frequency band of gravitational waves and potentially detect ultra-light dark matter, while demonstrating key technologies required for future space-based atomic interferometric detection missions like AEDGE [228], as illustrated in Figure 18. The ultimate goal of atomic interferometer gravitational wave detection is missions such as AEDGE, where atomic interferometry serves as the detection frequency standard.

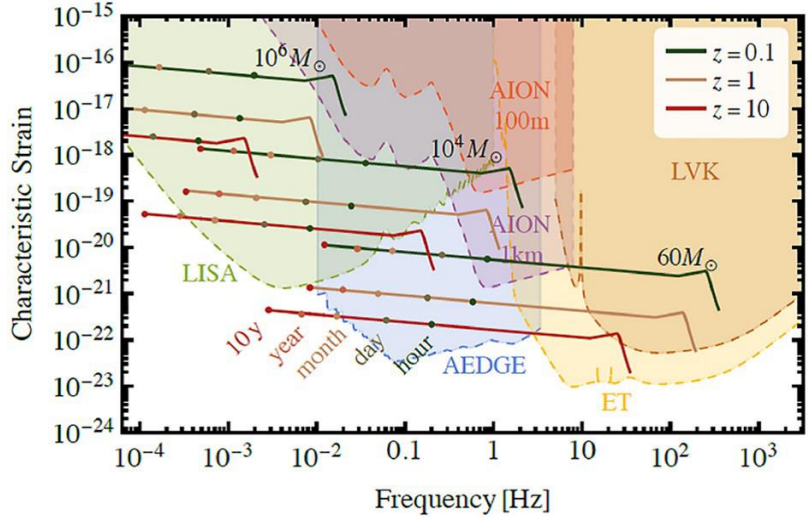

Figure 18 (Color online) The GW strain sensitivities and benchmark signals from BH binaries of different masses at different redshifts. The coloured dots indicate the times before mergers at which inspirals could be measured. The figure is from ref. [228].

Unlike photons, atoms come in many types. A significant advantage of atomic interferometry for detecting ultra-light/zero-mass physical fields and dark matter is its wide applicability: it can detect various potential physical fields and dark matter with different interactions. For example, scalar fields [229, 230, 231], pseudo-scalar fields [232, 233] (like axions [234, 235, 236]), vector fields [237], or dark photons [238, 239].

Regarding intermediate-frequency gravitational wave detection, apart from the space-based schemes discussed earlier and AEDGE, there are (i) AIGSO [240], which uses a 10-kilometer arm length atomic interferometer, with challenging technological development; (ii) various proposals using optical clocks, which could be prototyped first. Atomic interferometry can be applied as a gravity gradiometer, offering an option for separating gravitational wave signals.

# 6 Optical Path Noise, Inertial Sensor/Accelerometer Noise (Core Noise), and Frequency Noise Effects in Space Laser Interferometric Gravitational Wave Detectors

In the context of space-based laser interferometric gravitational wave detectors, there are two core noises: optical metrology (length) noise and inertial/accelerometer noise. Regarding laser frequency noise, its effect on the combined signal in the generalized Michelson configuration depends on the optical path difference of the two paths.

In space, due to the large distances involved, at the receiving end near the spacecraft, it is necessary to phase-lock the local laser oscillator to lock in the propagation (amplification) of the weak incoming light beam to another spacecraft or back; this measures the final interferometric phase difference between the two paths. The interferometric phase noise $\delta\varphi_{\text{interference}}$ at the final receiving satellite is given by:

$$\delta\varphi_{\text{interference}} = 2\pi \cdot \delta v(f) \cdot \Delta L/c + \text{phase locking noise} + \text{timing noise}$$
$$+ \text{accumulated other noises along the two paths}, \tag{40}$$

where $\delta v(f)$ is the frequency noise of the laser source at frequency $f$, $\Delta L$ is the optical path difference between the two chosen paths, and $c$ is the speed of light. Therefore, to reduce the interferometric phase noise $\delta\varphi_{\text{interference}}$, we must reduce the frequency noise of the laser source or decrease the optical path difference between the two paths, or both.

After traveling over long distances, the light received by the telescope on another spacecraft is greatly attenuated and needs to be amplified to transmit it to the next spacecraft. The amplification is achieved by phase-locking the local laser with the incoming weak laser either in phase or with a known frequency offset. Thus, the phase information is continuously propagated under conditions of phase coherence or known frequency offset. The phase records measured by the phase meter are tagged with time labels, commonly using pseudo-random codes, for subsequent propagation identification in data analysis. The method of obtaining the interferometric signal at the receiving satellite with time-tagged tracking analysis is known as Time-Delayed Interferometry (TDI). To meet the requirements of interferometric phase noise $\delta\varphi_{\text{interference}}$, the TDI configuration chosen for each space gravitational wave mission must satisfy the optical path difference conditions set by Equation (40).

As shown in Figure 19, simple TDIs include the first-generation Michelson TDI, typically referred to as X, Y, and Z TDI; the second-generation Michelson TDI, known as X1, Y1, and Z1 TDI; and the original (zeroth-generation) Michelson combinations, referred to as X0, Y0, and Z0 TDI. Discussions on various combinations of TDI are found in Section 7.2. Both phase locking and time tagging introduce noise. Other noises include Tilt-to-Length (TTL) noise, among others. Without orbital propulsion corrections, changes in arm lengths due to the motion of the solar system can reach 1% for missions similar to LISA, and 0.01-0.1% for missions similar to ASTROD-GW. Typically, appropriate TDI configurations (combinations) are selected to meet the requirement that the interferometric phase noise caused by laser frequency noise $\delta v(f)$ is smaller than the core noises (optical metrology noise and inertial/accelerometer noise) when discussing detection sensitivity and detection targets. For example, for a laser source with laser frequency noise $\delta v(f)$ of 30 Hz/√Hz, X, Y, and Z TDIs do not meet the requirements of LISA and ASTROD-GW, while X1, Y1, and Z1 TDIs do meet the requirements of LISA and ASTROD-GW.

Table 2 lists the core noises of current laser interferometric gravitational wave space mission projects and plans. Inertial/accelerometer noise determines the precision of strain measurement reference points, while optical metrology noise determines the precision of the measurements themselves. Laser frequency noise occurs during the measurement process, but after selecting appropriate TDI configuration combinations for measurement results, it is mitigated below the core noises.

The primary source of optical metrology noise is the shot noise in the long-arm laser link:

$$\theta_{\text{shot}} = (\hbar c/2\pi\lambda P)^{1/2} \text{ cycles} \cdot \text{Hz}^{-1/2}. \tag{41}$$

This equation refers to the RF noise caused by the detection of lasers from adjacent spacecraft. The higher the optical power $P$ received, the lower the RF noise. Using parameters estimated for LISA [50], this value is 6.9 μcycles·Hz$^{-1/2}$, which converts to an optical metrology noise of 7.34 pm·Hz$^{-1/2}$ for $\lambda$=1.064 μm. With other metrology noises sufficiently small, the requirement 10 (15) pm·Hz$^{-1/2}$ listed in Table 2 is feasible. Similar derivations of requirements apply to other space gravitational wave detection missions.

In December 2015, LISA Pathfinder was successfully launched, reducing the amplitude spectrum of inertial/accelerometer noise to [46,243]:

$$[P_a(f)]^{1/2} \leq 3 \text{ fm s}^{-2} \text{ Hz}^{-1/2} [1 + (10^{-4} \text{ Hz}/f)^2 + 16 (2 \times 10^{-5} \text{ Hz}/f)^{10}], (2 \times 10^{-5} \text{ Hz} \leq f \leq 1 \text{ Hz}). \tag{42}$$

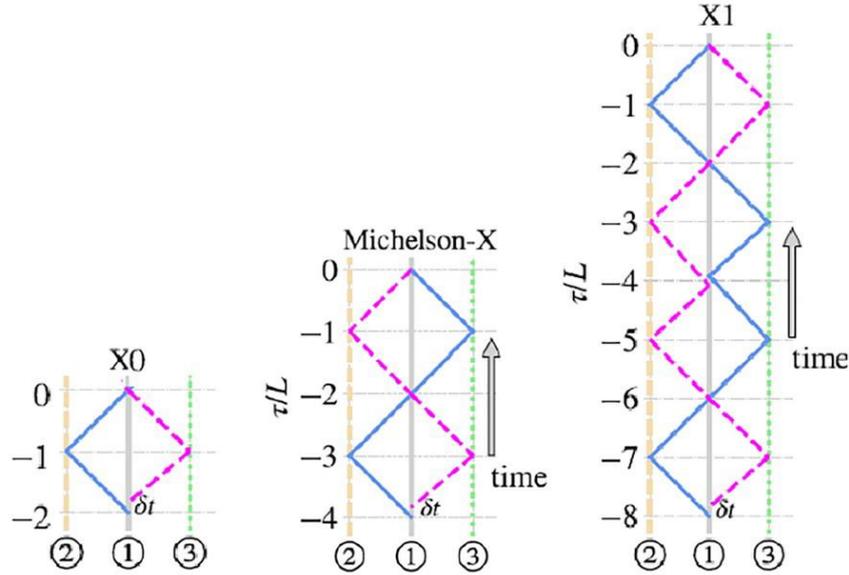

Figure 19 (Color online) The S/C layout-time delay diagrams for the zeroth-generation TDI channel Michelson X0, the first-generation TDI channel Michelson-X, and the second-generation TDI channel Michelson-X1. The vertical lines indicate the trajectories of S/C in the time direction (①–③ indicate S/C$i$, $i$=1, 2, 3), and the ticks on each $y$-axis show the value of time delay with respect to the TDI ending time $\tau$=0. The blue solid and red dashed lines show the paths of TDI channels, and indicate two groups of interfered laser beams. The figure (Wang diagrams) is adapted from refs. [241,242].

All space gravitational wave detection plans are based on this, each setting its specific goals. For instance, LISA adopts a slightly relaxed criterion:

$$[P_{a,\text{LISA}}(f)]^{1/2} \leq 3 \text{ fm s}^{-2} \text{ Hz}^{-1/2} [1 + (0.4 \text{ mHz}/f)^2]^{1/2} [1 + (f/8 \text{ mHz})^4]^{1/2}, (1 \times 10^{-4} \text{ Hz} \leq f \leq 1 \text{ Hz}), \tag{43}$$

Apart from the low-frequency red relaxed factor $[1 + (0.4 \text{ mHz}/f)^2]^{1/2}$, it also includes a blue relaxed factor $[1 + (f/8 \text{ mHz})^4]^{1/2}$. Similar considerations apply to the Taiji project. For discussions on other space gravitational wave missions, refer to Section 8.

# 7 Space Optical Interferometric Gravitational Wave Detection, Delay Lines, and Time-Delay Interferometry

There are two methods for space optical interferometric gravitational wave detection: one uses Michelson stellar interferometers, and the other uses generalized Michelson time-delay laser interferometers. The following two subsections discuss these methods. Section 3 discusses the potential development of continuously tunable fiber delay lines for use in both stellar interferometers and space-based time-delay interferometers.

## 7.1 Stellar Interferometers and Gravitational Wave Detection

In 1868, Fizeau [243, 244] proposed placing an aperture with two narrow slits in front of a telescope to use interferometry to overcome the broadening of stellar images caused by atmospheric disturbances and measure the angular size of stars. In 1873-74, Stephan [245], following Fizeau's idea, observed several bright stars using the 80 cm Foucault reflector at the Marseille Observatory and a double-slit aperture, and indeed saw interference fringes. When the distance $d$ between the double slits of the aperture increased to 1.22 $\lambda/\theta$, where $\theta$ is the angular diameter, the visibility of the fringes should have been zero. However, when the slit distance was increased to 65 cm, the interference fringes remained clearly visible, leading Stephan to conclude that the angular diameters $\theta$ of these bright stars were all less than 160 milliarcseconds (approximately 1.22 $\lambda/d$, where $\lambda$ is the wavelength).

In 1890, Michelson [246] used a stellar interferometer [247], the principle of which is schematically shown in Figure 20, to measure the angular sizes of celestial bodies. For a spherical star with an angular diameter $\theta$, the angular diameter measured when the interference fringes disappear (1.22 $\lambda/d$) corresponds to the measured angular diameter. In 1921, Michelson and Pease installed a 6-meter interferometer in front of the 2.5-meter Hooker telescope at the Wilson Observatory and measured the angular diameter of Betelgeuse (α Orionis) to be 0.47 arcseconds [250], based on its parallax value of 0.018 arcseconds, thereby determining a diameter of $3.84 \times 10^8$ km (2.58 AU), marking the first measurement of a star's diameter.

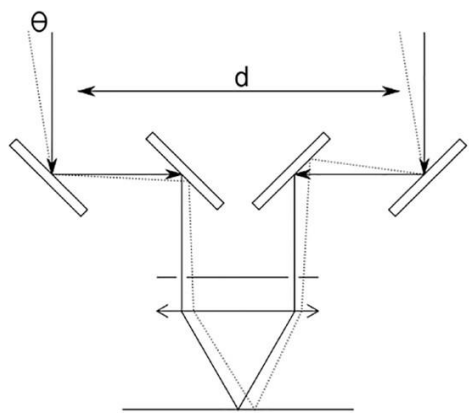

Figure 20 Schematic of the original Michelson stellar inferterometer. When the fringe visibility vanishes, the 1.22$\lambda/d$ is the apparent angular diameter. (https://www.wikiwand.com/en/Michelson_stellar_interferometer).

In 1954 at Jodrell Bank, after completing a radio stellar intensity interferometer together with students Jennison and das Gupta [248], Hanbury Brown and theoretical physicist Twiss established the mathematical foundation of optical intensity interferometry [249]. In 1955, Hanbury Brown and Twiss confirmed their concept in the laboratory. Subsequently, Hanbury Brown gathered and assembled equipment to measure the angular diameter of Sirius. Borrowing a 1.5-meter searchlight used during World War II, they replaced the lamp with a photomultiplier tube to receive and record starlight from Sirius. They amplified the fluctuations in anode current, and linked telescopes modified from two searchlights with a baseline separation of up to 10 meters to measure the current fluctuations. Over a cold period of four months, they accumulated 18 hours of integrated data, measuring Sirius' angular diameter to be 6.82±0.52 milliarcseconds (mas), marking the first direct measurement of the angular diameter of a main-sequence star [250].

The successful use of searchlights led to the establishment of a 10-meter prototype stellar intensity interferometer, prompting further development of an instrument dedicated to measuring main-sequence stars. After much effort, the Department of Scientific and Industrial Research (DSIR) in the UK partnered with the University of Sydney and the University of Manchester to construct the first astronomical instrument in Australia designed to measure the diameters of numerous stars—the Narrabri Stellar Intensity Interferometer (NSII). The NSII consisted of a large circular track with detector separations ranging from 10 to 188 meters. Operating from 1963 to 1974, it measured the angular diameters of 32 stars. Alongside measurements by Michelson and Pease, it preliminarily completed systematic measurements of the angular diameters of main-sequence stars and giants [251, 252].

In 1974, Labeyrie [253, 254] completed a starlight interferometer with a baseline of 12 meters, composed of a pair of separated 0.25-meter telescopes. Interference fringes produced by Vega (α Lyr) in the 500-750 nm spectral range were clearly visible, allowing determination that Vega's angular diameter was less than 5 mas (further confirming observations by Hanbury Brown et al. [255] using the Narrabri Stellar Intensity Interferometer, which measured Vega's 483.5 nm optical angular diameter to be 3.56±0.40 mas), marking the beginning of modern long baseline stellar interferometry.

Here, examples such as the Mark III stellar interferometer [256, 257] at Mt. Wilson Observatory in California and the Space Interferometry Mission (SIM) [30] are briefly mentioned to illustrate advancements. Astronomical observations require tracking celestial bodies, and Michelson interferometry requires that the optical path lengths of two interfering light beams are equal or differ within the coherence length. For ground-based astronomical facilities, variations in optical path length due to Earth's rotation need to be compensated for. The fringe tracking mechanism of Labeyrie et al.'s stellar interferometer involves moving optical devices on orbiting platforms between two astronomical telescopes to ensure that the optical path difference between two beams is equal or within the coherence length. The tracking method of the Mark III stellar interferometer uses a maneuverable variable optical delay line, as shown in Figure 21. Optical path difference compensation can also be achieved using continuously adjustable fiber delay lines, as detailed in Section 7.3.

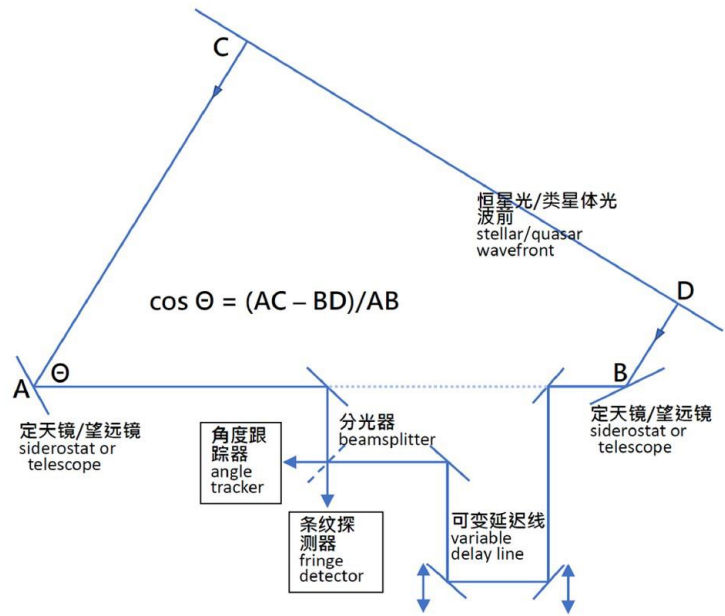

Figure 21 (Color online) Schematic diagram of a current Michelson stellar inferterometer using the delay line.

The Mark III stellar interferometer operated from 1987 to 1992, conducting measurements of celestial positions, stellar angular diameters, and binary star orbits within the wavelength range of 450-800 nm. The Mark I and Mark II were experimental setups that paved the way for the development of Mark III technology. Mark III consisted of three alt-azimuth telescopes (N, S, E) mounted on pedestals, forming two baselines NS and ES each of 12 meters for measuring celestial positions. Interferometric measurements with multiple baseline telescopes can synthesize imaging. Mark III also featured two telescopes in orbit, movable between 11 stations, with an effective aperture of 7.5 cm (sufficiently small), obviating the need for adaptive optics at Mt. Wilson. The United States Navy Precision Optical Interferometer (NPOI) [28, 29] is a follow-up to Mark III, comprising a Y-shaped three-arm imaging interferometer with arm lengths of 250 m and baselines up to 437 m, making it the highest-resolution optical imaging instrument for ground-based astronomical observations today. NPOI is a complex system capable of achieving milliarcsecond precision in astronomical angle measurements. The angular resolution of ground-based stellar interferometers is constrained by atmospheric disturbances, with limits in the tens of microarcseconds.

Another successor project of Mark III, the Space Interferometry Mission (SIM) [30], also known as SIM Lite (formerly SIM PlanetQuest), aimed primarily at detecting Earth-sized planets in the habitable zones of nearby stars outside the solar system. SIM experienced multiple delays and was ultimately canceled in 2010.

In Section 7.1.1, we discuss methods and conceptual tasks for detecting low-frequency gravitational waves using stellar interferometers. Gravity fields and gravitational waves permeate the universe, affecting both the light sources observed by stellar interferometers and the interferometers themselves. In extremely precise measurements, the influence of gravitational waves on the light paths and detectors must be considered.

Gravitational waves with periods longer than the observational timespan produce a simple apparent motion pattern in the sky [258]. Therefore, accurately measuring the proper motion of quasi-stellar objects will be a method for detecting ultra-low-frequency (10 fHz – 300 pHz) gravitational waves. We discuss this detection method in Section 7.1.2.

**7.1.1 Detection of Micro-Hertz and Sub-Micro-Hertz Low-Frequency Gravitational Waves**

If gravitational waves pass through the path of starlight reaching a stellar interferometer, they can affect the measurements of the angles of starlight sources made by the interferometer. The spectral changes in the measured celestial angles can reveal information about gravitational waves.

In 2021, Park et al. [259] proposed a method to detect microhertz low-frequency gravitational waves using a stellar interferometer. This method involves two coherent light beams from distant stars detected by two spacecraft at the ends of a space-based long baseline, suitable for amplitude or intensity interferometric measurements, as shown in Figure 22. Park et al. analyzed a stellar interferometer using visible light (550 nm) from the Crab Nebula received by spacecraft separated by 1 Gm (Gigameter), considering shot noise and possible acceleration noise achievable. They suggested that the sensitivity for detecting gravitational waves around 250 μHz could be comparable to that of LISA. In 2022, Fedderke et al. [260] further proposed and analyzed methods for using a stellar interferometer to detect sub-microhertz gravitational waves (10 nHz – 1 μHz) from planetary and white dwarf sources. These methods are still in early stages of discussion and require further research and analysis.

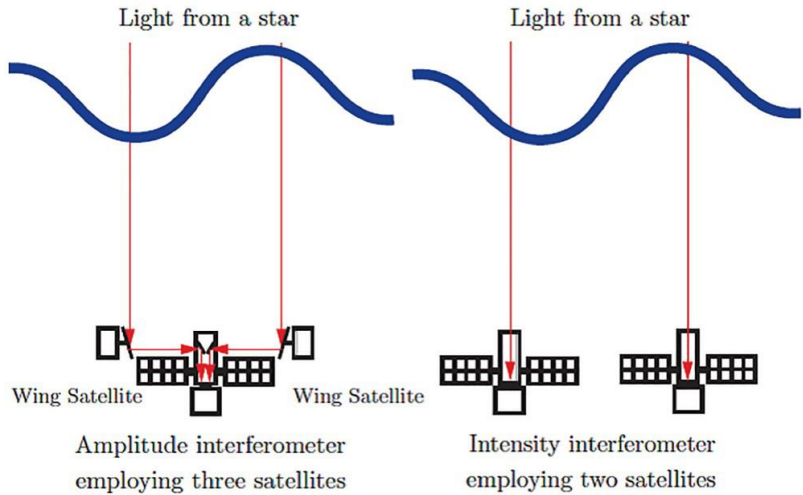

Figure 22 (Color online) A classical method with a three satellites configuration (left) and an intensity interferometer configuration of two satellites (right) for stellar interferometry of detecting gravitational waves. The figure is from ref. [259].

**7.1.2 Detection of Ultra-Low-Frequency (10 fHz–300 pHz) Gravitational Waves [126]**

The gravitational wave periods at 10 fHz are approximately $10^{14}$ seconds, or about 3.16 million years. At 300 pHz, the gravitational wave period is $3.3 \times 10^9$ seconds, roughly equivalent to 100 years. Gravitational waves with periods longer than the observation time span create a simple apparent proper motion pattern in the sky [258]. Therefore, precisely measuring the apparent proper motion of quasars and similar astrophysical objects can be a method to detect ultra-low-frequency gravitational waves (10 fHz – 300 pHz). Gwinn et al. [261] used this method to constrain the normalized energy density spectrum of stochastic gravitational waves, with frequencies below $2 \times 10^{-9}$ Hz and above $3 \times 10^{-18}$ Hz (including the ultra-low-frequency range), to be less than 0.11 $h^{-2}$ (95% confidence level) times the critical density of our universe. Using the Hubble constant value obtained from Planck 2018, $H_0 = (67.4 \pm 0.5)$ km s$^{-1}$Mpc$^{-1}$ [262, 263], gives $h = 0.674$, leading to a constraint of 0.24 according to its corresponding critical density value (labeled as QA [Quasar Astrometry] in Figure 4 of reference [126]). Long baseline optical interferometers/space-based stellar interferometers capable of microarcsecond and nanoarcsecond (nas) astrometry measurements are technically feasible [264]. Using such interferometers, precise astrometric measurements of quasar proper motions can be improved by four orders of magnitude, achieving accuracy on the order of nas/yr. In terms of sensitivity, measuring or constraining $\Omega$gw(f) can achieve a precision of $2.4 \times 10^{-9}$ or better (labeled as QAG [Quasar Astrometry Goal] in Figure 4 of reference [126]).

Using equation (1), we obtain the constraint on the characteristic strain $h_c(f)$:

$$h_c(f) < 4.2 \times 10^{-19} \text{ (Hz}/f\text{), for } 3 \times 10^{-18} \text{ Hz} < f < 2 \times 10^{-9} \text{ Hz.} \tag{44}$$

With a four-order improvement in angular resolution, the sensitivity reaches:

$$h_c(f) = 4.2 \times 10^{-23} \text{ (Hz}/f\text{), for } 3 \times 10^{-18} \text{ Hz} < f < 2 \times 10^{-9} \text{ Hz.} \tag{45}$$

Curves (44) and (45) are plotted in Figure 2 of reference [126], labeled as QA and QAG. Using equation (1), $\Omega_{gw}(f)$ can be converted to $[S_h(f)]^{1/2}$ and plotted in Figure 3 of reference [126].

**7.2 Generalized Michelson Interferometry and Time-Delay Interferometry**

In the Michelson interferometer, the wavefront is split into two parts to enter two different paths, and then the two wavefronts are recombined for interference. For white light, Michelson must precisely match the lengths of the two optical paths to produce interference fringes. The original Michelson interferometer used white light interference: the white light Michelson interferometer requires the two arms to be nearly equal in length, differing by no more than one wavelength. The earliest Michelson interferometer was used to compare the speed of light in different directions, such as in the famous Michelson-Morley experiment and in comparisons of the meter prototype and its replicas. In 1925, Michelson and Gale [265] used a carbon arc light source with good coherence to measure the Sagnac effect around a rectangular plot of 2010 feet × 1113 feet (latitude: 41°46'), obtaining 0.230 ± 0.005 interference fringes and an error in measuring the Earth's rotation speed of 2%.

After the invention of lasers, the coherence length became longer. People could construct unequal arm Michelson interferometers. Although the interference accuracy is not limited by frequency standards (such as clocks), it is still constrained by coherence length (frequency noise) [equation (40)]. Another configuration of the Michelson interferometer is the Mach-Zehnder interferometer. Doppler tracking requires measuring phase of beat to be generated by the local oscillator as a function of time. After appropriate consideration of heterodyne, equations (31) and (32) remain applicable.

Because the deviation of the electromagnetic wave velocity in plasma from vacuum velocity is inversely proportional to the square of the frequency, time uncertainties caused by solar wind or ionized gas in radio and microwave propagation are smaller in the Ka band (32 GHz) than in the X band (8.4 GHz), and smaller in the X band than in the S band (2.3 GHz). This is one of the motivations for using the Ka band and X band for Doppler tracking by the Cassini spacecraft to achieve better noise performance. Another motivation is the shorter wavelength, which improves measurement accuracy. At optical frequencies, wavelengths are more than four orders of magnitude smaller, and plasma effects are eight orders of magnitude smaller. Therefore, optical methods began to be used in the field of gravitational wave detection when higher optical path measurement sensitivity was required. As sensitivity improves, we need to suppress various noise below the target sensitivity level. This requires (i) reducing acceleration noise and achieving drag-free technology; (ii) minimizing laser noise as much as possible. LISA Pathfinder demonstrated basic drag-free technology [46, 47]. To reduce laser noise requirements, laser stability is needed. The best method is to achieve absolute stability; for example, by locking to iodine molecular lines. However, the stability currently achievable is not yet sufficient to meet the required sensitivity of $10^{-21}$ strain levels. To reduce laser noise requirements, the delayed interferometry measurement method (TDI) was developed.

For space-based laser interferometric gravitational wave antennas, arm lengths vary according to the dynamics of the solar system's orbit. To achieve the required sensitivity, laser frequency noise effects must be suppressed below core noise levels (such as optical path noise, acceleration noise, etc.) using the selected TDI configuration. As in equation (40), better matching of optical path lengths results in better elimination of laser frequency noise and makes it easier to achieve the desired sensitivity. With precise matching, laser frequency noise is almost completely eliminated, as in the original Michelson interferometer.

For interferometers with unequal arm lengths (paths), interference can only occur when the path length difference is less than the coherence length. For example, the coherence of a single-mode narrow linewidth (1 kHz) Nd:YAG laser exceeds 100 kilometers and can be used for interferometry with path differences less than 100 kilometers; for interferometric phase accuracy requirements, better matching of the two path lengths is needed, as in LISA's 25 meters and Taiji's 30 meters. The LISA and Taiji gravitational wave detection space missions involve three spacecraft moving along the solar orbit, with nominal arm lengths of 2.5 million kilometers and 3 million kilometers respectively. Orbital motion causes arm length differences of about 1%, approximately 20,000 to 30,000 kilometers, which prevents interference.

TDI design ensures that the optical path difference between the dual split laser beams is close enough to produce interference, as shown in Figure 23, where beams are split into two paths (input to output) labeled as 1 and 2, with their beat frequencies compared by the phase meter receiving and recording signals from the local oscillator.

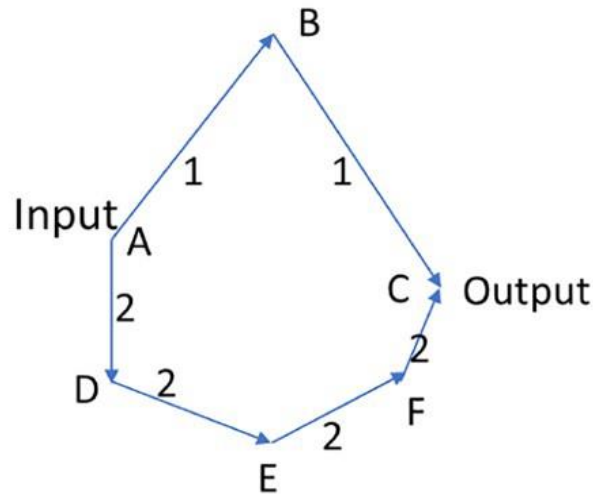

Figure 23 (Color online) Laser light is divided into two beams ABC (1) and ADEFC (2) from input A to output C. At each link, the phase of beat with the local laser oscillator is received and recorded in the phase meter to produce phase differences.

There are two methods to discuss interference noise:

(I) According to equation (40), the noise of the interferometric signals in arms 1 and 2 includes: (i) Frequency noise corresponding to the optical path difference, i.e., the first term of equation (40); (ii) Noise embedded during the propagation and amplification processes of the laser in arm 1 and arm 2. From the TDI configuration and the phase data structure at each connection point and receiving end, the interference signal can be directly explored. In principle, there is no need to use precise timing data in the link, but familiarity with the link structure is required (see Section 13).

(II) Consider the laser frequency noise effects (significant) at each node (A, B, C, D, E, F), along with the noise embedded during the laser propagation and amplification processes in arms 1 and 2, including timing noise. After canceling out the large number of frequency noise effects, the total noise is the interference noise at the receiving end.

The second method is commonly used in LISA to date, as seen in references [159, 266-282]. The first method is implicit in references [53, 55, 82, 105, 128, 145, 241, 242], based on the idea of splitting a laser into two beams for continuous propagation, without directly considering the large number of laser frequency noises in different optical paths.

The first method can determine the optical path difference and the local time of laser pseudo-code signal transmission in drag-free spacecraft (or standard equal-arm formation). The conversion between local time and the inherent time of the solar system center can be determined by laser pseudo-code ranging and its stability.

These two methods converge, as explained in Section 13. Since the laser and the laser-carrying pseudo-codes arrive at each spacecraft simultaneously, the time determined by the pseudo-code at each spacecraft is identical, i.e., the time recorded by the spacecraft clock or ultra-stable oscillator (the local time of the spacecraft). Therefore, the requirements for spacecraft clocks or ultra-stable oscillators can be relaxed. Pseudo-code ranging can be verified by Ka-band microwave communication distance measurement, spacecraft orbit determination, and planetary ephemeris checks. Moreover, this data can also be used to refine planetary ephemerides.

TDI was first studied in the concept research of the ASTROD mission [53, 55]. In deep space interferometric measurements, long distances are always involved. Due to the distance, the laser undergoes significant attenuation when it reaches the receiving spacecraft. Amplification is required to transmit the laser back or to another spacecraft. This process involves phase-locking the local laser to the incoming weak laser and transmitting the local laser back or to another spacecraft. Liao et al. [59, 60] demonstrated phase locking of local oscillators with 2 pW incoming laser light in the laboratory. Dick et al. [283] achieved phase locking for incoming weak lasers with 40 fW power. These developments meet the feasibility requirements for e-LISA/NGO [49] and ASTROD-GW [81] power demands. Subsequent experiments further improved feasibility [284, 285, 286, 287], with Sambridge et al. [287] recently demonstrating phase tracking of sub-fW lasers, averaging slip times exceeding 1000 seconds.

In the 1990s, two TDI configurations [53, 55] were used in concept studies of ASTROD interferometric measurements, with optical path differences calculated using Newtonian dynamics. These two TDI configurations are the unequal-arm Michelson TDI configuration and the Sagnac TDI configuration used for the three-spacecraft formation flight. The principle is to split two laser beams into paths 1 and 2, and interfere at their respective ends. For the unequal-arm Michelson TDI configuration (X configuration), as shown in Figure 19, one laser beam starts from spacecraft 1 (S/C1), points to spacecraft 2 (S/C2), and is received by S/C2, where it optically phase-locks the local laser; then the phase-locked laser beam points back to S/C1 and optically phase-locks another local laser in S/C1; and so on, returning to S/C1 via path 1:

Path 1: S/C1 → S/C2 → S/C1 → S/C3 → S/C1.  (46)

The second laser beam also starts from S/C1 but follows path 2:

Path 2: S/C1 → S/C3 → S/C1 → S/C2 → S/C1.  (47)

It returns to S/C1 and interferes coherently with the first laser beam. If the two paths have exactly the same optical path length, laser frequency noise cancels out; if the difference in optical path length is small, laser frequency noise is largely canceled out. In the Sagnac TDI configuration (Sagnac-α configuration), the two paths are:

Path 1: S/C1 → S/C2 → S/C3 → S/C1, Path 2: S/C1 → S/C3 → S/C2 → S/C1.  (48)

In the Sagnac-α2 configuration, the two paths I and II are I: Path 1 + Path 2 and II: Path 2 + Path 1. Since then, similar calculations have been carried out for the original LISA [45], eLISA/NGO [49], LISA-like space gravitational wave detection concepts with an arm length of $2 \times 10^6$ km, zero inclination

ASTROD-GW [77, 78], non-zero inclination ASTROD-GW [81], and current LISA and Taiji (using relativistic ephemerides calculations) [82, 145, 288, 289, 290, 291].

Since the proposal of the first and second generations of TDI in 1999 [159, 292], delayed interferometric measurements in LISA have been extensively studied. The first-generation TDI compensates for static unequal arm length differences, while the second-generation TDI compensates to some extent for motion. A configuration set (46) and (47) represents the Armstrong, Estabrook, and Tinto meaning of the first-generation TDI X configuration [159]. As with the TDI X configuration, if the laser emits from S/C 2, points to S/C 3 and S/C 1, it is TDI Y; if the laser emits from S/C 3, points to S/C 1 and S/C 2, it is TDI Z. TDI X, TDI Y, and TDI Z are channels available for data analysis. TDI X, TDI Y, and TDI Z form a basis in a linear space, linearly transforming into a base set that are uncorrelated in data analysis (under the equal-arm length approximation) [159, 292]:

$$A = Z - X;\ E = (X - 2Y + Z) / (6^{1/2});\ T = (X + Y + Z) / (3^{1/2}). \tag{49}$$

For the triangular formation of three spacecraft, there are still Relay-U, Beacon-P, and Monitor-D [277], as shown in Figure 24 [241, 242, 293].

Numerical TDI is discussed in Section 13. For many other aspects of TDI, readers can refer to the review [292].

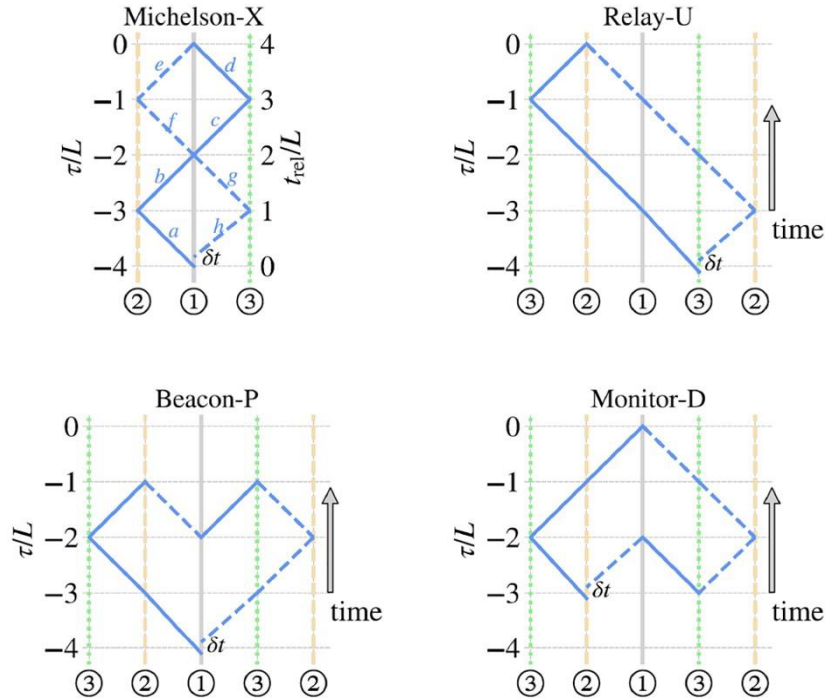

Figure 24 (Color online) The S/C layout-time delay diagrams for the first generation TDI channels Michelson-X, Relay-U, Beacon-P, and Monitor-D. The vertical lines indicate the trajectories of S/C in the time direction (①–③ indicate S/C$i$, $i$=1, 2, 3), and the ticks on each $y$-axis show the value of time delay with respect to the TDI ending time $\tau$=0. The $t$rel is the time with respect to the starting time $t$0 at starting S/C of TDI. To avoid the crossing caused by TDI paths at noninteger delay times and show the paths clearly, extra trajectory lines are plotted for S/C2 (dotted green lines) and S/C3 (dashed orange lines). The blue lines show the paths of TDI channels, the solid line and dashed line indicate two groups of interfered laser beams. The figure (Wang diagrams) is from refs. [241,242,293].

## 7.3 Continuously Tunable Fiber Delay Lines and Gravitational Wave Detection

Due to the precision requirements of detection and the limited coherence length of lasers, for the Taiji and LISA detectors, the requirements for the optical path difference between the two paths should be less than 30 m and 25 m respectively. The three-spacecraft formations of the Taiji and LISA detectors are influenced by gravitational forces such as those from the Sun and planets, causing arm lengths to vary by up to 1% over time. The variation in arm length for the Taiji detector can reach 30,000 km, and for the LISA detector, it can reach 25,000 km. Simple Michelson interferometric paths do not meet the requirement. For the first-generation Delayed Michelson double-interferometric paths X, Y, Z, the optical path differences can reach 370 m for the Taiji detector and 240 m for the LISA detector, both of which do not meet the conditions of being less than 30 m and 25 m [145]. For the second-generation Delayed Michelson interferometric fourfold paths X1, Y1, Z1, the optical path differences are less than 5 m for the Taiji detector and less than 3 m for the LISA detector, both of which meet the requirements [145].

Using continuously adjustable fiber delay lines in spacecraft can compensate for the optical path differences in the first-generation Delayed Michelson interferometric double paths X, Y, Z of the Taiji and LISA detectors, thereby meeting their requirements. The principles of manufacturing continuously adjustable fiber delay lines are now described according to reference [54]. This method can also be applied to other space gravitational wave detection missions such as LISAmax, TianQin, AMIGO, ASTROD-GW, Super-ASTROD, etc.

For spaceborne astronomical interferometers, an all-fiber delay line is an ideal choice [54, 300]. A schematic configuration of the fiber delay line is shown in Figure 24. The same configuration can also be used for ground-based astronomical interferometers and space gravitational wave detectors [54, 294]. As shown in Figure 25, we can use EOMs (Electro-Optic Modulators) or PZTs (Piezoelectric Transducers), or both, to achieve a change in optical path length from 0 to $\delta$. The difference in total path length between the lower and upper paths in loop k (k = 1, 2, 3, ..., n) is $2^{k-1} \delta$. To continuously change the optical delay from 0 to $\delta$, we can apply voltage to EOMs and/or PZTs. For reaching $\delta$ to $2\delta$, we can quickly turn off the voltage, switch the laser path of the first loop from the upper path to the lower path, and then increase the voltage applied to EOMs and/or PZTs, similar to what is done in laser length measurement [295] and fiber optic length measurement [296]. When the optical path difference reaches $2\delta$, to change the optical delay continuously from $2\delta$ to $3\delta$, we can quickly turn off the voltage applied to EOMs and/or PZTs simultaneously, switch the laser of the first loop to the upper path, switch the laser of the second loop to the lower path, and then increase the voltage applied to EOMs and/or PZTs again, like a binary counting circuit [295]. Through $n$ loops, we can continuously change the optical delay to a total length of $2^n \delta$ within the switching time precision. The voltage applied to EOMs and PZTs can be quickly turned on and off within 1 millisecond. In 1996, Tsinghua University was developing a time-adjustable directional coupler with a rapid adjustment time of 1 ms in the laboratory. Below is a report on the progress at that time.

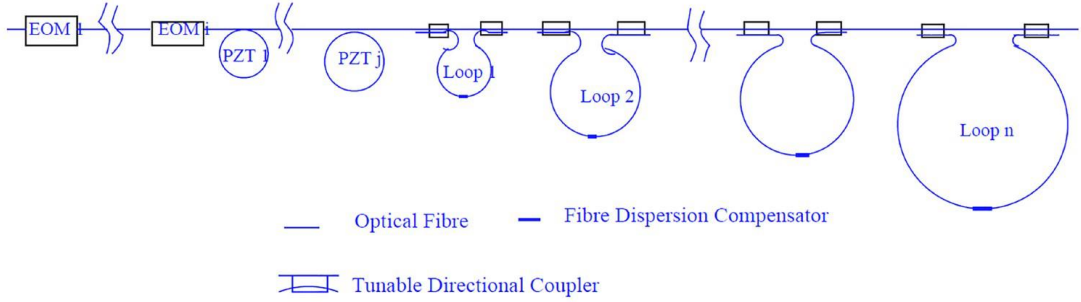

Figure 25 (Color online) A schematic configuration of fibre delay lines. The figure is from refs. [54,55].

Using silicon wafers as polishing substrates [297], we have improved the side polishing technique, simultaneously polishing more than eight optical fibers, achieving high sensitivity (85–90 dB) in droplet testing. Utilizing these side-polished optical fibers (semi-couplers) and liquid crystals, reference [298] manufactured polarizers with high extinction ratios. Due to this improved polishing technique, the effective interaction length of the semi-couplers manufactured in reference [298] has been quadrupled compared to those in reference [299]. Following the methods of Digonnet and Shaw [299], reference [298] produced mechanically adjustable fiber-directional couplers, achieving close to 100% power tunability in single-mode fiber couplers.

To meet the adjustable speed requirements of delay lines, the first step involves applying a voltage to change the refractive index of the matching liquid crystal between two semi-couplers [294]. The expected response time is 1 ms. This time scale is much shorter than characteristic atmospheric turbulence time scales and is sufficient for delay lines in ground-based interferometers. The second step involves planning research on delay lines made from mechanically adjustable directional devices, followed by the development of electrically adjustable directional couplers. Subsequently, we plan to create a prototype of a continuously variable all-fiber delay line with a 1 ms time scale and study its characteristics.

## 8 Detection Sensitivity and Noise Requirements

This section discusses the detection sensitivity density spectrum of various space laser interferometric gravitational wave detectors and their noise requirements.

In the frequency domain, for an incident gravitational wave

$$h(f) = h^{+}(f)\mathbf{e}_{+} + h^{\times}(f)\mathbf{e}_{\times}, \qquad (50)$$

the response $h(f)$ of a space-based gravitational wave detector is given by

$$\underline{h}(f) \equiv F^{+}(\theta, \varphi, \psi, f)\, h^{+}(f) + F^{\times}(\theta, \varphi, \psi, f)\, h^{\times}(f), \qquad (51)$$

where $F^+(\theta, \varphi, \psi, f)$ and $F^\times(\theta, \varphi, \psi, f)$ are complex response functions that depend on the spatial angles $(\theta, \varphi)$ of the gravitational wave source and its polarization angle $\psi$.

The strain power spectral density sensitivity of the space-based gravitational wave detector is defined by the effective noise power spectral density:

$$S_n(f) = P_n(f)/R^2(f), \tag{52}$$

where $R(f)$ is the average response function of the detector to the spatial direction and polarization of the gravitational wave signal. The gravitational wave response function $R^2(f)$ relates the power spectral density of the incident gravitational wave signal to the power spectral density of the recorded signal in the detector.

For any available TDI configuration, the average response function $R_{TDI}(f)$ can be numerically computed from equations (31) and (32) by calculating the phase difference between the two paths. For TDI X0, X, A, and E configurations, their average response functions $R_{X0}^2(f)$, $R_X^2(f)$, $R_A^2(f)$ and $R_E^2(f)$ can be approximated as follows according to [300, 301, 302]:

$$R_{X0}^2(f) \approx 0.3/(1+0.6\,x^2),$$
$$R_X^2(f) \approx (16\,x^2 \sin^2 x)\,R_{X0}(f) \approx (16 x^2 \sin^2 x)\cdot 0.3/(1+0.6\,x^2),$$
$$R_A^2(f) = R_E^2(f) \approx (9/20)\,(16\,x^2 \sin^2 x)\,/\,(1+0.6\,x^2), \tag{53}$$

where $x = 2\pi f L x = 2\pi f L$. At low frequencies, the gravitational wave response of X/A/E channels scales wiyh $x^4$, and the impact of 1% arm length variation in space missions on the gravitational wave response is minimal. At low frequencies, the response of TDI T channel to gravitational waves correlates with arm length differences [302]:

$$R_T^2(f) \approx 0.8 x^4\,[(L_{12} - L_{23})^2 + (L_{12} - L_{13})^2 + (L_{13} - L_{23})^2]. \tag{54}$$

Figure 26 illustrates the average response functions of LISA and Taiji X, A, E, and T TDI channels. The response of the T channel varies with arm length changes, where the dark gray region corresponds to a 50% reduction in arm length variation over 400 days, and the combined dark and light gray regions correspond to a 90% reduction in arm length over 400 days. Histograms on the right side of the upper and lower images show the T channel response at frequencies of 0.01 mHz, 0.1 mHz, and 1 mHz [302].

Optical measurement noise contributes four times to the Michelson X TDI channel, while test mass acceleration noise contributes sixteen times. By dividing displacement by twice the round-trip distance 2L, we convert it into strain, thus obtaining:

$$4P_{op,mission}/(2L_{mission})^2 = P_{op,mission}/(L_{mission})^2. \tag{55}$$

Similarly, the factor 1/(2L) is also applied to path length changes caused by gravitational waves: $16 P_{acc,mission}/(2L_{mission})^2 = 4 P_{acc,mission}/(L_{mission})^2$. The strain power spectral density $P_n(f)$ is therefore:

$$\begin{aligned} P_n(f) &= P_{12}^{op}(f) + P_{21}^{op}(f) + P_{13}^{op}(f) + P_{31}^{op}(f) \\ &\quad + \{4\cos^2(f/f^*) \times [P_{21}^{acc}(f) + P_{31}^{acc}(f)+] + 4 P_{12}^{acc}(f) + 4 P_{13}^{acc}(f)\}/[(2\pi f)^4 (L)^2] \\ &= 4 P^{op}(f) + 8 [1+\cos^2(f/f^*)] \times P^{acc}(f)/[(2\pi f)^4 (L)^2], \end{aligned} \tag{56}$$

In the above equations, assuming each detector in each spacecraft has the same noise spectrum $P^{op}(f)$ and $P^{acc}(f)$. The $\cos^2(f/f^*)$ term arises from the combination of inertial and accelerometer noise in times $t$ and $t-2L$ in spacecraft 1.

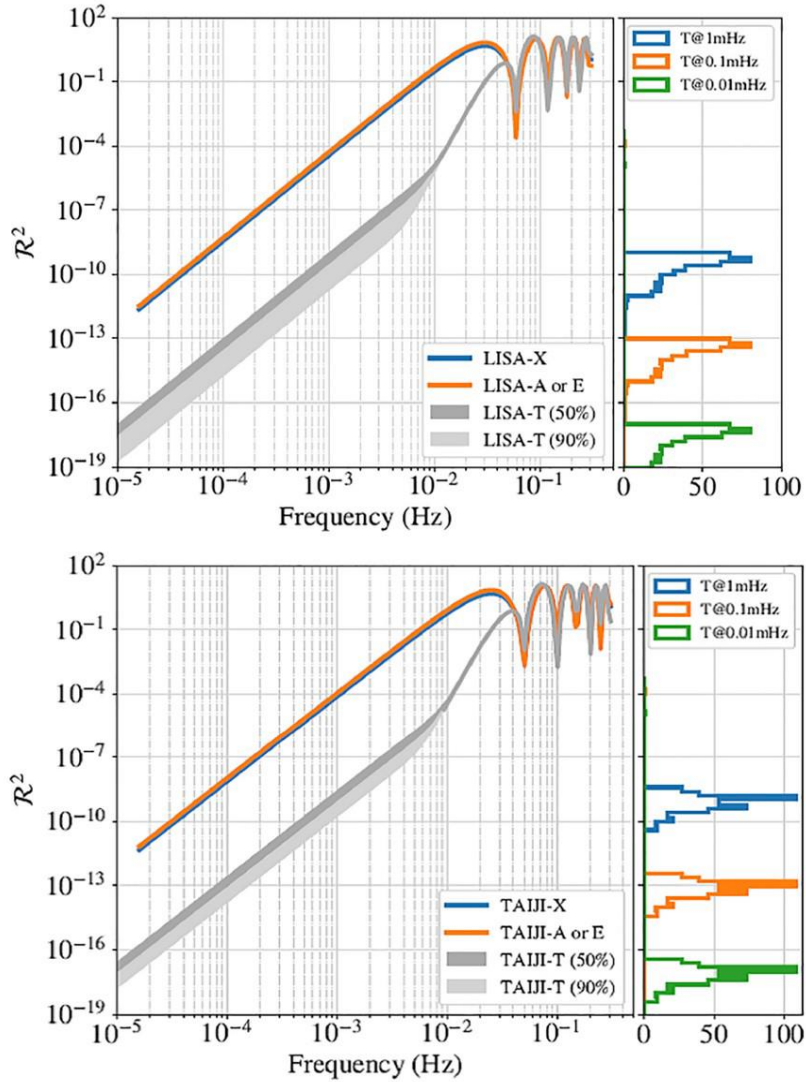

Figure 26 (Color online) The average responses of TDI X, A, E, and T channels for LISA (upper panel) and Taiji (lower panel). The T channel is sensitive to the variances of arm lengths, and the dark grey region shows the best 50% percentile in 400 days, and the dark and light grey areas together show the best 90% percentile. The right panel in each plot shows the histogram of the T channel's response at frequencies 0.01, 0.1, 1 mHz. The figure is from ref. [302].

Hence, the total noise in the X TDI channel is:

$$P_X(f) = 4 P^{op}(f)/(4L^2) + 8 [1 + \cos^2(f/f^*)]P^{acc}(f)/[(2\pi f)^4(2L)^2]$$
$$= P^{op}(f)/L^2 + 2 [1 + \cos^2(f/f^*)] P^{acc}(f) / [(2\pi f)^4(L)^2]. \tag{57}$$

The contribution of noise to the gravitational wave signal in the X channel PXn(f)P_{Xn}(f)PXn(f) is approximately:

$$P_{Xn}(f) \approx 16x^2\sin^2 x \, P_X(f). \quad \text{[related to } R_X^2(f) \approx (16 x^2 \sin^2 x) R_{X0}(f))]. \tag{58}$$

From equations (50), (51), (57), and (58), the strain power spectral density sensitivity of the space-based gravitational wave detector's Michelson X TDI channel, i.e., its effective noise power spectral density, is given by:

$$S_{Xn}(f) = P_{Xn}(f)/R_X^2(f) = 16x^2\sin^2 x \, P_X(f)/(16 x^2 \sin^2 x R_{X0}(f))$$
$$= P_X(f)/R_{X0}(f) \approx \{P_{op,mission}/(L_{mission})^2 + 2[1+\cos^2(f/f^*)]P_{acc,mission}/[(2\pi f)^4(L_{mission})^2\}/ [0.3/(1+0.6 \, x^2)]$$
$$= [10/(3L_{mission}^2)] [P_{op,mission}^2 + 2[1+\cos^2(f/f^*)]P_{acc,mission}/(2\pi f)^4] \times (1+0.6 \, x^2)$$
$$\approx [10/(3L_{mission}^2)] [P_{op,mission}^2 + 4P_{acc,mission}/(2\pi f)^4] \times (1+0.6 \, x^2). \tag{59}$$

If the contribution of frequency noise effects and other noise suppressions can be neglected below the core noise level, the gravitational wave sensitivity given below is excellent for most estimates and analyses [300]:

$$S_{Xn,mission}^{1/2}(f)=(20/3)^{1/2}(1/L_{mission})\times[(1+(f/(1.29f_{mission})^2]^{1/2}\times[(P_{op,mission}+4P_{acc,mission}/(2\pi f)^4)]^{1/2}\text{Hz}^{-1/2}, \tag{60}$$

where $L_{mission}$ is the arm length of the gravitational wave space mission, and $f_{mission}=1/(2\pi L_{mission})$ is the detector's critical (characteristic) frequency. Equation (60) is an excellent approximation of the gravitational wave sensitivity averaged over sky positions and polarizations for the Michelson interferometer X TDI channel in a nearly equilateral triangle formation. For derivation, please refer to references [300] and the cited references therein. It also serves as an excellent approximation for the classic Michelson interferometer in triangular formation.

As mentioned in Section 1.12 and Figure 12, the strain power spectral density amplitude used in this article corresponds to the gravitational wave sensitivity averaged over sky positions and polarizations for the nearly equilateral triangular formation as given by equation (60).

As described in Section 6, LISA's requirements for accelerometer noise spectrum amplitude are given by equation (43), i.e., $[P_{acc,LISA}(f)]^{1/2} \leq 3 \text{ fm s}^{-2} \text{ Hz}^{-1/2} [1 + (0.4 \text{ mHz} / f)^2]^{1/2} \times [1 + (f/8 \text{ mHz})^4]^{1/2}$, $(1 \times 10^{-4} \text{ Hz} \leq f \leq 1 \text{ Hz})$; its requirement for optical ranging noise (optical length noise) is

$$[P_{op,LISA}(f)]^{1/2} \leq 10 \text{ pm Hz}^{-1/2} [1 + (2 \text{ mHz} / f)^4]^{1/2}. \tag{61}$$

Substituting these two equations into equation (60), we obtain the strain power spectral density amplitude / gravitational wave sensitivity for the LISA detector:

$$S_{n,\text{LISA}}^{1/2}(f) = (20/3)^{1/2}(1/L_{\text{LISA}})\times[(1+(f/(1.29f_{\text{LISA}}))^2)]^{1/2}\times [(P_{\text{op,LISA}}+4P_{\text{acc,LISA}}/(2\pi f)^4)]^{1/2}\text{Hz}^{-1/2}, \qquad (62)$$

The strain power spectral density amplitude / gravitational wave sensitivity for LISA as depicted in Figure 12 is given by this equation. The coefficient for acceleration noise listed for LISA in Table 2 is the factor 3 fm s$^{-2}$ Hz$^{-1/2}$ from equation (43). The laser ranging noise (optical length noise) is the factor 10 pm Hz$^{-1/2}$ from equation (61). After careful estimation, LISA [51] relaxed this coefficient to 15 pm Hz$^{-1/2}$ in its Definition Study Report.

The Taiji mission has the same requirements for accelerometer noise spectrum amplitude as LISA, given by equation (43). Its requirement for optical ranging noise is:

$$[P_{\text{op,Taiji}}(f)]^{1/2} \leq 8 \text{ pm Hz}^{-1/2} [1 + (2 \text{ mHz}/f)^4]^{1/2}. \qquad (63)$$

The strain power spectral density amplitude / gravitational wave sensitivity for the Taiji detector as depicted in Figure 12 is given by equation (60) corresponding to Taiji.

According to equation (40), the laser frequency noise effect in the X channel is:

$$\delta\varphi_X \cong 2\pi\delta\nu(f) \times \delta t_X, \qquad (64)$$

where $\delta\nu(f)$ is the frequency noise of the laser source at frequency $f$, and $\delta t_X$ is the optical path delay between the two paths of the X channel. (64) also applies to other TDI channels.

Figure 27 shows the core noise and laser frequency noise effects for LISA (upper image) and Taiji (lower image) in the first 400 days of scientific mission. The core noise and laser frequency noise effects of the T channel vary due to changes in arm length.

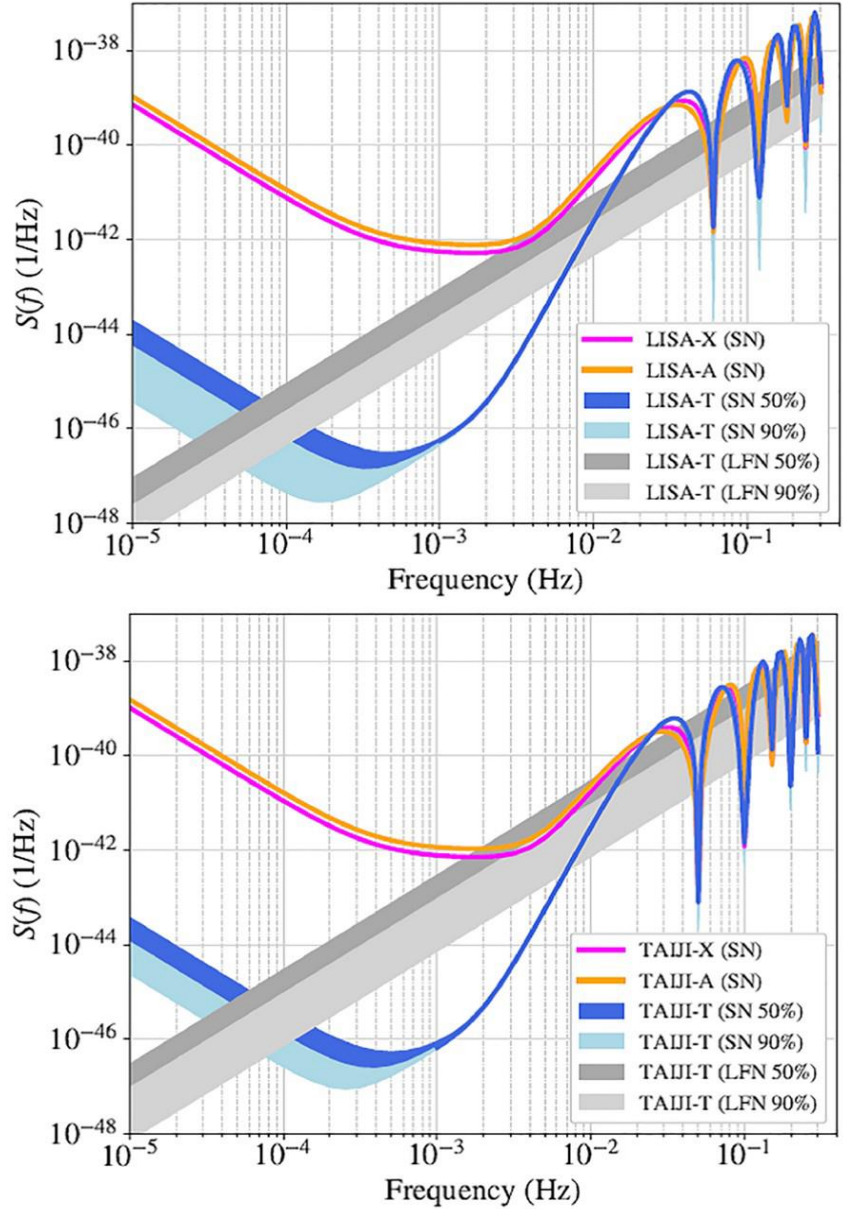

Figure 27 (Color online) The core noise (SN, acceleration noise +optical metrology noise) and the laser frequency noise (LFN) effect for LISA (upper panel) and Taiji (lower panel) in the first 400 days. The dark grey shows the highest laser noise effect in 50% of the first 400 days, and dark grey together with light grey show the laser noise effect 90% percentile in T channel. The dark blue shows the highest core noise in 50% of the first 400 days in T channel, and the dark blue together with light blue show the core noise in 90%. The figure is from ref. [302].

The strain power spectral density amplitude/gravity wave sensitivity of ALIA is plotted based on its parameters from Table 2 and its corresponding equation (60), excluding red and blue factors.

ASTROD-GW has requirements for the noise spectral amplitude of its inertial/acceleration sensor:

$$[S_{\text{acc,ASTROD-GW}}(f)]^{1/2} \leq 3 \text{ fm s}^{-2} \text{ Hz}^{-1/2} [1+(0.1\text{mHz}/f)^2]^{1/2}. \tag{65}$$

Its requirements for optical ranging noise are:

$$[S_{\text{op,ASTROD-GW}}(f)]^{1/2} \leq 1000 \text{ pm Hz}^{-1/2} [1+(0.2 \text{ mHz}/f)^4]^{1/2}. \tag{66}$$

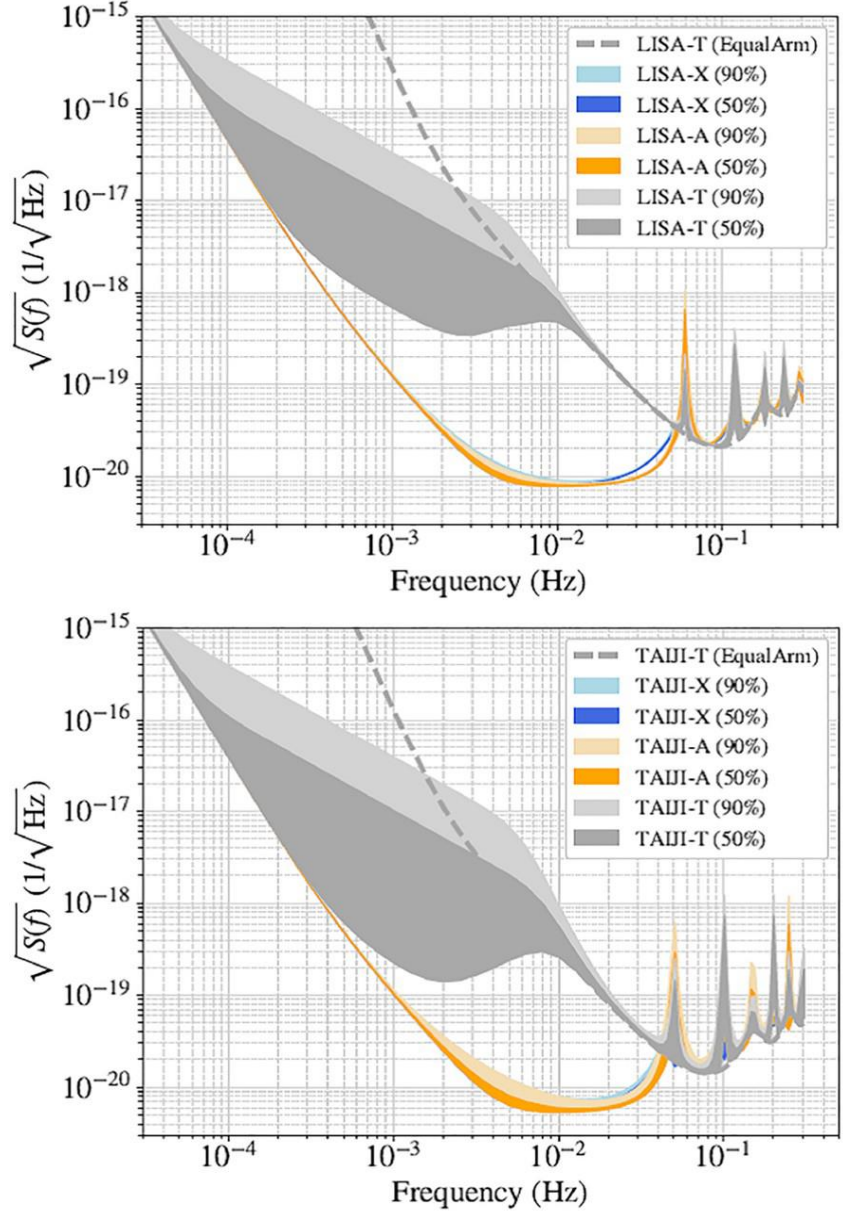

Figure 28 (Color online) The average sensitivities of TDI X, A, E, and T channels by considering the core noises and laser noise effect. The dark color areas show the best sensitivity in the 50% percentile of the first 400 days for X, A, and T channels, and the dark color together with the light color areas show the best sensitivity in 90% percentile. The sensitivities of T channels for equal-arm configuration are shown by dashed curves [303,304]. The figure is from ref. [302].

The strain power spectral density amplitude/gravity wave sensitivity of ASTROD-GW, according to equation (60), is plotted in Figure 12. Folkner's mission has noise requirements and gravity wave sensitivity similar to ASTROD-GW. The requirements for the inertial/acceleration sensor noise spectral amplitude and optical ranging noise of aASTROD-GW are both ten times smaller than those of ASTROD-GW. LISAmax has noise requirement for inertial/acceleration sensors similar to ASTROD-GW, and its optical ranging noise is similar to aASTROD-GW.

The strain power spectral density amplitude/gravity wave sensitivity of DECIGO and B-DECIGO in Figure 12 is plotted according to reference [87]. The Big Bang Observer is similar to DECIGO. The strain power spectral density amplitude/gravity wave sensitivity of AMIGO, b-AMIGO, and e-AMIGO is plotted according to reference [106]. The sensitivity curve of DO is similar to e-AMIGO.

Super-ASTROD has noise requirements for inertial/acceleration sensors similar to aASTROD-GW, and its optical ranging noise requirement is 5 nm/Hz$^{1/2}$. The dashed lines extending to lower frequencies around ~100 µHz in Super-ASTROD, aASTROD-GW, and ASTROD-GW represent the removal of the red factor $[1 + (0.1 \text{ mHz} / f)^2]^{1/2}$ from the requirements for inertial/acceleration sensor noise.

The strain power spectral density amplitude/gravity wave sensitivity of TianQin in Figure 12 is plotted according to reference [99].

The strain power spectral density amplitude/gravity wave sensitivity of gLISA in Figure 12 is plotted according to references [115-118].

## 9 Scientific Goals of Space Gravitational Wave Detection

This section reviews and discusses the scientific objectives of proposed and ongoing mid-frequency (decihertz) and low-frequency (millihertz, microhertz) space gravitational wave detection missions and projects (references 45, 50, 51, 78, 80, 81, 128). LISA was approved earlier this year, scheduled for launch in 2035. Further research over the next decade on scientific objectives and data analysis will be highly beneficial for preparing space gravitational wave detection missions.

### 9.1 Compact Binaries

From the understanding of astrophysical binary star evolution and the structure of the Milky Way, it is estimated that the Milky Way contains roughly ten million compact binary and multi-star gravitational wave sources. Most of these sources contribute only to the gravitational wave foreground, with only approximately tens of thousands of systems discernible by the millihertz space gravitational wave detectors like LISA or Taiji/TianQin, capable of detecting individual inspiral and merger-generated gravitational waves and studying the structure of the Milky Way (references 45, 49, 51). LISA and Taiji are expected to detect about 10,000 double white dwarf binaries, with the majority in the frequency range of 3-6 mHz (orbital periods of approximately 300-600 seconds) (references 45, 49, 51). These sources belong to the population of normal Type Ia supernovae and special supernova progenitors. For a review of electromagnetic counterparts of compact object gravitational wave mergers, refer to reference [305]. In the 3-6 mHz frequency band, LISA/Taiji are more sensitive compared to microhertz low-frequency gravitational wave detection missions like ASTROD-GW, Folkner's mission, and LISAmax (Figure 12). LISA/Taiji will launch first and, in addition to validating binary stars, these signals can also serve as calibration sources for low-frequency gravitational wave detection missions [306, 307, 308].

Below frequencies of a few millihertz, approximately ten million compact binaries will form the detectable foreground for LISA/Taiji and microhertz low-frequency gravitational wave detectors [111, 309, 310, 311, 312, 313, 314, 315). At these frequencies, microhertz low-frequency gravitational wave detectors are more sensitive compared to millihertz low-frequency gravitational wave detectors (Figure 12). More gravitational wave sources will be separately detected, and microhertz low-frequency gravitational wave detectors will improve the observational outcomes of millihertz low-frequency gravitational wave detectors like LISA and Taiji/TianQin [111].

Ground-based gravitational wave detectors have already detected neutron star mergers and over a hundred stellar-mass black hole mergers. Based on the inferred distribution densities from existing observations, it is anticipated that millihertz low-frequency gravitational wave detectors will detect tens to hundreds of events involving neutron stars, stellar-mass black holes, or white dwarf mergers. These events will further enhance our understanding of various merger processes [51].

Section 1.9, Table 1, and Figure 10 discuss multi-band observations simulations for millihertz, mid-frequency and high-frequency gravitational waves, illustrating that the observation of stellar-mass black hole mergers will achieve measurement accuracy approaching or surpassing one-thousandth in cosmological model parameters and the Hubble constant and will improve our understanding of physical cosmology and dark energy significantly [107].

**9.2 Massive Black Holes and Their Co-Evolution with Galaxies**

The astronomical community has discovered relationships between the masses of supermassive black holes and the masses of their host galaxy nuclei, as well as between the masses of supermassive black holes and the velocity dispersions of their host galaxies. These relationships indicate a connection between central supermassive black holes and the evolution of galaxy structure. Observational evidence suggests that supermassive black holes exist in the majority of local galaxy groups. Newly fueled quasars may arise from the merger of two gas-rich massive galaxies. Space interferometers in the low-frequency range (100 nHz - 100 mHz) for gravitational wave observations and pulsar arrays in the ultra-low-frequency range (300 pHz - 100 nHz) will become key tools for studying the coevolution of galaxies and supermassive black holes.

The standard theory for the formation of supermassive black holes involves the merger of various supermassive black hole binaries. Using space gravitational wave detectors and pulsar timing arrays, gravitational waves from these supermassive black hole binary mergers can be detected and explored across cosmic distances. Despite different merger tree models and merger tree models with seed black holes, they predict significant detection rates for low-frequency space gravitational wave detectors (LISA/Taiji/TianQin and ASTROD-GW/Folkner's space mission/LISAmax) and pulsar timing arrays [49, 179, 316, 317, 318]. Pulsar timing arrays are most sensitive in the 300 pHz - 100 nHz frequency range, LISA & Taiji/TianQin are most sensitive in the 3 mHz - 0.1 Hz frequency range, and microhertz low-frequency gravitational wave detectors are most sensitive in the 100 nHz - 3 mHz frequency range (Figures 12-14, references [126] Figures 2-4). Low-frequency gravitational wave detectors will directly observe how supermassive black holes form, grow, and interact throughout the history of galaxy formation. ASTROD-GW/LISAmax will detect stochastic gravitational wave backgrounds from supermassive black hole binary mergers in the 100 nHz - 100 μHz frequency range. LISA/Taiji can detect the spin, merger, and oscillation of double supermassive black hole sources with masses of $10^5$ - $10^7$ solar masses, and the spin of double black hole sources with masses of $10^4$ solar masses, at an expected rate of 10 to 1000 detections per year [51]. Similar numbers of sources are expected for microhertz low-frequency gravitational wave detection missions, but with better angular resolution (see section 10.3), covering gravitational wave event masses ranging from $10^5$ to $10^{10}$ solar masses (reference 81). These observational results are crucial for studying the coevolution of galaxies hosting supermassive black holes. Low-frequency gravitational wave detection missions will explore the cosmic horizon beyond z ~ 20.

Figures 12-14 depict trajectories of some examples of double supermassive black hole gravitational wave sources' inspirals, mergers, and oscillations on the plots. The black lines represent gravitational wave inspirals, mergers, and oscillations emitted by various equal-mass black hole binary mergers on circular orbits at different redshifts: solid line for z = 1, dashed line for z = 5, and long dashed line for z = 20. The trajectories of $10^6$ $M_\odot$ - $10^6$ $M_\odot$ double supermassive black hole mergers at z = 1, $10^5$ $M_\odot$ - $10^5$ $M_\odot$ double supermassive black hole mergers at z = 20, and $10^4$ $M_\odot$ - $10^4$ $M_\odot$ double supermassive black hole mergers at z = 20 are from reference [319]; others are scaled accordingly. Corresponding curves in Figures 13 and 14 are obtained by the transforming equation (1). Double supermassive black hole merger events have high signal-to-noise ratios for space detectors. Some equal-mass (ranging from $10^2$ $M_\odot$ to $10^{10}$ $M_\odot$) circular orbit events shown in Figures 12-14 are targets for space detectors, with lower mass events possibly detectable in early stages by future ground-based detectors.

By detecting double supermassive black hole merger events and their background gravitational waves, properties and distributions of supermassive black holes can be inferred and potential population models tested. For decades, pulsar timing arrays have been collecting data to detect stochastic gravitational wave backgrounds from double supermassive black hole mergers. Modeling of the isotropic stochastic background spectrum density for double supermassive black hole mergers has led many authors to obtain frequency attributes of equation (35), with recent observational results from pulsar timing arrays listed in Table 6. The fit of isotropic random background equation (35) model essentially yields a result of $A_{yr}$ = $2.4^{+0.7}_{-0.6}$, confirming the existence of stochastic gravitational waves in the ultra-low-frequency range at a 4σ uncertainty level, broadly consistent with astrophysical understanding.

**9.3 Extreme Mass Ratio Inspirals (EMRI)**

Extreme mass ratio inspirals are sources of gravitational waves for space gravitational wave detectors. LISA/Taiji are sensitive to central massive objects in the mass range of $10^4$ - $10^7$ solar masses. LISA/Taiji is expected to detect gravitational wave events ranging from hundreds to thousands per year [51]; ASTROD-GW is similar or more, particularly to larger central black holes, and it offers better angular resolution (section 10.3).

**9.4 Testing Relativistic Gravity**

An important scientific goal of millihertz low-frequency gravitational wave detection is to test general relativity and precisely study black hole physics under strong gravity [45, 51]. Microhertz low-frequency gravitational wave detection, with its higher precision in the frequency range of 100 nHz to 1 mHz, will further advance this goal in many aspects. This includes testing strong-field gravity, precise detection of Kerr spacetime properties, and measuring/constraining the mass of gravitons. Considerations related to these aspects have been discussed in reference [320]. The sensitivity at lower frequencies is crucial for enhancing the precision of various tests [320]. Further research in these areas will be highly valuable.

**9.5 Dark Energy and Cosmology**

In the context of dark energy, determining the value of $w$ in the dark energy equation of state

$$w = p/\rho \tag{67}$$

is crucial. Here, $w$ is a function of cosmic time, $p$ represents pressure, and $\rho$ denotes the density of dark energy. A cosmological constant for dark energy corresponds to $w=-1$. From observational cosmology, our universe appears to be nearly flat. In a flat Friedman-Lemaître-Robertson-Walker (FLRW) universe, the luminosity distance $d_L$ is given by:

$$d_L(z) = (1+z)(H_0)^{-1} \int_0^z dz' [\Omega_m(1+z')^3 + \Omega_{\text{DE}}(1+z')^{3(1+w)}]^{-\frac{1}{2}}, \tag{68}$$

where $H_0$ is the Hubble constant, $\Omega_{\text{DE}}$ is the current dark energy density parameter, and $w$ is the assumed constant state equation of dark energy. For non-constant $w$ and non-flat FLRW universes, similar but more complex expressions can be derived. Here, we present Equation (68) to illustrate the process.

By comparing the observed luminosity distance as a function of redshift $z$, one can determine the parameter $w$ as a function of redshift and compare it with various cosmological models. Dark energy cosmological models can be tested in this way. Supernova observations and gamma-ray burst observations, together with redshift measurements of luminosity distances, are currently the focus of dark energy detectors. Gravitational wave detectors will provide even more precise results.

Spacegravitational wave detectors observing the inspiral of binary supermassive black holes and extreme mass ratio inspirals are excellent tools for determining luminosity distances. Along with redshifts determined from related electromagnetic observations of galaxies or galaxy clusters, these space gravitational wave detectors also serve as dark energy probes. The gravitational waveforms produced by mergers of massive black holes in the co-evolution of galaxies provide accurately calibrated luminosity distances at high redshifts. These binary inspiral signals can serve as standard candles/whistles [321, 322]. With better angular resolution (Section 10.3), microhertz low-frequency gravitational wave detection will have better opportunities to identify associated electromagnetic redshifts, thus improving the determination of the dark energy equation of state [80, 81].

## 9.6 Primordial (Relic) Gravitational Waves

To directly detect primordial (or inflationary) gravitational waves, researchers can use frequencies lower or higher than the LISA band [78, 323]. This is because these frequencies may have fewer astrophysical foreground sources to mask the detection [324]. DECIGO [84] and Big Bang Observer [85] search for background gravitational waves at higher frequencies, while ASTROD-GW [78, 323] targets the lower frequency range. Their instrument sensitivity goals aim to reach the critical density level of $10^{-17}$. The main challenge lies in the magnitude of the foreground and whether it can be distinguished.

The straight line in the lower left corner of Figure 12 corresponds to the upper limit of $\Omega_{gw}=1\times10^{-16}$ for the cosmic microwave background of primordial gravitational waves. For ASTROD-GW, sensitivity can reach this level when using a 6-spacecraft formation for the stochastic detection of gravitational waves. However, the expected upper limit of massive black hole gravitational wave foregrounds is higher than the sensitivity of 3-spacecraft ASTROD-GW. Detecting this background depends on whether the "foreground" from these massive black hole gravitational waves can be separated by different frequency dependencies or other characteristics distinguishable by the 6-spacecraft ASTROD-GW [80].

Other potential sources of gravitational waves within related frequency bands, such as those generated by cosmic strings and phase transitions, are also worth exploring, as detailed in references [50, 51].

## 10 Basic Orbital Configurations, Angular Resolution, and Multiple-Formation Configurations

In the first section, we will first discuss the orbital configurations around Earth. Sections two and three provide an overview of basic orbits akin to LISA and ASTRO-GW. Sections four and five respectively discuss their angular resolution and configurations with multiple formations. These fundamental configurations will be used in section eleven for numerical design and orbital optimization for both types of missions.

### 10.1 Geocentric Orbital Configurations

These orbits are around-Earth orbits far from lunar orbits (either inside or outside). The orbital configuration of the OMEGA gravitational wave detection mission [325, 326] serves as an early example. Proposed to NASA as a candidate MIDEX mission in 1998, it was re-proposed in 2011 as a mission concept white paper. The mission consists of six identical spacecraft in a stable Earth orbit at an altitude of 600,000 kilometers, with two spacecraft at each vertex forming an approximately equilateral triangle. These orbits are stable, allowing for a three-year scientific operation plan, extendable if needed. The arms of the triangle are approximately 1 million kilometers (1 Gm) long. The mission formation is located beyond lunar orbit.

Currently and potentially selected space mission concepts using around-Earth orbital configurations actively under study include TianQin, gLISA, and B-DECIGO. Research has been conducted on the orbit of AMIGO [105], which has opted for a heliocentric orbit. Two mission proposals, GEOGRAWI [116]/gLISA [117] and GADFLI [118], use geostationary Earth orbits. These missions operate with three spacecraft in a near-equilateral triangle formation at a distance of approximately 73,000 kilometers. The orbit selection for B-DECIGO, whether it will use a heliocentric or geocentric orbit, has not been finalized.

TianQin is a gravitational wave detection mission in an orbit around Earth's center at a distance of 100,000 kilometers. Three spacecraft form a nearly equilateral triangle with an arm length of about 170,000 kilometers, orbiting Earth with a period of 44 hours [99].

During the Earth's orbit around the Sun, there are times when sunlight can shine into the telescope's line of sight. When the line of sight passes through the Sun, a sunshield is needed; OMEGA [326] proposed a solution that may also be applicable to other missions of this type. TianQin proposes using two groups of three spacecraft each, alternating observations every three months.

**10.2 Basic Orbital Configuration of LISA-Type Missions**

As shown in Figure 3, the basic LISA-type formation [38,45,293,327-332] orbits around the Sun with a circular orbit radius $R$ (= 1 AU). Since the distance (arm length) $L$ between spacecraft is much smaller than the circular orbit radius of 1 AU, we can consider the spacecraft orbits as perturbed orbits around the circular orbit. The orbital equations for the spacecraft relative to the formation center (center of mass) are known as Euler-Hill equations, Hill equations, or Clohessy and Wiltshire equations. Hill used these equations in the 19th century for lunar theory research [333]. Clohessy and Wiltshire [334], after the start of the space age in 1957, derived and used these equations to design terminal guidance systems for satellite rendezvous in 1960. They defined a coordinate system around the center-of-mass orbit of the formation (either around the Sun or Earth) called the CW coordinate system. The origin of this system lies on the circular reference (formation center) orbit, rotating at the same angular velocity $\Omega$ as the reference orbit.

To maintain a perturbed orbit at a first-order distance O($\alpha$) = O($L/(2R)$) from the origin and to remain stationary in the CW coordinate system, the orbital elements eccentricity $e$ and inclination $i$ must satisfy the following relationships:

$$e = 3^{-1/2} i + \mathrm{O}(\alpha), \quad i = \varepsilon \equiv \alpha \equiv L/(2R). \tag{69}$$

One method to form a nearly equilateral triangle configuration with side length or arm length $L(1 + \mathrm{O}(\alpha))$ is to space the orbital nodes by 120 degrees and select true anomaly and argument of perigee such that each spacecraft reaches its maximum altitude (northern hemisphere) above the ecliptic plane at the apogee (first configuration); another method is to place the apogee at the minimum altitude (southern hemisphere) below the ecliptic plane (second configuration) [38]. With these choices, the mission configuration plane forms a 60-degree angle with the ecliptic plane, tangential to the formation center circular orbit. For a square configuration, spacing the orbital nodes by 90 degrees suffices; similar constructions can be made for any regular polygon or planar configuration. The first configuration rotates clockwise; the second configuration rotates counterclockwise. Thus, in the CW coordinate system, there are only two planes at ±60 degrees to the reference orbit plane, where spacecraft (test particles) follow the CW equations, rotating rigidly around the origin at angular velocities of −Ω and Ω.

Following methods by Dhurandhar et al. [332] and Wang and Ni [289], the basic orbital equations for three spacecraft in a LISA-type configuration can be written. First, for any ellipse in the X-Y plane, the orbital equations are:

$$X = R(\cos \psi + e), \quad Y = R(1 - e^2)^{1/2} \sin \psi, \tag{70}$$

where $R$ is the semi-major axis of the ellipse, $e$ is the eccentricity, and $\psi$ is the true anomaly. Define $\alpha$ as the planned arm length $L$ of the orbital configuration relative to twice the average Earth orbit radius $R$ (1 AU), i.e., $\alpha = L/(2R)$. Using a heliocentric coordinate system $(X, Y, Z)$ and selecting initial time $t_0$ as a specific epoch in the Julian calendar, with $X$-axis in the direction of the vernal equinox. Define a set of spacecraft elliptical orbits as:

$$\begin{aligned}
X_f &= R(\cos\psi_f + e)\cos\varepsilon, \\
Y_f &= R(1-e^2)^{1/2}\sin\psi_f, \\
Z_f &= R(\cos\psi_f + e)\sin\varepsilon.
\end{aligned} \tag{71}$$

For LISA (2.5 Gm arm length), $R = 1$ AU; $e = 0.004811$; $\varepsilon = 0.008333$. The relationship between true anomaly $\psi_f$ and mean anomaly $\Omega(t - t_0)$ is:

$$\psi_f + e\sin\psi_f = \Omega(t - t_0). \tag{72}$$

where $\Omega$ is $2\pi$ divided by one sidereal year. $\psi_f$ can be solved numerically through iteration. Specifically, define $\psi_k$ given by:

$$\psi_k + e\sin\psi_k = \Omega(t - t_0) - 120°(k - 1), \text{ for } k = 1, 2, 3. \tag{73}$$

Define $X_{fk}$, $Y_{fk}$, $Z_{fk}$ (k = 1, 2, 3) to be:

$$\begin{aligned}
X_{fk} &= R(\cos\psi_k + e)\cos\varepsilon, \\
Y_{fk} &= R(1-e^2)^{1/2}\sin\psi_k, \\
Z_{fk} &= R(\cos\psi_k + e)\sin\varepsilon.
\end{aligned} \tag{74}$$

Define $\varphi_0 \equiv \psi_E - 10°$, where $\psi_E$ is the Earth's angle position relative to the X-axis at $t_0$. Define $X_{f(k)}$, $Y_{f(k)}$, $Z_{f(k)}$, ($k = 1, 2, 3$), i.e. $X_{f(1)}$, $Y_{f(1)}$, $Z_{f(1)}$; $X_{f(2)}$, $Y_{f(2)}$, $Z_{f(2)}$; $X_{f(3)}$, $Y_{f(3)}$, $Z_{f(3)}$ to be

$$X_{f(k)} = X_{fk} \cos[120°(k\text{-}1)+\varphi_0] - Y_{fk} \sin[120°(k\text{-}1)+\varphi_0],$$
$$Y_{f(k)} = X_{fk} \sin[120°(k\text{-}1)+\varphi_0] + Y_{fk} \cos[120°(k\text{-}1)+\varphi_0], \quad (75)$$
$$Z_{f(k)} = Z_{fk}.$$

The basic orbits for three spacecraft (for one-body central problem) are:

$$\mathbf{R}_{S/C1} = (X_{f(1)}, Y_{f(1)}, Z_{f(1)}),$$
$$\mathbf{R}_{S/C2} = (X_{f(2)}, Y_{f(2)}, Z_{f(2)}), \quad (76)$$
$$\mathbf{R}_{S/C3} = (X_{f(3)}, Y_{f(3)}, Z_{f(3)}).$$

Both LISA and TAIJI are space missions of LISA type located at an average distance of 1 AU from the Sun. The initial design positions can be obtained by selecting $t = t_0$. The initial design velocities can be obtained by calculating the position derivative at time $t = t_0$. For example, choosing $t_0$ = JD2461853.0 (March 22, 2028, 12:00:00), the initial design configurations (positions and velocities) of the three spacecraft of LISA at this time in J2000.0 ecliptic coordinate at the vernal equinox can be computed as described in Section 11, followed by optimization of arm length and velocity. The final design configuration of the spacecraft is listed in Table 8, column three, from reference [145]. The initial design configuration of TAIJI's three spacecraft at this time in J2000.0 ecliptic coordinate at the vernal equinox can be calculated and listed in Table 9, column three, from reference [145]. Starting from these initial conditions, the orbital configuration can be optimized using planetary and lunar ephemeris as shown in Section 11.2, and the optimized final spacecraft initial state column is listed in the fifth column of Table 4. For other options in different periods (for example, in the period of 2036 approaching the time when the TAIJI program reaches the scientific orbit), the approach is the same.and so forth for other choices at different periods.

Table 8 Initial conditions of three S/C for optimized LISA with arm length 2.5 Gm at epoch JD2461853.0 (2028-Mar-22[nd] 12:00:00) in J2000 equatorial (Earth mean equator and equinox) solar-system-barycentric coordinate system. The table is extracted from [145].

|  |  | Initial states of S/C after final optimization |
|---|---|---|
| S/C1 position (AU) | $X, Y, Z$ | −9.342355891858E−01, 3.222027047288E−01, 1.415510473840E−01 |
| S/C1 velocity (AU/d) | $V_x, V_y, V_z$ | −6.020533666442E−03, −1.471303796371E−02, −6.532104563056E−03 |
| S/C2 position (AU) | $X, Y, Z$ | −9.422917194822E−01, 3.075956329521E−01, 1.403200701890E−01 |
| S/C2 velocity (AU/d) | $V_x, V_y, V_z$ | −5.875601408922E−03, −1.480936170059E−02, −6.319195852807E−03 |
| S/C3 position (AU) | $X, Y, Z$ | −9.335382669969E−01, 3.132742531958E−01, 1.273476800288E−01 |
| S/C3 velocity (AU/d) | $V_x, V_y, V_z$ | −5.949351791423E−03, −1.490443611747E−02, −6.410590762560E−03 |

**10.3 Basic Orbital Configuration of ASTROD-GW**

ASTROD-GW's basic configuration consists of three spacecraft located near the Sun-Earth Lagrange points L3, L4, and L5, orbiting the Sun in nearly circular orbits, forming an equilateral triangle as shown in Figure 4 [76-81]. In the restricted three-body problem of the Earth-Sun-spacecraft system, the dominant force on the spacecraft comes from the Sun. Since the Earth-Sun orbit is elliptical, the Lagrange points are not stationary in the Earth-Sun rotating frame. The motion of test particles near L3,

L4, and L5 deviates from circular orbits by approximately O(e), where e (=0.0167) is the eccentricity of Earth's orbit around the Sun. However, spacecraft can be placed in halo orbits near each Lagrange point, largely compensating for the non-stationary motion of the Lagrange points to maintain nearly circular orbits around the Sun.

**Table 9** Initial states (conditions) of three S/C of Taiji at epoch JD2461853.0 (2028-Mar-22nd 12:00:00) for our initial choice (third column) and after optimizations (fifth column) in the J2000 equatorial (Earth mean equator and equinox) solar-system-barycentric coordinate system [145]

|  |  | Initial choice of S/C initial states |  | Initial states of S/C after final optimization |
|---|---|---|---|---|
| S/C1 Position (AU) | X | -9.337345684115E-01 | adjust to ==> | -9.337343160303E-01 |
|  | Y | 3.237549276553E-01 |  | 3.237548395220E-01 |
|  | Z | 1.426066025785E-01 |  | 1.426065637750E-01 |
| S/C1 Velocity (AU/day) | $V_x$ | -6.034814754038E-03 | = | -6.034814754038E-03 |
|  | $V_y$ | -1.469355864558E-02 |  | -1.469355864558E-02 |
|  | $V_z$ | -6.554198841518E-03 |  | -6.554198841518E-03 |
| S/C2 Position (AU) | X | -9.433977273640E-01 | = | -9.433977273640E-01 |
|  | Y | 3.062344469040E-01 |  | 3.062344469040E-01 |
|  | Z | 1.411270887844E-01 |  | 1.411270887844E-01 |
| S/C2 Velocity (AU/day) | $V_x$ | -5.861017364349E-03 | = | -5.861017364349E-03 |
|  | $V_y$ | -1.480919323217E-02 |  | -1.480919323217E-02 |
|  | $V_z$ | -6.298978166673E-03 |  | -6.298978166673E-03 |
| S/C3 Position (AU) | X | -9.328957809408E-01 | = | -9.328957809408E-01 |
|  | Y | 3.130424089270E-01 |  | 3.130424089270E-01 |
|  | Z | 1.255542247698E-01 |  | 1.255542247698E-01 |
| S/C3 Velocity (AU/day) | $V_x$ | -5.949486480991E-03 | = | -5.949486480991E-03 |
|  | $V_y$ | -1.492350292755E-02 |  | -1.492350292755E-02 |
|  | $V_z$ | -6.408454202380E-03 |  | -6.408454202380E-03 |

The circular orbits of spacecraft near L3, L4, and L5 are stable or quasi-stable within 20 years (their orbits relative to their respective Lagrange points are also stable or quasi-stable, with deviations from circular orbits of the order O($e^2$) AU). The deviation of the spacecraft triangle from an equilateral triangle is of the order O($e^2$) in terms of arm length. For non-precession configurations, angular resolution faces an issue of image ambiguity. To address this problem, it is necessary to tilt the orbits relative to the ecliptic. When the spacecraft orbits have an inclination $\lambda$ (in radians) relative to the ecliptic plane, the change in arm length is approximately O($\lambda^2$). Therefore, to match these two effects (achieving O($10^{-4}$)), $\lambda$ should be on the order of O(1°).

This section derives the inclined analytical orbits of spacecraft in the solar gravitational field and explains angular resolution and how to eliminate image ambiguity. In Section 11, we will use planetary ephemerides to design and optimize orbit configurations and see that perturbations from all planets except Earth are on the order of O($10^{-4}$). Since L3, L4, and L5 are actually stable within 20 years, the Earth's influence has been considered. Therefore, a suitable tilted circular orbit can serve as our base orbit, with deviations from the optimized actual orbit configuration on the order of O($10^{-4}$).

For Super-ASTROD [83], three spacecraft can also be placed near the Sun-Jupiter orbital plane with a small inclination angle at Sun-Jupiter L3, L4, and L5 points, while one or two additional spacecraft have a larger inclination.

For ASTROD-EM [105], three spacecraft will be placed near the Earth-Moon L3, L4, and L5 points. For spacecraft dynamics, we solve the restricted four-body (Earth, Moon, Sun, and spacecraft with negligible gravitational field) problem. The results are not ideal.

In the initial plan, ASTROD-GW's orbit selection is on the ecliptic plane, with an inclination $\lambda = 0$. The angular resolution of gravitational wave sources on the celestial sphere has ecliptic plane image ambiguity. Although resolution is good in most sky areas, it is poor near the ecliptic poles. After 2010, we redesigned ASTROD-GW's basic orbit with a small inclination angle to address these issues, while keeping arm length variations within the range of $O(10^{-4})$ [80, 81].

According to [80, 81], the basic idea is that if ASTROD-GW spacecraft orbits are slightly inclined by an angle $\lambda$, appropriately designed interferometric planes are also tilted by a similar angle. As the ASTROD-GW formation evolves, interferometric planes can be designed to precess on the ecliptic. Thus, there is no ambiguity in the angular position of gravitational wave sources near the polar regions, and angular resolution is not poor near the poles (see the next subsection). There is also a half-year modulation effect when measuring the angular momentum of the Sun and the Galaxy, which is beneficial.

The following reviews the derivation of the formation (configuration) of ASTROD-GW spacecraft orbits tilted relative to the ecliptic plane. First, consider the circular orbits of spacecraft in the Newtonian gravitational-central problem (one-body central problem) in spherical coordinates $(r, \theta, \varphi)$:

$$r = a, \theta = 90°, \varphi = \omega t + \varphi_0, \tag{77}$$

where $a$, $\omega$, and $\varphi_0$ are constants. For the spacecraft considered here, we have $a = 1$ AU, $\omega = 2\pi/T_0$, $T_0 = 1$ sidereal year, and $\varphi_0$ is the initial phase in the coordinates under consideration. (For Super-ASTROD spacecraft, $a = 5$ AU, $\omega = 2\pi/T_0$, $T_0 = 11$ sidereal years; for μAries spacecraft, $a =$ Mars orbit radius, $\omega = 2\pi/T_0$, $T_0 =$ Mars year). The spacecraft orbit in Cartesian coordinates at time $t$ is:

$$x = a \cos\varphi = a \cos(\omega t + \varphi_0); y = a \sin\varphi = a \sin(\omega t + \varphi_0); z = 0. \tag{78}$$

Now, the orbit is transformed into an orbit with inclination angle $\lambda$, and the orbital plane intersects the $xy$ plane (ecliptic plane) at the line $\varphi = \Phi_0$ in the $xy$ plane. The active transformation matrix is

$$\mathbf{R}(\lambda; \Phi_0) = \begin{bmatrix} \cos^2\Phi_0 + \sin^2\Phi_0 \cos\lambda & \sin\Phi_0 \cos\Phi_0 (1-\cos\lambda) & \sin\Phi_0 \sin\lambda \\ \sin\Phi_0 \cos\Phi_0 (1-\cos\lambda) & \sin^2\Phi_0 + \cos^2\Phi_0 \cos\lambda & -\cos\Phi_0 \sin\lambda \\ -\sin\Phi_0 \sin\lambda & \cos\Phi_0 \sin\lambda & \cos\lambda \end{bmatrix}. \tag{79}$$

Th new spacecraft orbit is

$$\begin{pmatrix} x' \\ y' \\ z' \end{pmatrix} = \begin{pmatrix} a[1-\sin^2\Phi_0(1-\cos\lambda)]\cos\varphi + a\sin\Phi_0\cos\Phi_0(1-\cos\lambda)\sin\varphi \\ a\cos\Phi_0\sin\Phi_0(1-\cos\lambda)\cos\varphi + a[1-\cos^2\Phi_0(1-\cos\lambda)]\sin\varphi \\ -a\sin\Phi_0\sin\lambda\cos\varphi + a\cos\Phi_0\sin\lambda\sin\varphi \end{pmatrix}. \tag{80}$$

For the three orbits with an inclination angle of $\lambda$ (in radians), choose:

S/C I: $\Phi_0(I) = 270°$, $\varphi_0(I) = 0°$;

S/C II: $\Phi_0(\text{II}) = 150°$, $\varphi_0(\text{II}) = 120°$;

S/C III: $\Phi_0(\text{III}) = 30°$, $\varphi_0(\text{III}) = 240°$. (81)

Define

$$\xi \equiv 1 - \cos\lambda = 0.5\lambda^2 + O(\lambda^4), \tag{82}$$

From formulas (80) and (81), we obtain

(i) S/C I orbit

$$\begin{Bmatrix} x^{\text{I}} \\ y^{\text{I}} \\ z^{\text{I}} \end{Bmatrix} = \begin{Bmatrix} a\cos\omega t - \xi a\cos\omega t \\ a\sin\omega t \\ a\cos\omega t \sin\lambda \end{Bmatrix}. \tag{83}$$

(ii) S/C II orbit

$$\begin{Bmatrix} x^{\text{II}} \\ y^{\text{II}} \\ z^{\text{II}} \end{Bmatrix} = \begin{pmatrix} a[(-1/2)\cos\omega t - (\sqrt{3}/2)\sin\omega t] + (a/2)\xi[(3^{1/2}/2)\sin\omega t - (1/2)\cos\omega t] \\ a[(-1/2)\sin\omega t + (3^{1/2}/2)\cos\omega t] + (3^{1/2}/2)a\xi[(3^{1/2}/2)\sin\omega t - (1/2)\cos\omega t] \\ a\sin\lambda[(3^{1/2}/2)\sin\omega t - (1/2)\cos\omega t] \end{pmatrix}. \tag{84}$$

(iii) S/C III orbit

$$\begin{Bmatrix} x^{\text{III}} \\ y^{\text{III}} \\ z^{\text{III}} \end{Bmatrix} = \begin{pmatrix} a[(-1/2)\cos\omega t + (3^{1/2}/2)\sin\omega t] + (a/2)\xi[(3^{1/2}/2)\sin\omega t - (1/2)\cos\omega t] \\ a[(-1/2)\sin\omega t - (3^{1/2}/2)\cos\omega t] - (3^{1/2}/2)a\xi[(-3^{1/2}/2)\sin\omega t - (1/2)\cos\omega t] \\ a\sin\lambda[(-3^{1/2}/2)\sin\omega t - (1/2)\cos\omega t] \end{pmatrix}. \tag{85}$$

One can easily check that $[(x^{\text{I}})^2 + (y^{\text{I}})^2 + (z^{\text{I}})^2]^{1/2} = [(x^{\text{II}})^2 + (y^{\text{II}})^2 + (z^{\text{II}})^2]^{1/2} = [(x^{\text{III}})^2 + (y^{\text{III}})^2 + (z^{\text{III}})^2]^{1/2} = a$ hold fpr consistency.

Calculate the arm vectors $\mathbf{V}_{\text{II-I}} = \mathbf{r}^{\text{II}} - \mathbf{r}^{\text{I}}$, $\mathbf{V}_{\text{III-II}} = \mathbf{r}^{\text{III}} - \mathbf{r}^{\text{II}}$ 和 $\mathbf{V}_{\text{I-III}} = \mathbf{r}^{\text{I}} - \mathbf{r}^{\text{III}}$:

$$\mathbf{V}_{\text{II-I}} = \begin{pmatrix} a[-(3/2)\cos\omega t - (3^{1/2}/2)\sin\omega t] + a\xi[(3^{1/2}/4)\sin\omega t + (3/4)\cos\omega t] \\ a[-(3/2)\sin\omega t + (3^{1/2}/2)\cos\omega t] + a\xi[(3/4)\sin\omega t - (3^{1/2}/4)\cos\omega t] \\ a\sin\lambda[(3^{1/2}/2)\sin\omega t - (3/2)\cos\omega t] \end{pmatrix}, \tag{86}$$

$$\mathbf{V}_{\text{III-II}} = \begin{pmatrix} 3^{1/2}a\sin\omega t - (3^{1/2}/2)a\xi\sin\omega t \\ -3^{1/2}a\cos\omega t + (3^{1/2}/2)a\xi\cos\omega t \\ -3^{1/2}a\sin\lambda\sin\omega t \end{pmatrix}, \tag{87}$$

$$\mathbf{V}_{\text{I-III}} = \begin{pmatrix} a[(3/2)\cos\omega t - (3^{1/2}/2)\sin\omega t] + a\xi[(3^{1/2}/4)\sin\omega t - (3/4)\cos\omega t] \\ a[(3/2)\sin\omega t + (3^{1/2}/2)\cos\omega t] + a\xi[-(3/4)\sin\omega t - (3^{1/2}/4)\cos\omega t] \\ a\sin\lambda[(3^{1/2}/2)\sin\omega t + (3/2)\cos\omega t] \end{pmatrix}, \tag{88}$$

The closure relation $\mathbf{V}_{\text{II-I}} + \mathbf{V}_{\text{III-II}} + \mathbf{V}_{\text{I-III}} = \mathbf{0}$ is checked. The arm lengths are calculated to be

$|\mathbf{V}_{\text{II-I}}| = 3^{1/2}a[(1-\xi/2)^2 + \sin^2\lambda\sin^2(\omega t - 60°)]^{1/2}$,

$|\mathbf{V}_{\text{III-II}}| = 3^{1/2}a[(1-\xi/2)^2 + \sin^2\lambda\sin^2(\omega t)]^{1/2}$,

$|\mathbf{V}_{\text{I-III}}| = 3^{1/2}a[(1-\xi/2)^2 + \sin^2\lambda\sin^2(\omega t + 60°)]^{1/2}$. (89)

The fractional arm length variation is within $(1/2)\sin^2\lambda$ which is about $10^{-4}$ for $\lambda$ around 1°.

The cross-product vector $\mathbf{N}(t) \equiv \mathbf{V}_{\text{III-II}} \times \mathbf{V}_{\text{I-III}}$ is normal to the orbit configuration plane and has the following components:

$$\mathbf{N} = [(3^{3/2}/2)(1-\xi/2)a^2]\begin{bmatrix}-\sin\lambda\cos2\omega t\\ -\sin\lambda\sin 2\omega t\\ (1-\xi/2)\end{bmatrix}. \tag{90}$$

The normalized unit normal vector $\underline{n}$ is then:

$$\mathbf{n} = [\sin^2\lambda + (1-\xi/2)^2]^{1/2}\begin{pmatrix}-\sin\lambda\cos 2\omega t\\ -\sin\lambda\sin 2\omega t\\ (1-\xi/2)\end{pmatrix}. \tag{91}$$

The geometric center $\mathbf{V}_c$ of the ASTROD-GW spacecraft configuration is

$$\mathbf{V}_c = \begin{pmatrix}-(1/2)\xi a\cos\omega t\\ (1/2)\xi a\sin\omega t\\ 0\end{pmatrix}. \tag{92}$$

ASTROD-GW configuration consists of 3 sets of 2-arm interferometers. The geometric centers of these 3 interferometers are approximately 0.25 astronomical units from the Sun. Reference [82] considers planetary perturbations and uses planetary ephemerides to numerically simulate and optimize orbital configurations with inclinations of 0.5°, 1°, 1.5°, 2°, 2.5°, and 3°. In Section 11.3, we use the 1° inclination case as an example to explain. When the LISA configuration orbits the Sun, it resembles multiple detector arrays distributed along a 1 AU orbit, suitable for long-term observation of gravitational wave sources. Although the ASTROD-GW configuration extends up to 1.73 AU, its operation around the Sun also approximates multiple detector arrays on a 1 AU orbit.

**10.4 Angular Resolution**

Considering coherent gravitational wave sources' angular resolution. Taking the LISA mission as an example, LISA's detector configuration modulates azimuthal angles by $2\pi$ radians and inclination angles by 1.05 radians (60°) due to its orbital motion around the Sun, causing the antenna (for plane waves) response pattern to sweep across the sky once per year. The antenna response is not isotropic, but for single-frequency gravitational wave sources, LISA's average linear angular resolution over a year differs by less than a factor of 3 across all directions [45]. This holds true for all LISA-like gravitational wave detectors. Angular resolution is essentially inversely proportional to the strain signal-to-noise ratio. If LISA's inclination is 0.017-0.052 radians (1-3°), the angular resolution degrades by a factor of 30-10 (approximately the ratio of 1.05 radians sine to 0.03-0.1 radians sine); solid angle localization in the celestial sphere deteriorates by the square of this factor. Away from the polar regions ($\theta \gg 0.017$-0.052 radians), the spherical degree localization in the celestial sphere becomes $\sin^2\theta$. If the signal-to-noise ratio decreases by a factor of $p$, the linear angular resolution worsens by $p$ times. Compared to LISA, ASTROD-GW has lower sensitivity above 1 mHz, which worsens angular resolution due to these factors. In the 100 nHz-1 mHz range, ASTROD-GW has better sensitivity than LISA by a factor of 104. Hence, polar resolution is similar to LISA's for ASTROD-GW, while in other sky regions, linear resolution is approximately increased by $104 \times \sin\theta$ (increasing by 104 times, but solid angle decreasing $\sin^2\theta$ [linear angular resolution decreases sin θ]). Although there is a slight dependence on the configuration inclination λ, beyond a factor of 3, ASTROD-GW's average antenna pattern is 52 × sin θ times better than LISA's in regions away from the poles. From Equation (91), the average time required is half a year rather than a full year [80, 81].

For more complex sources, such as chirping gravitational waves from binary black holes, fitting is required for accuracy in parameter estimation. However, the trend in parameter accuracy remains the same: it is inversely proportional to the strain signal-to-noise ratio for similar cases.

For Super-ASTROD, which has a strain signal-to-noise ratio five times better than ASTROD-GW in the low-frequency region, the angular resolution is also five times better. For polar resolution, ASTROD inclination strategy can be applied. However, due to Super-ASTROD's one or two S/C off the ecliptic, this may not be necessary. For ASTROD-EM, with a lunar orbit inclined about 5° to the ecliptic, and its intersection moving period is 18.61 years, the Lagrange point of the Earth-Moon also moves. Depending on the mission's time and duration, a slightly inclined orbit may not be necessary [105].

Most Earth-orbiting gravitational wave missions suffer from dipole smearing and poor polar resolution. However, this is not a major issue as we only need to consider two polarities to identify electromagnetic counterparts, and the polar regions are only a small part of the sky.

**10.5 Six/Twelve Spacecraft Formations**

To detect relic gravitational waves more sensitively using correlated detection, the proposals for Big Bang Observer [85] and DECIGO [84] suggest placing 12 spacecraft in Earth orbit, divided into 3 groups spaced 120 degrees apart on the orbit. Two groups have 3 spacecraft each arranged in a triangular configuration similar to LISA, and the third group has 6 spacecraft arranged in two LISA-like triangles forming a star-shaped structure (Figure 6). Another configuration is to have each group with 4 spacecraft forming an approximately square configuration (also tilted relative to the ecliptic by 60 °) (Figure 6).

To detect background or relic gravitational waves more sensitively, two triangular ASTROD-GW arrays (i.e., 6-S/C constellation) are needed for correlated detection. The second near-triangular formation can again be placed near L3, L4, and L5, but with corresponding S/C separated from the first formation by $1 \times 10^6$ km to $5 \times 10^6$ km [81].

**11 Orbit Design and Optimization Using Ephemerides**

Despite the Sun's dominance in the solar system, other planets and celestial bodies influence spacecraft orbits, especially Jupiter, Venus, and Earth. Ephemerides are essential for orbit design. Currently, there are three complete fundamental ephemerides for the solar system: DE ephemerides (Development Ephemerides) [24], EPM ephemerides (Ephemerides of Planets and Moon) [25], and INPOP ephemerides (Intégrateur Numérique Planétaire de l'Observatoire de Paris) [26]. Any of these ephemerides can be used for orbit design and optimization. For ease of numerical handling, we usually use the CGC (Center for Gravitation and Cosmology) ephemeris framework, combined with initial conditions obtained from the DE ephemerides at a certain epoch, and subsequently evolve using post-Newtonian approximations.

**11.1 CGC Ephemerides**

In 1998, simulations and parameter determinations for ASTROD were started [335, 336]. We compiled a post-Newtonian ephemeris of the solar system, including the Sun's quadrupole moment, the eight planets, Pluto, the Moon, and the three largest asteroids. We refer to this working ephemeris as CGC 1. We used this ephemeris as a base and added random terms to simulate noise, generating simulated range data. For the ASTROD mission, we used Kalman filtering to determine the fitting accuracy of relativistic gravitational parameters and solar system parameters over 1050 days.

To better assess the measurement of $\dot{G}/G$ accuracy, we also need to monitor the masses of other asteroids. For this, we considered all known 492 asteroids with diameters greater than 65 km to obtain an improved ephemeris framework—CGC 2—and calculated the disturbances these 492 asteroids cause to the ASTROD spacecraft [337, 338].

In constructing the CGC ephemeris framework, we used the parametrized post-Newtonian (PPN) framework from Brumberg's book [339], with parameters $\beta$ and $\gamma$ parameterizing the post-Newtonian framework and its solar system barycenter metric and dynamic equations (where gauge parameter $\alpha$ is set to zero). These equations were used to construct numerical ephemerides for the eight planets, Pluto, the Moon, and the Sun (PPN parameters $\gamma = \beta = 1$, Sun quadrupole moment parameter $J_2 = 2 \times 10^{-7}$). Initial positions and velocities at 0:00 on June 10, 2005 were taken from the DE403 ephemeris. Evolution was solved using a 4th-order Runge-Kutta method with a step size $h = 0.01$ days. In reference [336], the evolution of 11 bodies was expanded to 14 bodies, including the three largest asteroids Vesta, Juno, and Ceres (CGC 1 ephemeris). Due to the Sun's quadrupole axis being slightly inclined to the ecliptic plane (7°), this tilt was neglected in the CGC 1 ephemeris. In CGC 2 ephemeris, disturbances from an additional 489 asteroids were added.

In the first optimization of the ASTROD-GW orbit [340-342], we used the CGC 2.5 ephemeris, considering only the three largest asteroids but adding Earth's precession and nutation; also considering solar and Earth's second harmonic to fourth harmonic. In subsequent simulations, we added perturbations from an additional 349 asteroids, termed CGC 2.7 ephemeris [289-291]. The differences in Earth's orbit evolution compared to DE405 within 3700 days from JD2461944.0 (June 21, 2028) are shown in figure 5 of reference [81]. The difference in Earth-Sun radial distance is less than approximately 200 m. Differences in latitude and longitude of the Earth are less than 1 mas.

**11.2 Numerical Orbit Design and Orbit Optimization for LISA and Taiji**

The mission orbit formation configuration for LISA is shown in Figure 3. Initially, the arm length was chosen to be 5 Gm, but due to possible budget constraints, it was later reduced to 1 Gm. After the LISA Pathfinder met all requirements, it was changed to 2.5 Gm, with the project being officially approved on January 25th of this year. The arm lengths of the three spacecraft must be kept as equal as possible, and the relative Doppler velocities between any two spacecraft must be less than 10 m/s. Various previous works have analyzed and numerically studied different arm length configurations for LISA [327-332]. For LISA-like space gravitational wave detection missions, we follow Dhurandhar et

al.'s analysis procedure [332] in Section 10.2 for initial selection. Through this orbital selection, numerical orbits are computed and optimized using CGC ephemerides, similar to what was done for the ASTROD-GW orbit design (see the next subsection). The final optimized initial conditions for LISA are listed in Table 3 of Section 10.2, column 3. The final optimized initial conditions for Taiji are listed in Table 4 of Section 10.2, column 5; Figure 29 illustrates changes in the spacecraft configuration within 2200 days, including arm length, line-of-sight velocities, angle between the two arms, and the angle between the spacecraft and the Sun.

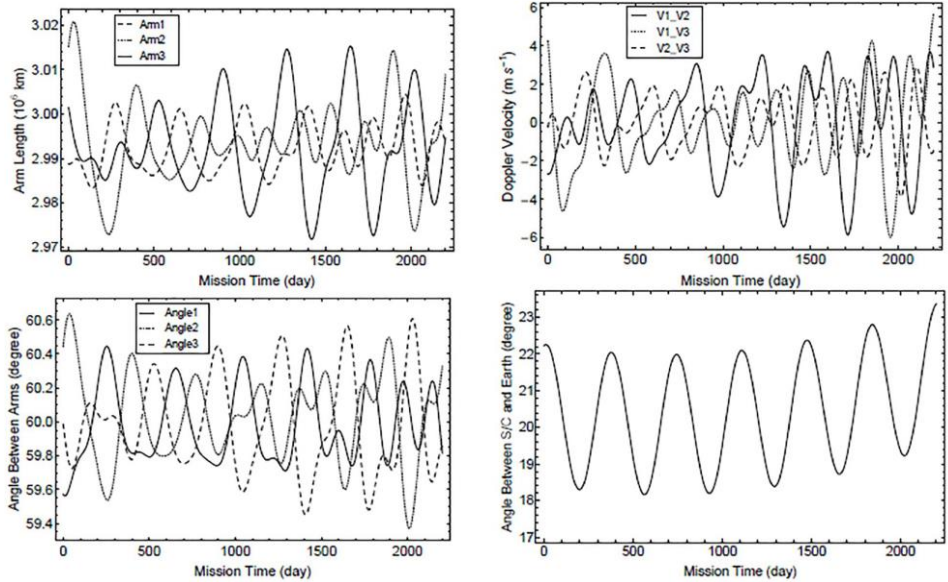

Figure 29 Variations of the arm lengths, velocities in the line-of-sight direction, formation angles and angle between barycenter of the S/Cs and Earth in 2200 d for the Taiji S/C configuration with initial conditions given in column 5 (after optimization) of Table 9.

**11.3 Orbit Optimization of ASTROD-GW**

The goal of orbit optimization for the ASTROD-GW mission is to equalize the three arm lengths of the ASTROD-GW formation as much as possible and reduce the relative line-of-sight velocities between the three pairs of spacecraft. In our initial optimization, the start time for the scientific phase of the mission was chosen as noon on June 21, 2025 (JD2460848.0), and optimization was carried out using CGC 2.5 ephemerides for 3700 days [340-342]. The preparation time for the mission may be longer, with the possibility of extending the mission to exceed 10 years in duration. In subsequent optimizations [290, 291, 343], we began optimization from noon on June 21, 2028 (JD2461944.0) using CGC 2.7 ephemerides for 20 years, which includes more asteroids than CGC 2.5. In both optimizations, the orbital configurations are set within the ecliptic plane with an inclination angle $\lambda = 0$. As the basic configuration of ASTROD-GW transitions to inclined precession orbits, we have redesigned and optimized our orbital configurations. The orbital configurations, starting from noon on June 21, 2035 (JD2464500.0), are worked out using CGC 2.7.1 for 10 years, with inclinations of 0.5°, 1°, 1.5°, 2°, 2.5°, and 3° [82].

In this section, we outline the design and optimization methods for the precession orbit formation (configuration) according to [82], using CGC 2.7.1 with an inclination of 1° as an example. The differences between CGC 2.7.1 and CGC 2.7 are detailed in Section 11.3.1. In Section 11.3.2, we review how the initial conditions of the spacecraft are selected as the starting point for numerical optimization. In Section 11.3.3, we discuss optimization methods, present optimized results in tabular form, and illustrate the characteristics of the orbits after optimization.

**11.3.1 CGC 2.7.1 Ephemerides**

In the CGC 2.7.1 ephemerides framework, in addition to Vesta, Ceres, and Pallas, we selected 340 asteroids from the Lowell database. The masses of these 340 asteroids are provided by Lowell data [344] rather than estimated based on classification as in CGC 2.7 [288, 289, 291]. The orbital elements of these asteroids are also updated from the Lowell database.

Over the period of 10 years starting from June 21, 2035, the differences in Earth's heliocentric distance between CGC 2.7.1 and DE430 calculations are within 150 m, while differences in longitude and latitude are within 1.4 mas and 0.45 mas, respectively. These differences do not affect our TDI computation results.

**11.3.2 Preliminary Selection of Spacecraft Initial Conditions**

The right ascension of the Earth, R.A., at JD2464500 (June 21, 2035, 12:00:00) according to DE430 ephemerides is 17h57m45.09s, or 269.438°. The initial positions of the three spacecraft are derived by selecting $\omega t$ as 89.44° in Equation (81) (where $\varphi = \omega t + \varphi_0$), and velocities are obtained by computing the derivative of $t$ from Equation (80). S/C1 near the Lagrange point L3 is partially shadowed by the Sun within the Earth's line of sight (Figure 30, upper image), which obstructs communication with ground stations. To avoid this shadowing, we rotate forward the initial angles $\Phi_0$ and $\varphi_0$ for this instance (with an inclination of 1.0°) by 2.0°. The orbit of S/C1 is depicted in the lower image of Figure 30. The initial selections of the initial states of the three S/Cs under this scenario are listed in Table 10, column 3.

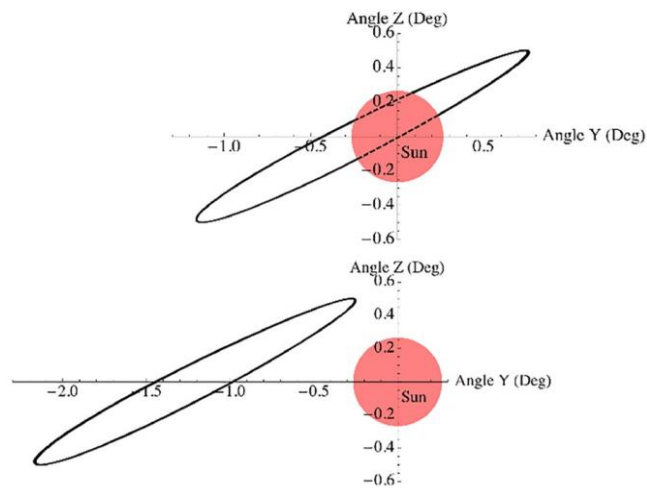

Figure 30 (Color online) S/C1 view from Earth before rotating the initial conditions by an angle (upper diagram) and after rotating by an angle 2.0° (lower diagram) for the case of inclination angle 1.0° (The rotation of 2.0° of S/C1 in orbit is equivalent to a rotation of 1.0° viewed from Earth). The figure is from ref. [82].

**Table 10** Initial states of S/Cs for the configuration with an inclination angle of 1° at epoch JD2464500.0 for initial choice, after period optimization, and after all optimizations in J2000 equatorial solar-system-barycentric coordinate system [82]

| $\lambda=1.0°$ | | Initial choice of S/C initial states | Initial states of S/Cs after period optimization | Initial states of S/C after final optimization |
|---|---|---|---|---|
| S/C1 Position (AU) | X | −2.8842263289715×10⁻² | −2.8842263289715×10⁻² | −2.8842514605546×10⁻² |
| | Y | 9.1157742309044×10⁻¹ | 9.1157742309044×10⁻¹ | 9.1158659433458×10⁻¹ |
| | Z | 3.9552690922456×10⁻¹ | 3.9552690922456×10⁻¹ | 3.9553088730467×10⁻¹ |
| S/C1 Velocity (AU/day) | $V_x$ | −1.7188548244458×10⁻² | −1.7188535691176×10⁻² | −1.7188363750567×10⁻² |
| | $V_y$ | −2.8220395391983×10⁻⁴ | −2.8220375159556×10⁻⁴ | −2.8220098038726×10⁻⁴ |
| | $V_z$ | −4.4970276654173×10⁻⁴ | −4.4970243993363×10⁻⁴ | −4.4969796642665×10⁻⁴ |
| S/C2 Position (AU) | X | 8.7453598387569×10⁻¹ | 8.7453598387569×10⁻¹ | 8.7453598387569×10⁻¹ |
| | Y | −4.3802677355114×10⁻¹ | −4.3802677355114×10⁻¹ | −4.3802677355114×10⁻¹ |
| | Z | −2.0634980179207×10⁻¹ | −2.0634980179207×10⁻¹ | −2.0634980179207×10⁻¹ |
| S/C2 Velocity (AU/day) | $V_x$ | 8.2301784322477×10⁻³ | 8.2301033726700×10⁻³ | 8.2301033726700×10⁻³ |
| | $V_y$ | 1.3797379424198×10⁻² | 1.3797253460590×10⁻² | 1.3797253460590×10⁻² |
| | $V_z$ | 6.1425805519808×10⁻³ | 6.1425244722884×10⁻³ | 6.1425244722884×10⁻³ |
| S/C3 Position (AU) | X | −8.5683596527799×10⁻¹ | −8.5683596527799×10⁻¹ | −8.5679330969623×10⁻¹ |
| | Y | −4.8998222347472×10⁻¹ | −4.8998222347472×10⁻¹ | −4.8995800210059×10⁻¹ |
| | Z | −1.9592963105165×10⁻¹ | −1.9592963105165×10⁻¹ | −1.9591994878015×10⁻¹ |
| S/C3 Velocity (AU/day) | $V_x$ | 8.9788714330506×10⁻³ | 8.9787977300014×10⁻³ | 8.9792464008067×10⁻³ |
| | $V_y$ | −1.3530263187520×10⁻² | −1.3530152097744×10⁻² | −1.3530828362023×10⁻² |
| | $V_z$ | −5.6998631854817×10⁻³ | −5.6998163886731×10⁻³ | −5.7001012664635×10⁻³ |

**11.3.3 Optimization Methods and Post-Optimization Characteristics**

Our optimization approach involves modifying initial velocities and heliocentric distances to achieve (i) as balanced as possible arm lengths of the ASTROD-GW formation and (ii) reducing the relative Doppler velocities between the three pairs of spacecraft as much as possible.

During the actual optimization process, we use the following equation to modify the orbital mean period:

$$V_{new} = V_{prev} + \Delta V \approx [1 - (1/3)(\Delta T/T)] V_{prev}. \tag{93}$$

For the case with an inclination of 1°, we calculate the initial selections of the three spacecraft orbits listed in Table 10, column 3, using CGC 2.7.1 ephemerides. The 10-year mean periods for the three S/Cs are 365.256 days (S/C1), 365.267 days (S/C2), and 365.266 days (S/C3). We adjust the initial velocities using Equation (93) to set the mean periods of S/C1, S/C2, and S/C3 to 365.255 days, 365.257 days, and 365.257 days, respectively. The initial conditions after this step are listed in Table 10, column 4. In the next step, we adjust the spacecraft eccentricity to near-circular using the following equations:

$$R_{new} = R_{prev} + \Delta R \approx (1 - \Delta R/R) R_{prev}, \quad V_{new} = V_{prev} + \Delta V \approx (1 - \Delta R/R) V_{prev}. \tag{94}$$

Here, $R$ is the spacecraft's initial heliocentric distance. Adjusting $\pm(\Delta R/R)$ modifies the eccentricity without altering the orbital period. Through iterative optimization, the final optimized initial conditions are listed in Table 10, column 5. Figure 31 illustrates the initial and final optimized configuration of the spacecraft with a 1° inclination over 10 years, showing (a) arm lengths, (b) differences in arm lengths, (c) formation angles, (d) line-of-sight velocities, (e) inclination of the formation plane normal relative to the ecliptic plane normal, and (f) direction angles of the formation plane normal.

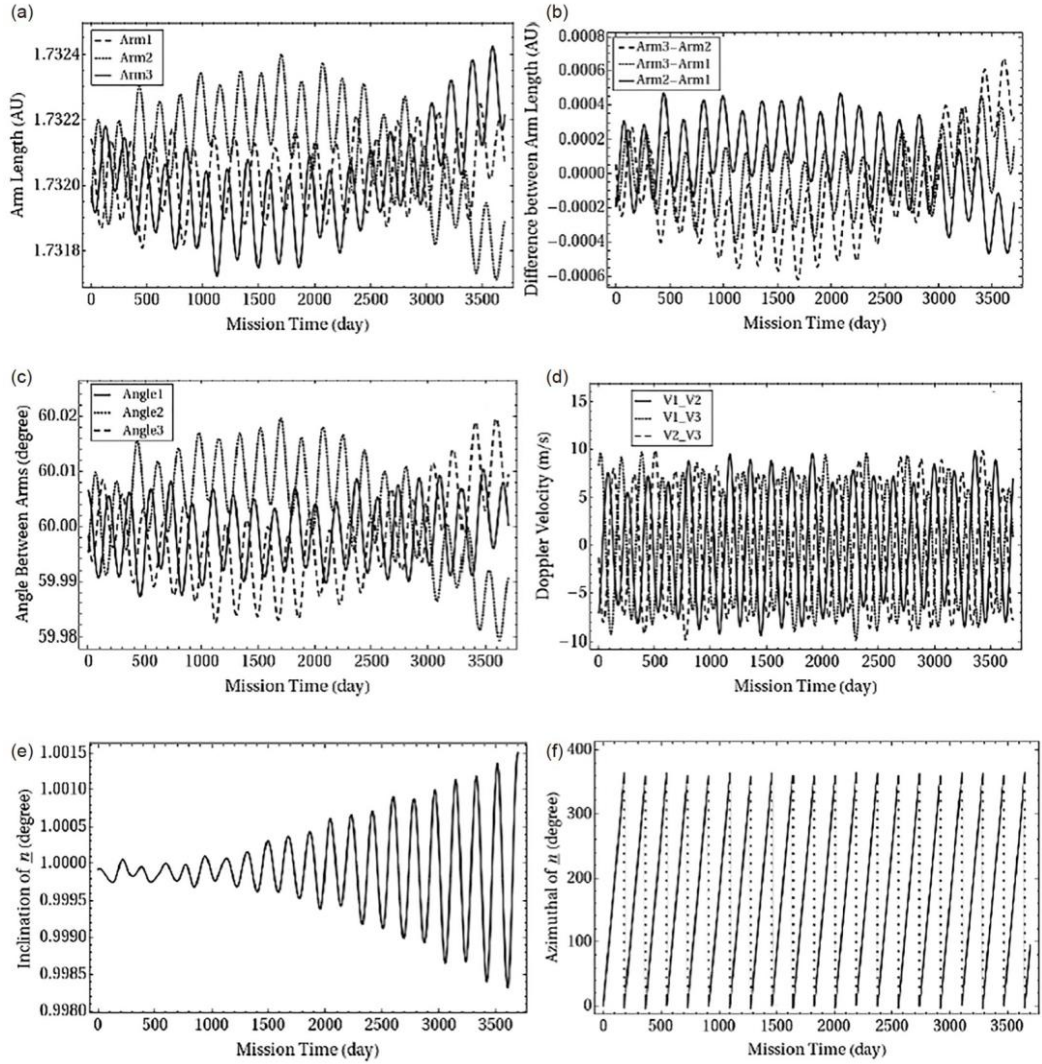

Figure 31 The variation of armlengths (a), difference of armlengths (b), angles between arms (c), velocities in the measure direction (d), inclination of the unit normal **n** of the ASTROD-GW formation (e) and azimuthal angle of **n** (f) in 10 years for the case with the nominal formation inclination angle 1°.

For inclinations of 0.0°, 0.5°, 1.5°, 2°, 2.5°, and 3°, the optimization process is similar to that of 1.0°, and the results can be found in [82].

## 12 Deployment of Earth-like Solar Orbit Formations

Extensive studies have been conducted on the deployment of spacecraft to Earth orbit, Earth-Moon Lagrange point halo orbits, and Sun-Earth L1 and L2 Lagrange point halo orbits. Here, we briefly discuss the methods for deploying spacecraft to various positions in Earth-like solar orbits. References [81, 345] provide preliminary designs for the transfer orbits of spacecraft from separation from the launch vehicle to near the L3, L4, and L5 points of the mission orbits.

In the mission study of ASTROD I, the spacecraft obtains suitable delta-V before the final stage of separation from the launcher in low Earth orbit (LEO) to enter solar orbits via a ballistic (geodesic) Venus flyby. Employing a similar strategy, ASTROD-GW spacecraft can be directly launched into a designated

Hohmann transfer orbit or near Venus flyby orbit. Thus, the major delta-V required for each spacecraft to reach its destination occurs near the destination, propelling the spacecraft to stay near the designated Lagrange points (or other Earth-centric solar orbit points). In Table 11, rows 2-4, we list the types of transfer orbits, transfer times, delta-V values for transfers, and specific impulse mass ratios for the three ASTROD-GW S/C. These estimates apply to any mission spacecraft deployed to the same positions.

**Table 11**. Estimated Delta-V and Propellant Mass Ratio for Solar transfer of S/C

| Angle ahead of Earth in solar orbit | Transfer Orbit | Transfer Time | Solar Transfer Delta-V after injection from LEO to solar transfer orbit | Solar Transfer Propellant Mass Ratio (Isp=320 s) |
|---|---|---|---|---|
| 180° (near L3) | Venus flyby transfer | 1.3-1.5 yr | 2.2-2.5 km/s | 0.50-0.55 |
| 60° (near L4) | Inner Hohmann, 2 Revolutions | 1.833 yr | 1.028 km/s | 0.280 |
| 300° (− 60°) (near L5) | Outer Hohmann, 1 Revolutions | 1.167 yr | 2 km/s | 0.47 |
| 0 - 60° | Inner Hohmann, ≤ 2 Revolutions | Less than 1.833 yr | Less than 1.028 km/s | Larger than 0.280 |
| 60° - 300° | Venus flyby transfer | 1.3-1.5 yr | 2.2-2.5 km/s | 0.50-0.55 |
| 300° - 360° | Outer Hohmann, 1 Revolutions | 1.167 yr | 2 km/s | 0.47 |

The specific impulse mass ratios for spacecraft 1, 2, and 3 are approximately 0.5-0.55, 0.280, and 0.47, respectively. Relative to 500 kg dry mass, the total masses of the three ASTROD-GW spacecraft at different positions in the solar orbit range from 1111-1266 kg, 723 kg, and 1035 kg (including propellant and propulsion modules with a propellant mass ratio of 10% of dry mass).

Estimations have been made for deploying spacecraft to other positions in solar orbits, listed in Table 11, rows 5-7. The basic setup includes: (i) spacecraft propelled by efficient propulsion modules (with a specific impulse Isp of 320s) and a propulsion module mass ratio of 10% of total propellant mass, suitable for high delta-V operations upon reaching the destination; (ii) separation of the propulsion module upon achieving the target state.

Further optimization studies are underway for the deployment of formations with a 20-year mission duration and inclinations relative to the ecliptic plane similar to LISA and ASTROD-GW missions, following separation from the launcher [346].

**13 Time Delay Interferometry (TDI)**

In spacecraft formation flight without drag, maintaining precise formation is challenging, and laser frequency stability is limited. As analyzed in Section 6, the noise in interferometric signals scales with laser frequency noise and path differences. To meet noise requirements, selecting appropriate paths for time delay interferometry (TDI) becomes crucial. Further discussions on first and second-generation TDI are provided in Section 7.2. Section 8 reviews the sensitivity of TDI to gravitational waves. Based on precise orbit calculations and optimizations from the previous section, variations in arm lengths are estimated to be tens of thousands of kilometers for LISA and Taiji, and up to hundreds of thousands of kilometers for ASTROD-GW. TDI indeed presents a natural solution. Another approach is using equilateral configurations like DECIGO and B-DECIGO, necessitating methods to maintain propulsion acceleration without affecting accelerometer sensitivity, see references [347] for more details.

From measurements to data suitable for theoretical work, it is divided into the following three steps according to reference [51]: L0: Raw data processed onboard the spacecraft. L0.5: Processed on the ground and publicly available. L1: Processed from L0.5 data into time delay interferometry measurement data, used for scientific purposes, forming L2 individual gravitational wave source waveforms, signals, and their parameters, and can be compiled into an L3 catalog.

When tracking lasers, pseudo-random codes [348] are commonly used. In recent analyses, reference [348] demonstrated the use of data structure and less stringent observational constraints/analysis of Taiji simulated data for distance measurement, successfully replacing pseudo-random code methods. This means that even without using pseudo-random codes, it is possible to extract and track from data structures without strict timing requirements or the need for separate timing if there is strong computing power. This implies that interference signals are embedded in the data stream already.

There are still many areas for further research regarding TDI configurations and data structures, such as optimizing configurations/combinations for higher frequencies [349,350].

**14 Payload**

Space-based gravitational wave detection primarily measures changes in the distance between two S/C (or celestial bodies) as gravitational waves pass through. The two S/C (or celestial bodies) must undergo geodesic motion (or such motion can be inferred). Due to the weakness of gravitational waves, distance measurements must be super-sensitive. Typical implementations (missions) consist of three spacecraft forming an almost equilateral triangle formation. Three spacecraft use interferometry to measure distances among each other. Each spacecraft carries a payload consisting of two mass blocks, two telescopes, two lasers, a weak light detection and processing system, a laser stabilization system, and a drag-free system. Precision/optical clocks or absolute laser stabilization systems and absolute laser metrology systems can be used for the lower-frequency part of the space gravitational wave band or for higher accuracy.

*Weak light phase locking and processing*: This is crucial for solar orbit missions. For ASTROD-GW, located at a distance of 260 Gm (1.73 AU), and other microhertz gravitational wave detection space missions, it is necessary to lock local lasers to incoming light at 100 fW for amplification and manipulation. For 100 fW ($\lambda = 1064$ nm) weak light, there are $5 \times 10^5$ photons per second. This is suitable for 100 kHz frequency tuning. For LISA, 85 pW weak light locking is required, which has been successfully demonstrated (Section 7.2). Sambridge et al. [287] were able to phase track sub-fW lasers with average slip times exceeding 1000 seconds, a major breakthrough. Frequency tracking, modulation/demodulation, and encoding are the key focuses, making it a mature experimental technique that will benefit future microhertz gravitational wave detection. This is also crucial for deep space optical communication.

*Drag-free system design and development*: The drag-free system consists of high-precision accelerometers/inertial sensors for detecting non-drag-free motion and micropropulsion systems for feedback to maintain spacecraft drag-free. LISA Pathfinder successfully demonstrated drag-free technology in the frequency range above 100 μHz, meeting not only the requirements of LISA Pathfinder but also those of LISA. This success paved the way for all proposed space missions in 2017 (Table 2). However, more work is needed for the lower frequency range (100 nHz to 100 μHz). We discussed frequency sensitivity spectra and red factors in Section 6. To suppress red factors, position sensor noise must be flat to 100 nHz, requiring small gravitational accelerations generated by the spacecraft on the test masses and capable of modeling them to required levels for low frequencies. The self-gravity acceleration needs to be stabilized or measured in real time. For this, absolute laser metrology systems can be used to monitor the position of major mass distributions in S/C. To completely eliminate or exceed this factor, optical sensing and optical feedback control may be required. A laser metrology system based on absolute laser measurements is proposed to reduce noise for ASTROD's accelerometer/inertial sensor design, especially in the low-frequency range. In addition, ASTROD also needs to monitor the positions of spacecraft components for easy modeling [56,57].

*Micropropulsion systems*: Micropropulsion is needed for drag-free feedback control. Field emission electric propulsion (FEEP) systems have high thrust-to-weight ratios and are good candidates for micropropulsion systems. FEEP systems exhibit good sensitivity in the μN range. The main issue with FEEP is its lifespan. Cold gas thrusters have emerged as an alternative due to technical issues encountered in FEEP technology development. The GAIA mission carries cold gas thrusters used for AOCS (Attitude and Orbit Control System) [351]. MICROSCOPE [352] (https://microscope.cnes.fr/en/MICROSCOPE/GP_mission.htm) and LISA Pathfinder are equipped with cold gas thrusters based on GAIA technology. Compared to FEEP, the main drawback of cold gas thrusters is the higher mass required per delta-V. The total mission duration is limited by the quantity of propellant stored in tanks. Therefore, if FEEP technology can be used later, it would be the preferred choice.

*Laser system*: Nd lasers with non-planar ring oscillators pumped by laser diodes are available, outputting 2 W of power. Frequency noise must be suppressed to 30 Hz/Hz$^{-1/2}$ in the relevant frequency range. The laser is the source of measurement, and a more stable laser system can simplify and improve the measurement and analysis processes, aiding in achieving higher precision. Recently, commercialized optical clocks with an accuracy of $10^{-15}$ have become available [353-355]. Developing space laser systems with frequency noise of 3 Hz/Hz$^{-1/2}$ or 0.3 Hz//Hz$^{-1/2}$ should be explored experimentally.

*Laser frequency standards/clocks*: Space optical clocks and optical frequency comb synthesizer technology are essential for achieving and simplifying the sensitivity goals of low-frequency gravitational wave detection missions. Another use of optical clocks and optical frequency comb synthesizers is calibrating optical metrology for missions like ASTROD-GW. This is crucial for laser metrology inertial sensors and monitoring internal distances within spacecraft, correcting local gravity variations caused by thermal effects, etc. All these measurements use lasers as standards. They require calibration using optical frequency standards or absolute stable laser frequencies based on atomic or molecular lines. The advent of space optical clocks and optical combs may simplify the experimental design of missions like ASTROD-GW.

Currently, laboratory optical clocks [207,209] have achieved fractional errors at the $10^{-19}$ level. Such clock accuracy or higher will be used for extremely sensitive measurements of gravity, motion, and inertial navigation. The use of such clocks will facilitate the detection of microhertz low-frequency gravitational waves and inspire the demand for redesigned low-frequency microhertz space gravitational wave detection schemes. Exploring these schemes in more detail is worthwhile.

*Absolute laser metrology system*: Using ultra precision laser frequency standards/clocks, an absolute laser metrology system can be constructed to monitor the position of various parts of the spacecraft for gravity modeling.

*Radiation Monitor*: A small radiation detector on the spacecraft will monitor the test mass charging of inertial/drag-free sensors. This radiation monitor can also be used to measure solar high-energy particles (SEP) and galactic cosmic rays (GCR) in the fields of solar and galactic physics, and applied to space weather [356, 357].

**15 Summary and Outlook**

The first artificial satellite, Sputnik, was launched in 1957. On December 3, 2015, the gravitational wave technology verification mission, LISA Pathfinder, was launched. This mission successfully tested and demonstrated drag-free technology, meeting the requirements of not only LISA Pathfinder but also those of LISA. This success paved the way for all proposed space gravitational wave missions (Table 2). Space gravitational wave missions are expected to be launched in eleven years. Weak light phase locking has been validated in the laboratory, and weak light technology is rapidly advancing. With the first direct detection of gravitational waves by LIGO and the success of the LISA Pathfinder mission, the future of space gravitational wave detection is bright.

The primary scientific goals of space gravitational wave detectors are to detect (i) gravitational waves from massive black hole mergers; (ii) extreme mass ratio inspirals; (iii) intermediate mass black holes; (iv) galactic compact binaries; (v) relic gravitational wave background; and (vi) to test theories of gravity. From Figures 12-14, it is evident that the signal-to-noise ratio (S/N) for double massive black hole mergers detected by gravitational waves is very high, and lower frequency bands (100 nHz – 100 μHz) are required for detecting larger-scale mergers with high S/N. Therefore, longer arms have an advantage. In exploring the co-evolution of black holes and galaxies, long-arm missions complement PTA effectively. Long-arm missions with better angular resolution are also more effective in determining the state equation of dark energy, testing relativistic gravity, and potentially exploring inflation physics. Deploying S/C to any position on a heliocentric orbit takes less than 1.8 years with a specific impulse mass ratio of less than 0.55, within the practical range achievable by launch vehicles.

Now is the time to seriously consider the second generation of space-based gravitational wave detectors. Laboratory optical clocks have achieved an error level of $10^{-19}$ and continue to improve. This level of precision in clocks will be used for space applications. This advancement benefits the laser pulse ranging schemes for long-arm configurations. The T2L2 on the JASON2 satellite has achieved a laser pulse timing accuracy of 3 picoseconds (ps). 0.9 mm (3 ps) at a distance of 1300 Gm (8.6 AU) corresponds to $7\times10^{-16}$, which is at the level of strain acceleration noise for some lower frequency ranges. Pulse timing accuracy is still being enhanced. Further detailed research into such schemes is worthwhile.

**Acknowledgements** I would like to thank An-Ming Wu and Gang Wang for many helpful discussions during long-time colaborations I would also like to thank Yong Tang for helps in referencing and Chinese nomenclature translation.